\documentclass[review]{elsarticle}
\usepackage{framed}
\usepackage{multicol}
\usepackage{subcaption}
\usepackage{graphicx}
\usepackage{amsmath}
\usepackage{color,soul}

\usepackage{nomencl}
\makenomenclature

\usepackage{lineno,hyperref}
\modulolinenumbers[5]

\usepackage{geometry}
\makeatletter
\def\ps@pprintTitle{%
	\let\@oddhead\@empty
	\let\@evenhead\@empty
	\def\@oddfoot{}%
	\let\@evenfoot\@oddfoot}
\makeatother

\begin{document}

\begin{frontmatter}

\title{1-D semi-analytical modeling and parametric study of a single phase rectangular Coupled Natural Circulation Loop}

\author{Akhil Dass, Sateesh Gedupudi \footnote{Corresponding author. Tel.: +91 44 2257 4721, Email: sateeshg@iitm.ac.in}}
\address{Heat Transfer and Thermal Power Laboratory, Department of Mechanical Engineering, IIT Madras, Chennai 600036, India}

\begin{abstract}
The study of heat exchangers with both the hot and cold fluid sides driven by buoyancy forces is an area of considerable interest due to their inherent passivity and non-existence of moving parts. The current study aims to study such heat exchange devices employing the basic Coupled Natural Circulation Loop (CNCL) systems. A 1-D Fourier series based semi-analytical model of the basic CNCL system is proposed. A 3-D CFD validation is performed to validate the developed 1-D model. The non-dimensional numbers such as Grashof number, Fourier number, Stanton number and Reynolds number, which determine the system behavior are identified and a detailed parametric study is performed. Both vertical and horizontal CNCL systems are considered along with the parallel and counter flow configurations. The heater-cooler location greatly influences the behavior of CNCL system. The vertical CNCL always exhibits counter flow configuration whereas the horizontal CNCL system may exhibit parallel or counter flow arrangement depending on the heater-cooler location and initial flow conditions.
\end{abstract}


\end{frontmatter}


\section{Introduction}

\begin{figure}[!h]
	\centering
	\includegraphics[width=0.8\linewidth]{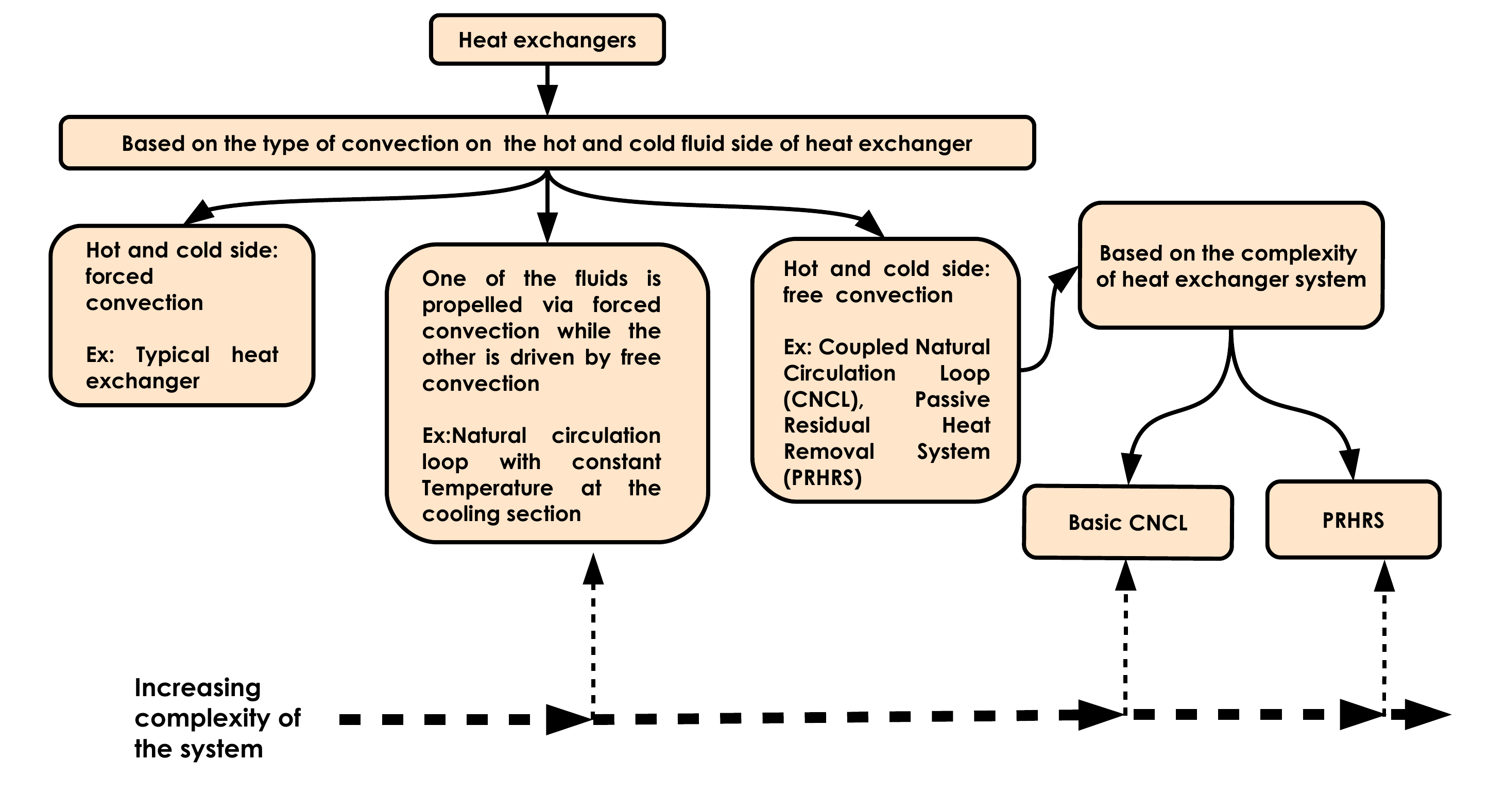}
	\caption{Classification of the heat exchangers based on the type of convection occurring at the hot and cold fluid sides.}
	\label{fig:fig1intro-flowchart}
\end{figure}

Heat exchange devices, also known as heat exchangers are heavily used in the industry and many daily life applications. The heat exchangers may be classified into various categories based on their size, geometry, and complexity. In the present study, the classification of heat exchangers based on the type of convection occurring on the hot and cold fluid sides is of prime interest. Figure \ref{fig:fig1intro-flowchart} represents the flowchart of the aforementioned classification.

The study of typical heat exchangers with forced convection has been thoroughly performed and has become part and parcel of many engineering textbooks, whereas the study of heat exchangers with at least one of the fluids driven by buoyancy forces is still an area of active research interest.
From Fig.\ref{fig:fig1intro-flowchart} we observe that a Natural Circulation Loop (NCL) , basic Coupled Natural Circulation Loop (CNCL) and Passive Residual Heat Removal Systems (PRHRS) are devices which employ buoyancy forces to initiate and maintain the flow. Natural circulation is a buoyancy-driven phenomenon that occurs in a closed conduit present in a body force field (gravitational, magnetic or centrifugal) when subjected to an external thermal stimulus. 

 A Natural Circulation Loop (NCL) is a device with circular or rectangular geometry that operates based on the natural circulation process when thermally stimulated at the heating and cooling sections (heat flux or internal heat generation or temperature boundary condition) of the loop, with the rest of the loop generally being insulated from the surroundings. The past five decades have witnessed a surge in the amount of literature pertaining to NCLs, which can directly be attributed to their applications in several domains such as solar heaters, turbine blade cooling, geothermal energy extraction, nuclear power generation, electronic chip cooling, chemical process industries, closed loop pulsating heat pipes, refrigeration, ship propulsion etc. \cite{Basu2014}.

\begin{figure}[!h]
	\centering
	\includegraphics[width=0.9\linewidth]{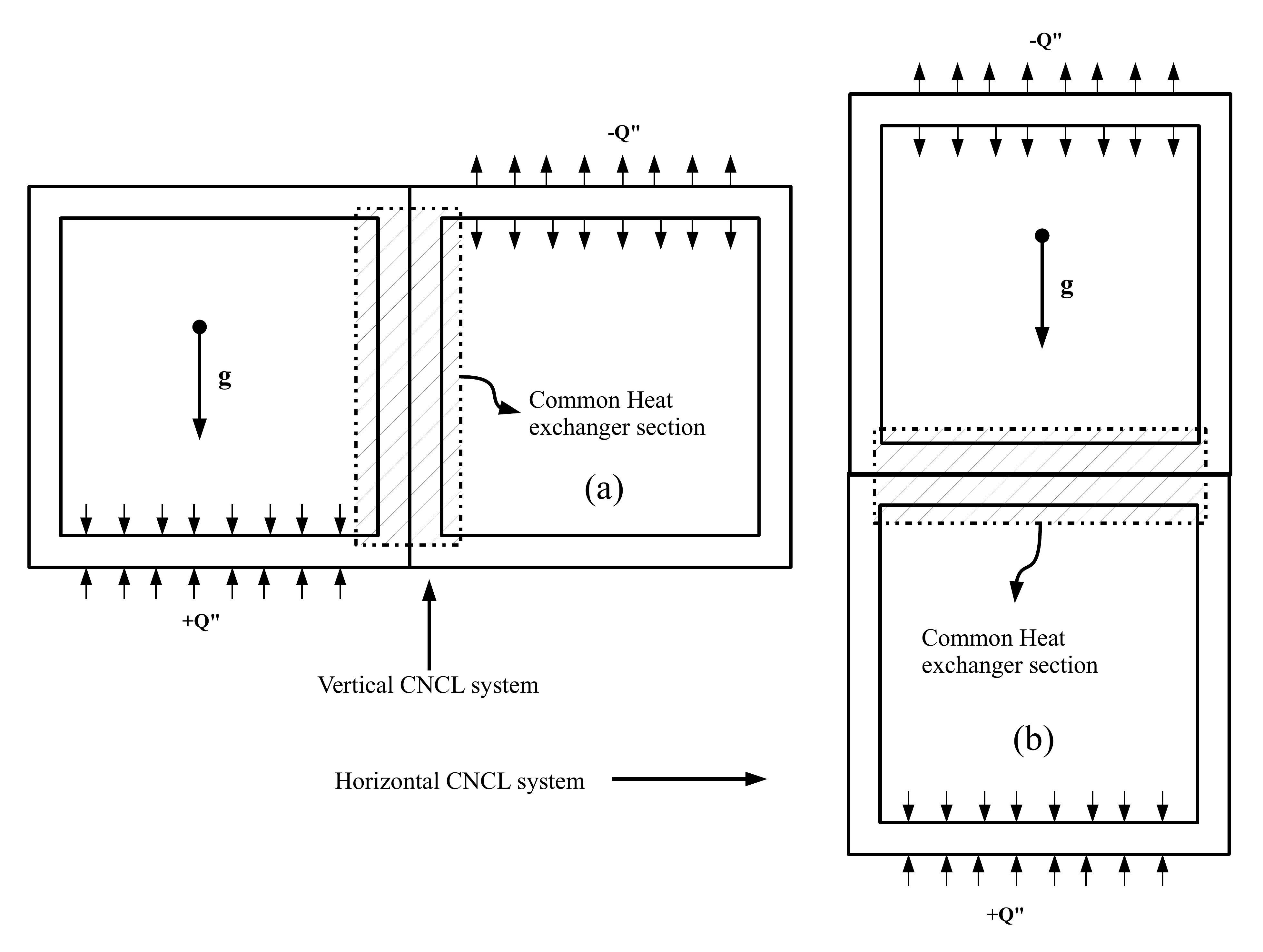}
	\caption{Basic CNCL systems considered for the current study. The CNCL systems are categorized as vertical or horizontal based on the common heat exchanger orientation w.r.t gravity.(a) Vertical CNCL system (b) Horizontal CNCL system.}
	\label{fig:fig2horizontal-and-vertical-cncl}
\end{figure}

\renewcommand{\arraystretch}{1.5}
\begin{table}[h!]
	\begin{center}
		\caption{Chronological summary of the literature pertaining to basic CNCL systems.}
		\scalebox{0.8}{
			\begin{tabular}{|p{0.5cm}|p{1cm}|p{1.5cm}|p{1.8cm}|p{1.5cm}|p{1.5cm}|p{2cm}|p{2cm}|p{6cm}|}\hline
				\textbf{Si No} & \textbf{Year}  & \textbf{Author}             & \textbf{CNCL Geometry}  & \textbf{Type of CNCL }                      & \textbf{Type of analysis   }  & \textbf{Method of \newline analysis}              & \textbf{Type of heat \newline exchanger}   & \textbf{Summary of salient points of the \newline literature}                                                                \\ \hline
				1     & 1987 & Davis and Roppo \cite{Davis1987}  & Toroidal      & Vertical                           & Steady               & 1D analytical                  & Point contact           & 1. Although the time dependent ODE system was presented no attempt was made to obtain the transient solution. \\ 
				&      &                   &               &                                    &                      &                                &                         & 2. No validation of the model presented.                                                                  \\
				&      &                   &               &                                    &                      &                                &                         & 3. Primary focus on the stability of steady states.                                                       \\ \hline
				2     & 1988 & Ehrhard     \cite{Ehrhard1988}      & Toroidal      & Vertical with square cross section & Transient and Steady & 1D analytical and experimental & 1D model-Point contact        & 1. Experimental validation of the 1-D CNCL model developed by Davis and Roppo was performed employing bifurcation maps.    \\
				&      &                   &               &                                    &                      &                                &     Experimental-Non point contact                    & 2. Transient validation of the 1-D model not performed.                                                   \\ \hline
				
				3     & 1988 & Salazar et al. \cite{salazar1988flow}   & Rectangular   & Vertical                           & Steady               & 1D analytical                  & Flat plate              & 1. Demonstrated the possibility of multiple steady state solution in rectangular CNCL.                    \\
				&      &                   &               &                                    &                      &                                &                         & 2. No validation of the model was performed.                                                                  \\ \hline
				4     & 2015 & Xun et al.  \cite{zhang2015analysis}      & Rectangular   & Verical                            & Transient            & 1D numerical and 2D CFD        & Flat plate              & 1. Validation of the 1-D numerical model with 2D CFD is performed.                                        \\
				&      &                   &               &                                    &                      &                                &                         & 2. Effect of heat transfer on oscillations of the CNCL is studied.                                        \\ \hline
				5     & 2016 & Duffey and Hughes \cite{DUFFEY2016455} & Rectangular   & Vertical and Horizontal            & Steady               & 1D analytical                  & General study           & 1. Importance of the CNCL systems is presented.                                                           \\
				&      &                   &               &                                    &                      &                                &                         & 2. The basic Horizontal and Vertical CNCL systems are depicted.                                           \\
				&      &                   &               &                                    &                      &                                &                         & 3. Links the basic CNCL and PRHRS.     \\ \hline                                                                  
		\end{tabular}}
	\end{center}
\label{tab1}
\end{table}

Some general observations and characteristics of single-phase Natural Circulation Loops are listed below:
\begin{enumerate}
	\item The NCL can be of any irregular geometry, but the toroidal and rectangular geometries have been extensively studied because of their simplicity and practical relevance \cite{Basu2014}.
	\item The heater and cooler orientations influence the transient and steady-state behavior of the NCL \cite{Vijayan2007}.
	\item The NCL is a dynamical system, which exhibits chaotic behavior at high power loads \cite{Fichera2003}.
	\item The NCL system is also very sensitive to the initial conditions at high power loads \cite{Fichera2003}.
	\item The power supplied to the NCL determines its dynamic characteristics. For lower power inputs, the NCL reaches a steady state, while increasing the power input leads to flow pulsations and/or reversals \cite{Fichera2003}.
	\item The NCL has a Lorenz like attractor for higher power inputs \cite{Fichera2003}.
	\item The viscous and buoyancy forces dictate the dynamic behavior of the natural circulation system \cite{Welander1967}.
\end{enumerate}

The above-mentioned characteristics and applications in numerous domains have augmented the research conducted in the field of NCLs. 
The domain of interest of the current paper is the heat exchanger wherein both the hot and cold fluid sides are driven by buoyancy forces. A basic Coupled Natural Circulation Loop (CNCL) is an ideal device which can be used to investigate the characteristics of such systems. CNCL is a device that is constructed from two NCLs, which are coupled thermally via a common heat exchanger section. The component NCLs are not hydraulically linked; hence, the energy transfer between the two NCLs is solely due to thermal coupling at the heat exchanger section.

A considerable amount of literature has been published on the complex CNCL systems such as the PRHRS (both primary and secondary loops driven by buoyancy) but the literature pertaining to basic CNCL systems such as those represented by Fig.\ref{fig:fig2horizontal-and-vertical-cncl} is limited. Table.1 represents the literature available on the basic CNCL systems.

Thus to add to the existing literature and provide a link between the NCL and PRHRS systems the following objectives are proposed for the present study:
\begin{enumerate}
	\item Develop a 1-D semi-analytical mathematical model for the transient analysis of the basic rectangular CNCL system.
	\item Conduct a 3-D CFD study of the basic rectangular CNCL system to validate the developed 1-D model and understand the behavior of systems represented in Fig.\ref{fig:fig2horizontal-and-vertical-cncl}. 
	\item Non-dimensionalize the 1-D model to identify the non-dimensional parameters which determine the CNCL behavior and thus by extension the PRHRS system.
	\item Detailed parametric study of the CNCL system behavior by varying the non-dimensional numbers.
	\item Study to determine the effect of  heater and cooler configurations on the CNCL system behavior.
	\item Study to determine the effect of parallel and counter flow orientations on the CNCL system response.
	\item Study to determine the common heat exchanger orientation of a CNCL system w.r.t gravity.
\end{enumerate}

The current study employs Fourier series to convert the Partial Differential Equations (PDE) of the CNCL system to Ordinary Differential Equations (ODE). Hart\cite{hart1984new} was first to employ this method to toroidal NCL systems and truncate the terms to obtain the transient dynamics of the system. Davis and Roppo \cite{Davis1987} utilized the methodology to model coupled toroidal NCLs and the point coupling was achieved by using the Dirac Delta function. Axial conduction effects were neglected in the models developed.

Bernal and Van Vleck \cite{rodriguez1998diffusion} proposed a mathematical model for generic NCL geometry and incorporated the axial conduction effects of the fluid in the model. Fichera and Pagano \cite{Fichera2003} utilized this method to develop the model of a rectangular NCL with heat flux boundary conditions. They also validated the model against experimental results. The model of the rectangular NCL was capable of capturing both the steady and chaotic behavior of the NCL system.
Salazar et al. \cite{salazar1988flow} proposed a 1-D steady-state model of a generic CNCL system with non-point contact, but the axial conduction effects were not considered for the study. The existence of multiple steady-state solutions was attributed to the nonlinear convective term of the energy equation. A similar approach is utilized in the current paper to derive the transient PDE of the rectangular basic CNCL system.

The 1-D model developed in the paper incorporates the following effects, making the model more general:
\begin{enumerate}[a)]
	\item     The axial conduction of the fluid is considered.
	\item    The Fourier series is employed to model the geometry and boundary conditions.
	\item     A non-point contact heat exchanger section is modeled.
	\item    Transient behavior of the system can be computed after truncation of the infinite series.
	\item    Bend loss effects are introduced in the model, which play a very vital role in the laminar regime.
\end{enumerate}

Transient CFD studies on NCLs in the laminar regime have been conducted by Kudariyawar et al. \cite{kudariyawar2016computational} and Louisos et al \cite{louisos2013chaotic}. Kudariyawar et al. performed transient CFD simulations using the laminar viscous model and observed that there was a good agreement with the experimental data, the deviation observed in the initial phase of the transient trend may be attributed to the non-consideration of wall thickness. Louisos et al. performed 2D CFD simulations of the toroidal NCL in the laminar regime. The higher Rayleigh number ranges were achieved by modifying the gravity. It is observed from the literature that the time taken to approach steady state from transience for fluids such as water is of the order of  $10^4s$, thus to minimize the time taken to reach steady state and also reduce the computational requirements fictitious fluids were chosen for the present study.  

The paper begins with an introduction to the 1-D semi-analytical model of the CNCL, which includes governing equations, initial and boundary conditions and solution methodology. This is followed by a 3-D CFD study of the vertical and horizontal CNCL systems and validation of the 1-D CNCL model for all the CFD cases considered. Mesh and time step independence studies were performed and fictitious fluids were employed to reduce the computational load. The governing equations of the CNCL are non dimensionalized to obtain the non-dimensional numbers which characterize the CNCL system. This is followed by the results section which includes a detailed parametric study of the CNCL employing the non-dimensional numbers. Finally, conclusions are drawn from the study.

\section{Contributions and benefits of the present study}

\begin{enumerate}
	\item It provides an analytical backbone to the study of basic CNCL systems: The current analysis provides an exact solution to the CNCL considered for the present study in terms of Fourier series, but because solving the resulting infinite dimensional system obtained is impractical, a truncated system is used with enough number of nodes to simulate the system dynamics.
	\item The 1-D methodology developed can be used to model multiple coupled NCL systems with non-point area contact. 
	\item The method developed is extremely user-friendly, modifiable and does not require the use of any commercial software.
	\item With extensions to the 1-D model developed in the current paper even complex systems may be modeled.
	\item The stability analysis of the CNCL systems can be performed.
\end{enumerate}

	\section{Mathematical modelling of CNCL system}

Salazar et al. \cite{salazar1988flow}  were the first to perform a steady state analysis of the CNCL system with a flat plate heat exchanger. The following simplifications were incorporated by Salazar et al. to model the CNCL:
\begin{enumerate}[a)]
	\item     The fluids in the primary and secondary loop of the CNCL are assumed to be driven exclusively via natural circulation.
	\item    The Intermediate Heat Exchanger section of the CNCL is a flat plate heat exchanger.
	\item    Heat transfer in the heat exchange zone is directly proportional to the local temperature difference between the primary loop and secondary loop.
	\item All thermophysical properties are assumed to have a constant magnitude.
	\item The Boussinesq hypothesis is employed to model the buoyancy forces.
	\item    One dimensional NCL momentum and energy equations are used for the study.
	\item    The axial heat conduction term of the energy equation is disregarded in the analysis performed.
	\item    Frictional forces are considered to be a linear function of velocity.
\end{enumerate}

The simplifications from (a) to (f) are considered in the present study with the inclusion of the axial heat conduction term and the implementation of a nonlinear function of velocity for the frictional forces. This study extends the analysis by enabling the modeling of the transient behavior of the CNCL system containing different fluids in each of the loops. Both ‘Loop 1’ and ‘Loop 2’ have identical dimensions and have a square cross-section. The current section only examines the 1-D modeling of Vertical CNCL, as the procedure is similar for the Horizontal CNCL configuration. To model the Horizontal CNCL, modifications are made to the piecewise functions representing the geometry and boundary conditions of the loops.

It is to be noted that employing the $-Q^{\prime\prime}$ on Loop 1 and $+Q^{\prime\prime}$ on Loop 2 at the CNCL common heat exchanger section results in a constant average temperature (w.r.t. time) in both loops thus decoupling the two loops. Hence the heat flux boundary condition cannot be employed at the common heat exchanger section, therefore a convection boundary condition is employed at the common heat exchanger section to model the CNCL system and ensure coupling amongst the constituent NCLs.

\subsection{ Governing equations of the CNCL}

\begin{figure}[!h]
	\centering
	\includegraphics[width=0.8\linewidth]{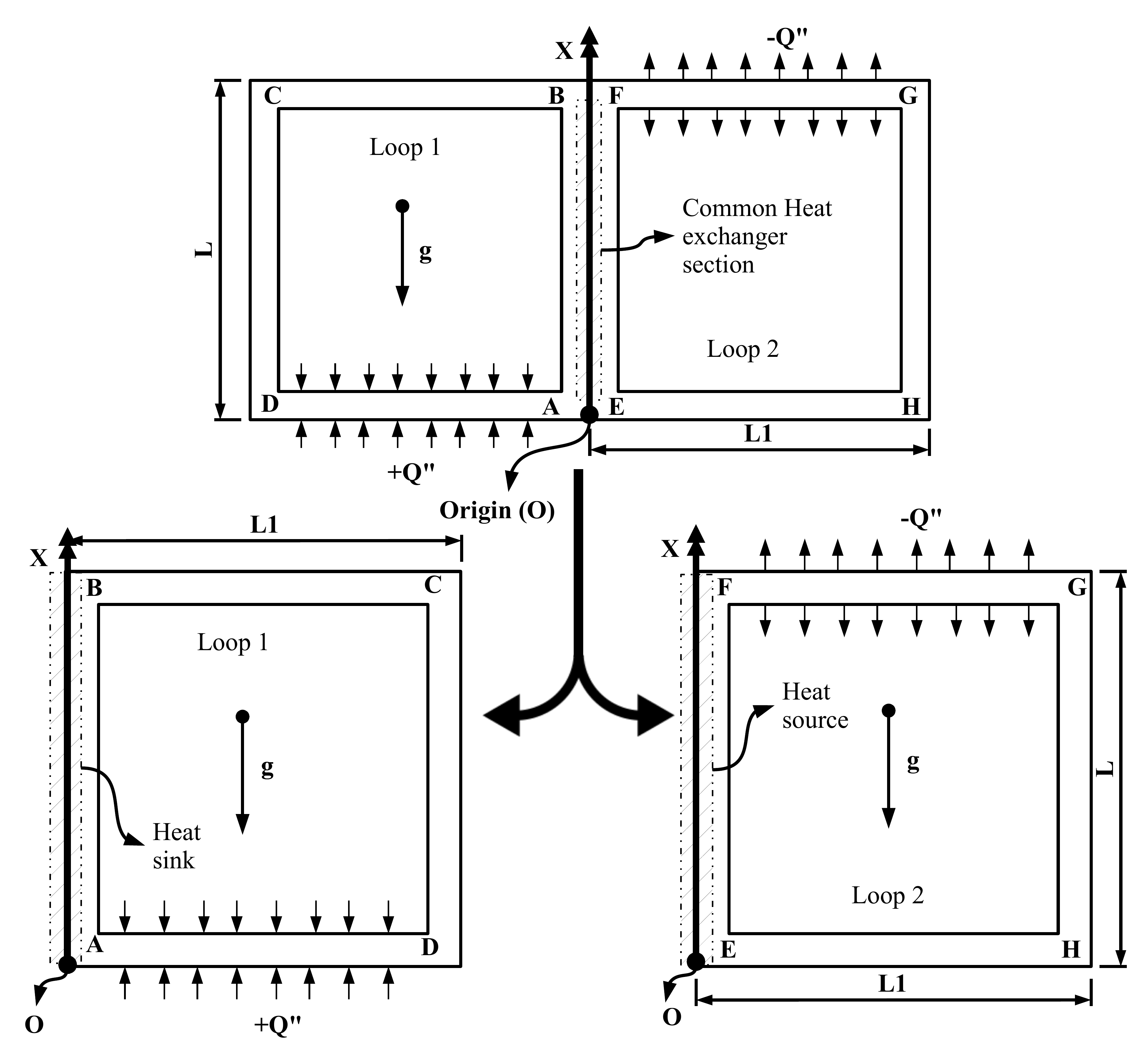}
	\caption{CNCL modeling approach.}
	\label{fig:fig3splitting-the-cncl}
\end{figure}

The CNCL has two NCLs: Loop 1, Loop 2, which are coupled at the common heat exchanger section. Corners of ’Loop 1’ are denoted using alphabets (A-D) and that of ‘Loop 2’ are denoted using alphabets (E-H), as shown in Fig.\ref{fig:fig3splitting-the-cncl}. To derive the governing equations of the CNCL system, the CNCL is divided into two parts. Thereby ‘Loop 1’ and ‘Loop 2’ become two distinct NCL’s with the heat exchanger section transforming into the heat sink for the NCL formed by ‘Loop 1’ and heat source for the NCL formed by ‘Loop 2’. Figure \ref{fig:fig3splitting-the-cncl} clearly depicts this process and labeling of the system. The objective behind flipping the NCL formed by ‘Loop 1’ vertically is to have a common sign for velocity vector (+ve in the clockwise direction) and have the origin $O$ at the left bottom corner. Apart from the common heat exchanger section and limbs of the CNCL where the constant heat flux addition or removal condition is enforced, the rest of the CNCL is insulated from the surroundings.

The governing equations of the CNCL can be derived from the simple force and energy balance applied on an infinitesimal 1-D element of the constituent NCLs and using appropriate piecewise functions to represent the complete system.

The governing equations of the CNCL system are:

\begin{equation}
\rho_1\frac{d\omega_{1}(t)}{dt}+\frac{4\tau_{1}}{D_h}=\frac{\rho_1 g\beta_1}{2(L+L1)}(\oint\!(T_{1}-T_{0})f_1(x)dx\,) - \frac{nK\rho_1 \omega_1 ^2}{4(L+L1)}
\end{equation}

\begin{equation}
\frac{\partial T_{1}}{\partial t}+\omega_{1}(t)(\frac{\partial T_{1}}{\partial x})=H_{1}(x,t)+a_1\frac{\partial^{2}T_{1}}{\partial x^{2}}
\end{equation}

\begin{equation}
\rho_2\frac{d\omega_{2}(t)}{dt}+\frac{4\tau_{2}}{D_h}=\frac{\rho_2 g\beta_2}{2(L+L1)}(\oint\!(T_{2}-T_{0})f_2(x)dx\,) - \frac{nK\rho_2 \omega_2 ^2}{4(L+L1)}
\end{equation}

\begin{equation}
\frac{\partial T_{2}}{\partial t}+\omega_{2}(t)(\frac{\partial T_{2}}{\partial x})=H_{2}(x,t)+a_2\frac{\partial^{2}T_{2}}{\partial x^{2}}
\end{equation}

where, 
\begin{equation}
\tau_{1}=\frac{\rho_1 \omega_{1}^{2}f_{F1}}{2}
\end{equation}

\begin{equation}
\tau_{2}=\frac{\rho_2 \omega_{2}^{2}f_{F2}}{2}
\end{equation}

Equations 1 and 3 represent the momentum equations of the Loop 1 and Loop 2, respectively.
Equations 2 and 4 represent the energy equations of the Loop 1 and Loop 2, respectively.
$\tau_1$ and $\tau_2$ represent the shear stresses acting on the fluid elements of Loop 1 and Loop 2, respectively. `$n$' represents the number of bends and `$k$' represents the bend loss coefficient.

\begin{equation}
H_{1}(x,t)=h_{1}(x)-\frac{U}{\rho_1 C_{p1}D_h}\lambda(x)(T_{1}-T_{2})
\end{equation}

\begin{equation}
H_{2}(x,t)=h_{2}(x)+\frac{U}{\rho_2 C_{p2}D_h}\lambda(x)(T_{1}-T_{2})
\end{equation}

$H_1$ and $H_2$ are functions which signify the net amount of power transferred to each of the constituent NCLs: Loop 1 and Loop 2, respectively.

\begin{equation}
f_1(x)=f_2(x)=\left\{ \global\long\def\arraystretch{1.2}
\begin{array}{@{}c@{\quad}l@{}}
1 & {0<x<L}\\
0 & {L<x<L+L1}\\
-1 & {L+L1<x<2L+L1}\\
0 & {2L+L1<x<2(L+L1)}
\end{array}\right.
\end{equation}

\begin{equation}
\lambda(x)=\left\{ \global\long\def\arraystretch{1.2}
\begin{array}{@{}c@{\quad}l@{}}
1 & {0<x<L}\\
0 & {L<x<L+L1}\\
0 & {L+L1<x<2L+L1}\\
0 & {2L+L1<x<2(L+L1)}
\end{array}\right.
\end{equation}

$f_i(x)$ is the function which represents the rectangular geometry of both ’Loop 1’ and ’Loop 2’. $\lambda(x)$ is the function which couples both Loop 1 and Loop 2. 

\subsection{Friction factor correlations}
The frictional forces direct the dynamic behavior of the CNCL system; hence, it is pivotal to employ proper correlations to determine their magnitude as the flow develops with time.  The current study involves flow in a square duct, thus the suitable friction factors for square duct are utilized. The general Fanning friction factor correlation is given by equation 11. In the laminar regime the magnitude of C=14.23 and d=1 for a square duct \cite{Yunus2006}.
\begin{equation}
f_F=C/Re^d
\end{equation}

\subsection{Bend loss Correlations}

The bend loss at the $90^{\circ} $ elbow bends of the CNCL plays a very vital role in the transient evolution of the velocity and temperature distributions in the CNCL system. The bend loss coefficient (K) for the current study is evaluated using the 3K method. Equation (12) provides the expression of the 3K method bend loss correlation which is valid across the laminar, transition and turbulent regime \cite{Darby2001}. The values of $K_1$,$K_\infty$ and $K_d$ used for elbow bend with $R/D_h=1$ are 800, 0.14 and 4 respectively and $D_n$ represents the hydraulic diameter of the elbow in inches \cite{Darby2001}.

\begin{figure}[!h]
	\centering
	\begin{subfigure}[b]{0.49\textwidth}
		\includegraphics[width=1\linewidth]{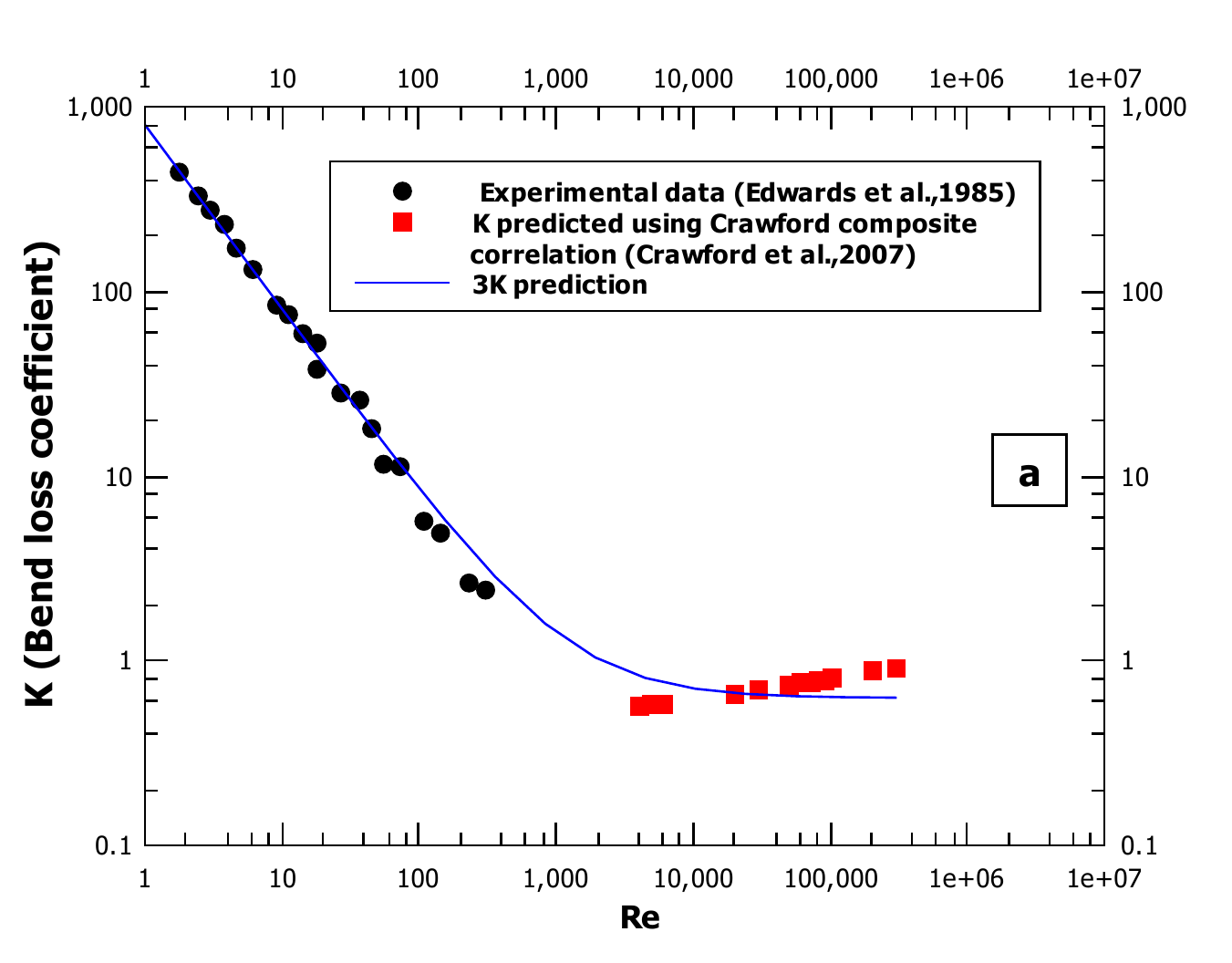}
	\end{subfigure}
	\hspace{\fill}
	\begin{subfigure}[b]{0.49\textwidth}
		\includegraphics[width=1\linewidth]{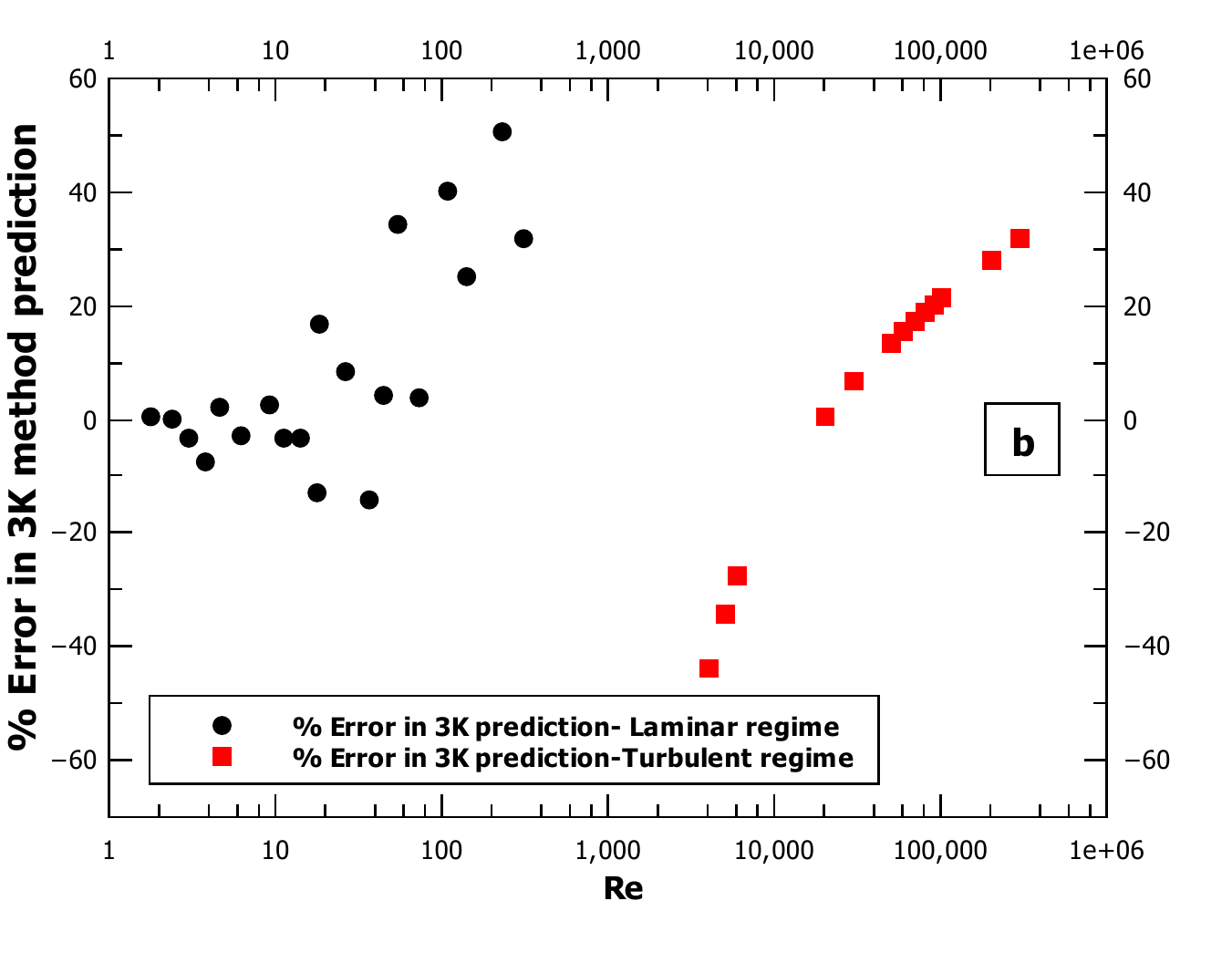}
	\end{subfigure}
	
	\caption{(a) Comparison of experimental data and 3K method prediction (b) Error introduced by the 3K method. }
	\label{Bend loss coefficient}
\end{figure}

\begin{equation}
K=\frac{K_1}{Re} + K_\infty(1+\frac{K_d}{D_n^{0.3}})
\end{equation}

Figure \ref{Bend loss coefficient}a compares the values of the 3K method prediction with experimental data. We observe that the 3K method provides a decent estimation of the bend loss over a wide range of Reynolds numbers. Figure \ref{Bend loss coefficient}b represents the corresponding error of the 3K method prediction with experimental data. We observe that the maximum error associated with the 3K prediction is roughly $\pm 50\% $. The experimental bend loss data in laminar regime was taken from the work of Edwards et al.\cite{Edwards1985} and the data for the turbulent regime was obtained employing the Crawford composite correlation from the work of Crawford et al.\cite{Crawford2007}.

\subsection{Initial and boundary conditions of the CNCL system}

\subsubsection{Initial conditions}

The initial condition of the CNCL system is represented by the initial conditions of Loop 1 and Loop 2. The initial conditions for the momentum and energy equations of the CNCL system are depicted by equations 13 and 14.
\begin{equation}
w_{1}(0)=w_{0}\ ,\ w_{2}(0)=w_{0}\ 
\end{equation}

\begin{equation}
T_{1}(x,0)=T_{0}\ ,\ T_{2}(x,0)=T_{0}
\end{equation}

\subsubsection{Boundary conditions}
$h_1(x)$ is the function which represents the location of the positive heat flux  boundary condition of Loop 1. $h_2(x)$ is the function which represents the location of the negative heat flux boundary condition of Loop 2.

\begin{equation}
h_{1}(x)=\left\{ \global\long\def\arraystretch{1.2}
\begin{array}{@{}c@{\quad}l@{}}
0 & {0<x<L}\\
0 & {L<x<L+L1}\\
0 & {L+L1<x<2L+L1}\\
\bigg(\frac{4Q^{\prime\prime}}{\rho_1 C_{p,1} D_h}\bigg) & {2L+L1<x<2(L+L1)}
\end{array}\right.
\end{equation}

\begin{equation}
h_{2}(x)=\left\{ \global\long\def\arraystretch{1.2}
\begin{array}{@{}c@{\quad}l@{}}
0 & {0<x<L}\\
\bigg(\frac{-4Q^{\prime\prime}}{\rho_2 C_{p,2} D_h}\bigg) & {L<x<L+L1}\\
0 & {L+L1<x<2L+L1}\\
0 & {2L+L1<x<2(L+L1)}
\end{array}\right.
\end{equation}


\subsection{ODEs which represent the 1-D single phase CNCL system}

To simplify the task of obtaining the solution to the aforementioned (Equation 2 and Equation 4) partial differential equations (PDEs), we employ the Fourier series expansions of the temperature and boundary conditions to convert the PDEs to ordinary differential equations (ODEs).

\begin{equation}
T_1(x,t)= \sum \limits_{k = -\infty }^\infty \alpha_k (t) e^{{ik\pi x}/(L+L1)}
\end{equation}

\begin{equation}
T_2(x,t)= \sum \limits_{k = -\infty }^\infty \beta_k (t) e^{{ik\pi x}/(L+L1)}
\end{equation}

\begin{equation}
h_1(x) = \sum \limits_{k = -\infty }^\infty \gamma_k e^{{ik\pi x}/(L+L1)} 
\end{equation}

\begin{equation}
h_2(x) = \sum \limits_{k = -\infty }^\infty \delta_k e^{{ik\pi x}/(L+L1)}
\end{equation}

\begin{equation}
\lambda(x) = \sum \limits_{k = -\infty }^\infty \zeta_k e^{{ik\pi x}/(L+L1)}
\end{equation}

\begin{equation}
f(x) = \sum \limits_{k = -\infty }^\infty A_k e^{{ik\pi x}/(L+L1)} 
\end{equation}

where, 
\begin{equation}
\alpha_0(t)=\frac{1}{2(L+L1)} \int_0^{2(L+L1)} T_1(x,t) dx 
\end{equation}

\begin{equation}
\beta_0(t)=\frac{1}{2(L+L1)} \int_0^{2(L+L1)} T_2(x,t) dx 
\end{equation}

for all $t$. In addition $  \overline{\alpha_k}=\alpha_{-k} $, the same applies for all aforementioned complex Fourier coefficients.    

Substituting the above expressions in equations (1-4) results in the following set of equations after restricting the number of Fourier nodes to three:

\begin{equation}
\frac{d\omega_1(t)}{dt}+\frac{2C}{D_h}(\frac{\nu}{D_h})^{d}(\omega_1)^{2-d}= g\beta\sum \limits_{n = -3 }^3 \alpha_n A_{-n} - \frac{nK \omega_1 ^2}{4(L+L1)}
\end{equation}

\begin{equation}
\frac{d (\alpha_0 (t))}{dt}+= \gamma_0 -  \sum \limits_{n = -3 }^3 \frac{U}{\rho_1 C_{p1} D_h} \zeta_{-n} (\alpha_n (t)-\beta_n (t) )
\end{equation}

\begin{equation}
\frac{d (\alpha_1 (t))}{dt}+\frac{i\pi}{L+L1} w_1(t) \alpha_1 (t)= \gamma_1 - \frac{a\alpha_1 (t)\pi^2}{(L+L1)^2}-  \sum \limits_{n = -3 }^3 \frac{U}{\rho_1 C_{p1} D_h} \zeta_{1-n} (\alpha_n (t)-\beta_n (t) )
\end{equation}

\begin{equation}
\frac{d (\alpha_2 (t))}{dt}+\frac{2i\pi}{L+L1} w_1(t) \alpha_2 (t)= \gamma_2 - \frac{4a\alpha_2 (t)\pi^2}{(L+L1)^2}-  \sum \limits_{n = -3 }^3 \frac{U}{\rho_1 C_{p1} D_h} \zeta_{2-n} (\alpha_n (t)-\beta_n (t) )
\end{equation}

\begin{equation}
\frac{d (\alpha_3 (t))}{dt}+\frac{3i\pi}{L+L1} w_1(t) \alpha_3 (t)= \gamma_3 - \frac{9a\alpha_3 (t)\pi^2}{(L+L1)^2}-  \sum \limits_{n = -3 }^3 \frac{U}{\rho_1 C_{p1} D_h} \zeta_{3-n} (\alpha_n (t)-\beta_n (t) )
\end{equation}

\begin{equation}
\frac{d\omega_2(t)}{dt}+\frac{2C}{D_h}(\frac{\nu}{D_h})^{d}(\omega_2)^{2-d}= g\beta\sum \limits_{n = -3 }^3 \beta_n A_{-n} - \frac{nK\omega_2 ^2}{4(L+L1)}
\end{equation}

\begin{equation}
\frac{d (\beta_0 (t))}{dt}= \delta_0 +  \sum \limits_{n = -3 }^3 \frac{U}{\rho_2 C_{p2} D_h} \zeta_{-n} (\alpha_n (t)-\beta_n (t) )
\end{equation}

\begin{equation}
\frac{d (\beta_1 (t))}{dt}+\frac{i\pi}{L+L1} w_2(t) \beta_1 (t)= \delta_1 - \frac{a\beta_1 (t)\pi^2}{(L+L1)^2}+  \sum \limits_{n = -3 }^3 \frac{U}{\rho_2 C_{p2} D_h} \zeta_{1-n} (\alpha_n (t)-\beta_n (t) )
\end{equation}

\begin{equation}
\frac{d (\beta_2 (t))}{dt}+\frac{2i\pi}{L+L1} w_2(t) \beta_2 (t)= \delta_2 - \frac{4a\beta_2 (t)\pi^2}{(L+L1)^2}+  \sum \limits_{n = -3 }^3 \frac{U}{\rho_2 C_{p2} D_h} \zeta_{2-n} (\alpha_n (t)-\beta_n (t) )
\end{equation}

\begin{equation}
\frac{d (\beta_3 (t))}{dt}+\frac{3i\pi}{L+L1} w_2(t) \beta_3 (t)= \delta_3 - \frac{9a\beta_3 (t)\pi^2}{(L+L1)^2}+  \sum \limits_{n = -3 }^3 \frac{U}{\rho_2 C_{p2} D_h} \zeta_{3-n} (\alpha_n (t)-\beta_n (t) )
\end{equation}

The equations [27-29] and [32-34] are separated into real and imaginary parts, thereby obtaining a set of 16 ordinary differential equations which need to be solved simultaneously to obtain $T_1(x,t)$ and $T_2(x,t)$. The Fourier coefficients are plugged into equations [35,36] to obtain temperature distributions in Loop 1 and Loop 2.

\begin{equation}
T_1(x,t)= \sum \limits_{k = -3 }^3 \alpha_k (t) e^{{ik\pi x}/(L+L1)}
\end{equation}

\begin{equation}
T_2(x,t)= \sum \limits_{k = -3 }^3 \beta_k (t) e^{{ik\pi x}/(L+L1)}
\end{equation}

\subsection{Initial conditions of the ODE system}

To obtain the transient behavior of the CNCL system we need to integrate the ODEs represented by equations (25-34) for which the initial conditions for each of the Fourier nodes need to be determined.
The initial conditions of the ODEs are derived from the initial conditions of the momentum and energy equation for both Loop 1 and Loop 2. Since the number of nodes is limited to three, we place the nodes equidistantly along the loop and apply the initial conditions mentioned in equations (13) and (14). This results in a matrix of 4 equations and 4 unknowns for each of the loops, as represented below.

$$
\begin{bmatrix}
1    &   1    &   1     &   1  \\

1  &  e^{{i2\pi }/3} &  e^{{i4\pi }/3}  &  e^{{i6\pi}/3}\\

1  &  e^{{i4\pi }/3} &   e^{{i8\pi }/3}  &   e^{{i12\pi}/3}\\

1  &   e^{{i6\pi }/3} &   e^{{i12\pi }/3}  &  e^{{i18\pi }/3}\\

\end{bmatrix} \times \left[ \begin{array}{c} \alpha_{0} \\ \alpha_{1} \\ \alpha_{2}\\ \alpha_{3}\end{array}\right] = \left[ \begin{array}{c} T_{0} \\ T_{0} \\ T_{0}\\ T_{0}\end{array}\right]
$$

$$
\begin{bmatrix}
1    &  1   & 1    & 1  \\

1  &  e^{{i2\pi }/3} &  e^{{i4\pi }/3}  &  e^{{i6\pi}/3}\\

1  &  e^{{i4\pi }/3} &   e^{{i8\pi }/3}  &   e^{{i12\pi}/3}\\

1  &   e^{{i6\pi }/3} &   e^{{i12\pi }/3}  &  e^{{i18\pi }/3}\\
\end{bmatrix} \times \left[ \begin{array}{c} \beta_{0} \\ \beta_{1} \\ \beta_{2}\\ \beta_{3}\end{array}\right] = \left[ \begin{array}{c} T_{0} \\ T_{0} \\ T_{0}\\ T_{0}\end{array}\right]
$$

Solving the matrix equations yields the initial value of the Fourier nodes of Temperature.
The solution is $\alpha_{0}=\beta_{0}=T_{0} $, whilst the values at all other nodes is zero.

\subsection{Numerical integration}

The set of ODEs obtained by using the Fourier series expansion cannot be solved analytically. Thus to obtain the dynamics of the system we need to use numerical integration techniques such as the Runge-Kutta method which has been employed for the current study using the MATLAB solver 'ode45'.

	\section{3-dimensional CFD study of a single phase CNCL system}

A detailed CFD investigation of the single-phase CNCL system has been conducted to explore the physics and dynamic characteristics. The main objective of this CFD study is to validate the 1-D CNCL model rigorously and probe its efficacy. The 3-D CFD study has been performed to study the single-phase CNCL system for the set of conditions specified in Table.\ref{tab:table2}.

\begin{table}[h!]
	\centering
	\caption{Cases considered for the CFD simulation.}
	\label{tab:table2}
	\scalebox{0.9}{
		\begin{tabular}{  l  l  l  l l  }
			
			\textbf{Case}            & \textbf{Loop 1}   &  \textbf{Loop 2} &  \textbf{CNCL Orientation} & \textbf{Description} \\ \hline
			CNCL-(a)        & FF1      &  FF1   &    Vertical       & Same fluid in both loops with counter flow configuration\\
			CNCL-(b)        & FF1      & FF2    &    Vertical       & Different fluids in the loops with counter flow configuration\\
			CNCL-(c)        & FF1      & FF1    &    Horizontal     & Same fluid in both loops with parallel flow configuration\\
			CNCL-(d)        & FF1      & FF1    &    Horizontal     & Same fluid in both loops with counter flow configuration\\ \hline
	\end{tabular}}
\end{table}

The developed 1-D model has been validated with the aforementioned CFD cases to display its robustness for use in prediction of the behavior of CNCL system. The CFD investigation of the CNCL system has been done using ANSYS Fluent, which is a commercial finite volume software. The pre-processing, processing and post-processing were performed employing the same software. FF1 and FF2 listed in Table.\ref{tab:table2} are the fluids used in this study, the thermophysical properties of which are elaborated in section 4.3.1.

\subsection{Geometry}
\begin{figure}[!h]
	\centering
	\begin{subfigure}[b]{0.55\textwidth}
		\includegraphics[width=1\linewidth]{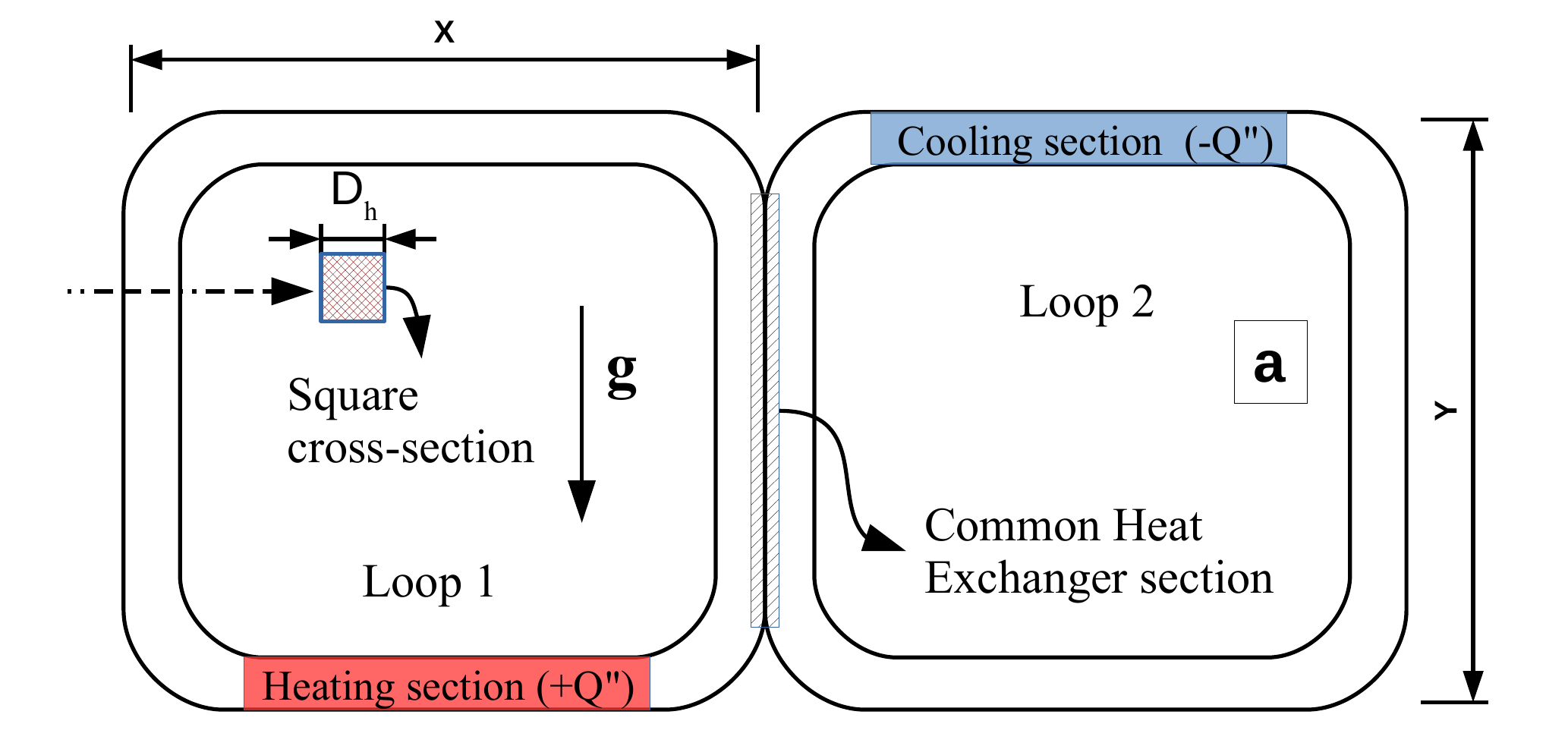}
	\end{subfigure}
	\hspace{\fill}
	\begin{subfigure}[b]{0.35\textwidth}
		\includegraphics[width=0.8\linewidth]{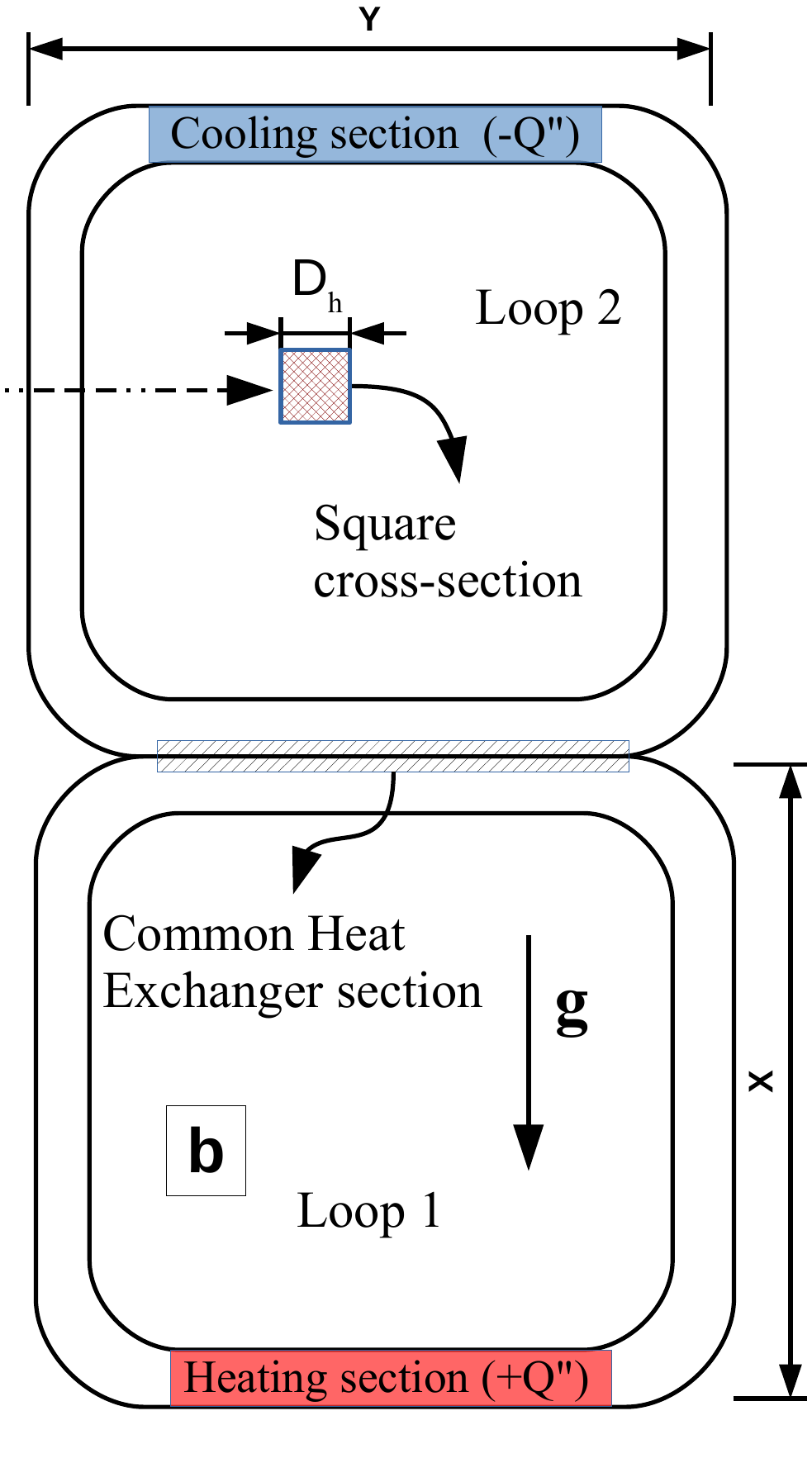}
	\end{subfigure}
	
	\caption{Schematic of the CNCL geometry made of square NCLs Loops 1 and 2 with a square cross section considered for CFD study. (a) Geometry of vertical CNCL cases CNCL-(a) and CNCL-(b).
		(b) Geometry of horizontal CNCL cases CNCL-(c) and CNCL-(d). }
	\label{fig:fig5cfdcnclgeometry}
\end{figure}

The geometry comprises of two parts: Loop 1 and Loop 2 which are coupled at the common heat exchanger section (flat plate heat exchanger). Loop 1 \&\ 2 are square NCLs with a square cross-sectional area. Figure \ref{fig:fig5cfdcnclgeometry} represents the CNCL geometry considered for the study. The 90-degree bends are chamfered to reduce the bend losses encountered by the flow. Apart from the cooling and heating sections, the rest of the loop is completely insulated. The dimensions of the geometry considered for the study are listed in Table.\ref{tab:table1}.

\begin{table}[h!]
	\centering
	\caption{Dimensions of the 3-D geometry}
	\label{tab:table1}
	\scalebox{0.9}{
		\begin{tabular}{  l  l  l  }
			
			\textbf{Parameter}  & \textbf{Description}   &  \textbf{Dimensions} \\ \hline
			X          & Height/Width of the CNCL &1 $m$ \\
			Y         & Height/Width of the CNCL &1 $m$ \\
			$D_h$          & Length of square cross section   & 0.04 $m$ \\ \hline
	\end{tabular}}
\end{table}

\subsection{Meshing}

\begin{figure}[!h]
	\centering
	\includegraphics[width=0.7\linewidth]{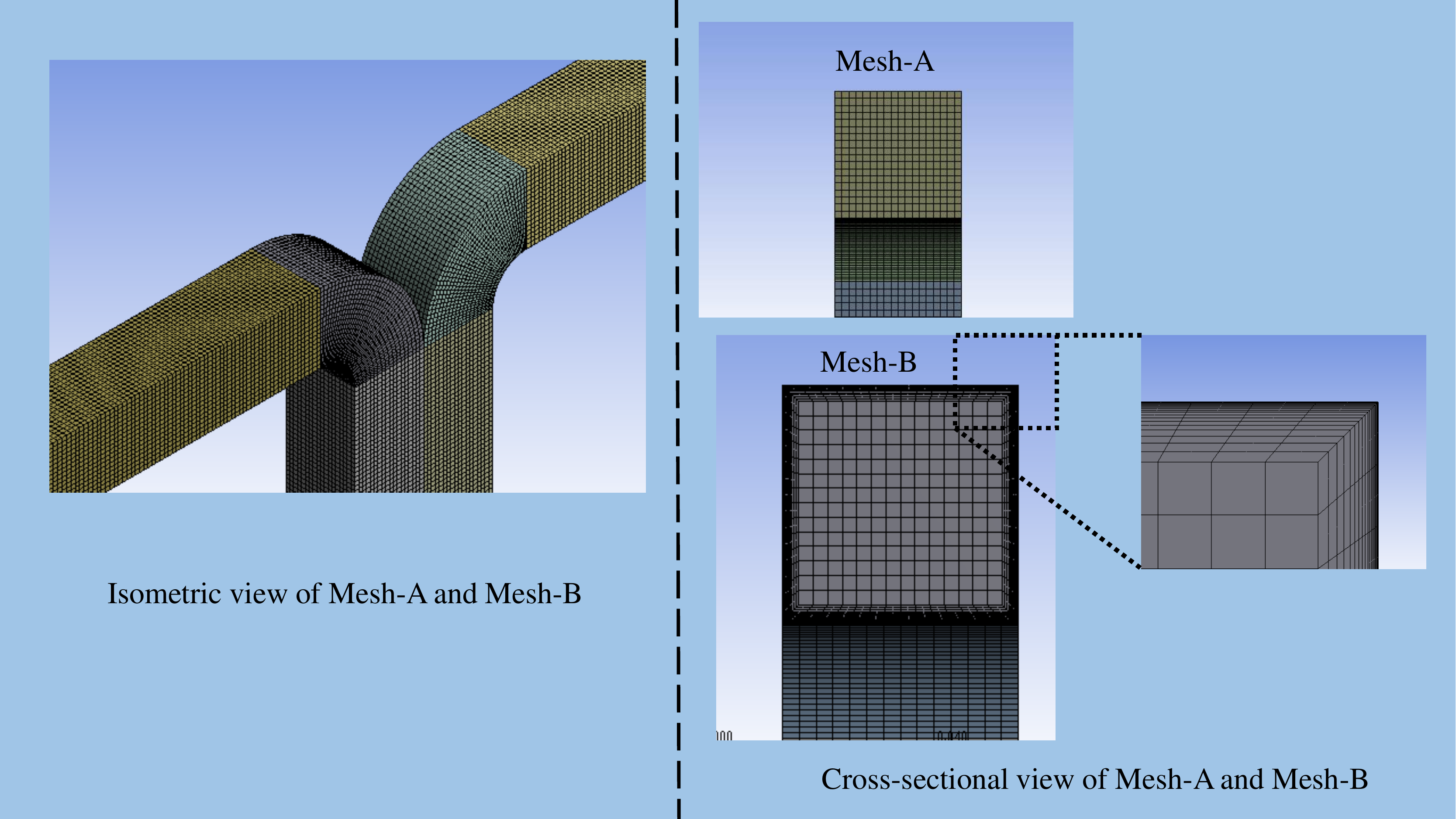}
	\caption{Schematic of the mesh used for the CFD study. Mesh-A represents the mesh without inflation at the walls and Mesh-B represents the mesh with inflation.}
	\label{fig:fig6mesh-details}
\end{figure}

The ANSYS Design Modeler was used to generate a structured 3-D mesh from the geometry, as shown in Fig.\ref{fig:fig6mesh-details}. The Multi-zone method was used along with the body sizing option to generate a hex-dominant high-quality mesh. Two meshes are considered, one without inflation called as Mesh-A and one with inflation as Mesh-B.

\subsection{Case setup}

A transient pressure based solver was used to simulate the transient dynamics of the CNCL system considered. The energy and momentum equation were solved simultaneously on the 3-D grid to obtain the numerical solution at each time step. The laminar flow model was used for the current simulation with the gravity set to $9.81$ $m/s^2$.

The Boussinesq approximation has been used to model the buoyancy forces. The operating temperature at startup is set to 300 $K$ and the operating pressure is set to 1 $atm$.

\subsubsection{Fluids used for the CFD study}

It is observed that the fluids which are normally used, have very low values of thermal diffusivity. This becomes an issue when one tries to numerically solve the coupled momentum and energy equations of buoyancy-driven flows. The low values of thermal diffusivity impose the selection of an extremely fine mesh and a small time step to get a numerically stable CFD solution. Thus if we employ common fluids such as water, for a given heat input they might require about 50 $min$ to reach the steady state (\cite{Fichera2003}). Besides, the extremely fine mesh and minute time step requirements make the computational process extremely expensive both in terms of memory and in terms of the clock time if we simulate until the fluid reaches a steady state. Since one of the main objectives is to validate the 1-D semi-analytical model developed we employ a fictitious fluid (a fluid with assumed thermophysical properties) for the 3-D CFD simulations to relax the mesh and time step conditions for numerical stability and hasten the time required to reach the steady state. The fluid chosen for simulation must be irrelevant as long as the underlying physics is the same. The thermophysical properties of the fictitious fluids are listed in Table.\ref{tab:table4}.

\begin{table}[h!]
	\centering
	\caption{Fluids used for the CFD study.}
	\label{tab:table4}
	\scalebox{0.85}{
		\begin{tabular}{  l  l  l  }
			
			\textbf{Property}              & \textbf{Fictitious fluid 1 (FF1)}  & \textbf{Fictitious fluid 2 (FF2)}    \\ \hline
			$\rho_{0} $   &      70 $kg/m^3$          &  50 $kg/m^3$    \\
			$C_p$         &      100 $J/kgK$          &   70 $J/kgK$    \\
			$\beta$       &      0.01 $1/K$           &   0.08 $1/K$    \\
			$a$           &      0.4 $m^2/s$          &   0.8 $m^2/s$   \\
			$\mu$         &    0.0007 $kg/ms$         &  0.005 $kg/ms$ \\ \hline
	\end{tabular}}
\end{table}

\subsubsection{Initial and boundary conditions}

At time $t=0$ the operating temperature is $T_0$ and the operating pressure is $P_0$. Apart from the heating, cooling and coupled heat exchanging section, the rest of the loop is completely insulated. We apply a constant heat flux $Q^{\prime\prime}$ at the heating section and the corresponding negative value at the cooling section. The coupling that was done in the geometry section results in the formation of a shadow for the coupled wall region. Table.\ref{tab:table5} lists the initial and boundary conditions of the CFD system used for the current simulation for all the cases CNCL-(a) to CNCL-(d) mentioned in Table.\ref{tab:table2}.

\begin{table}[h!]
	\centering
	\caption{Initial and boundary conditions}
	\label{tab:table5}
	\scalebox{0.85}{
	\begin{tabular}{  l  l  l  }
		
		\textbf{Parameter}     & \textbf{Description}  & \textbf{Values}    \\ \hline
		$T_{0} $      & Temperature at time $t=0$ &  300 $K$    \\
		$P_0$         & Pressure at time $t=0$    &   1 $atm$    \\
		$Q"$          & Constant heat flux supplied or extracted &  2000 $W/m^2$    \\
		$g$           & Gravitational constant    &  9.81 $m/s^2$   \\ \hline
	\end{tabular}}
\end{table}

\subsubsection{Solver settings and discretization schemes}
The PISO scheme was used for pressure-velocity coupling. The least squares cell-based scheme was used for the gradient spatial discretization. A second-order scheme for pressure discretization and second-order upwind scheme for the momentum and energy discretization was used respectively. Default solution controls with residuals of the order $10^{-3}$ were used and the average velocity and temperature were used as parameters for judging the convergence. The velocity field was initialized to zero and the temperature in the entire domain at time $t = 0$ $s$ was set to 300 K. A fixed time stepping method was used and the number of iterations per time step was set to 200.

\subsection{Grid independence and time step independence tests}

\begin{figure}[!h]
	\centering
	\begin{subfigure}[b]{0.49\textwidth}
		\includegraphics[width=1\linewidth]{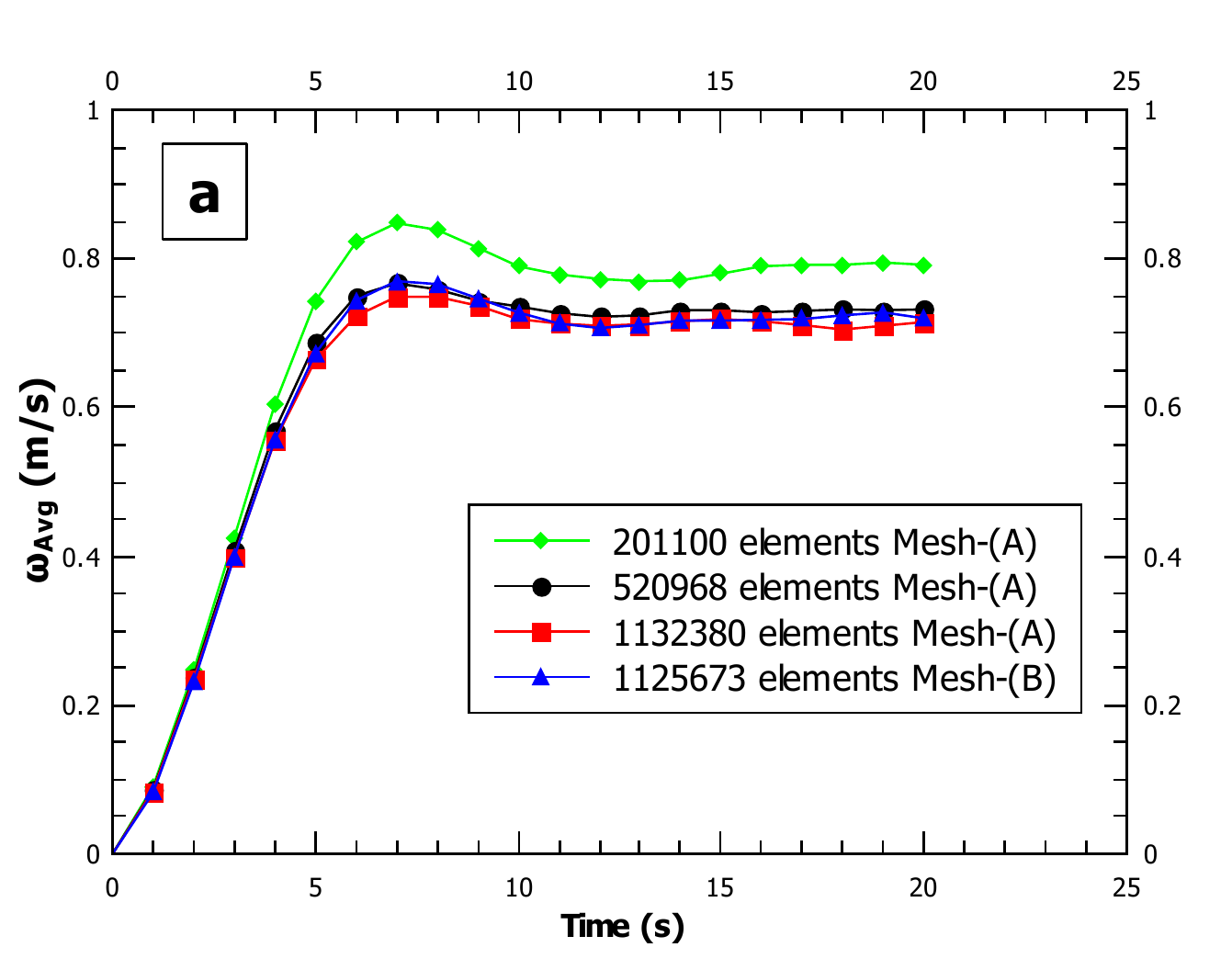}
	\end{subfigure}
	\hspace{\fill}
	\begin{subfigure}[b]{0.49\textwidth}
		\includegraphics[width=1\linewidth]{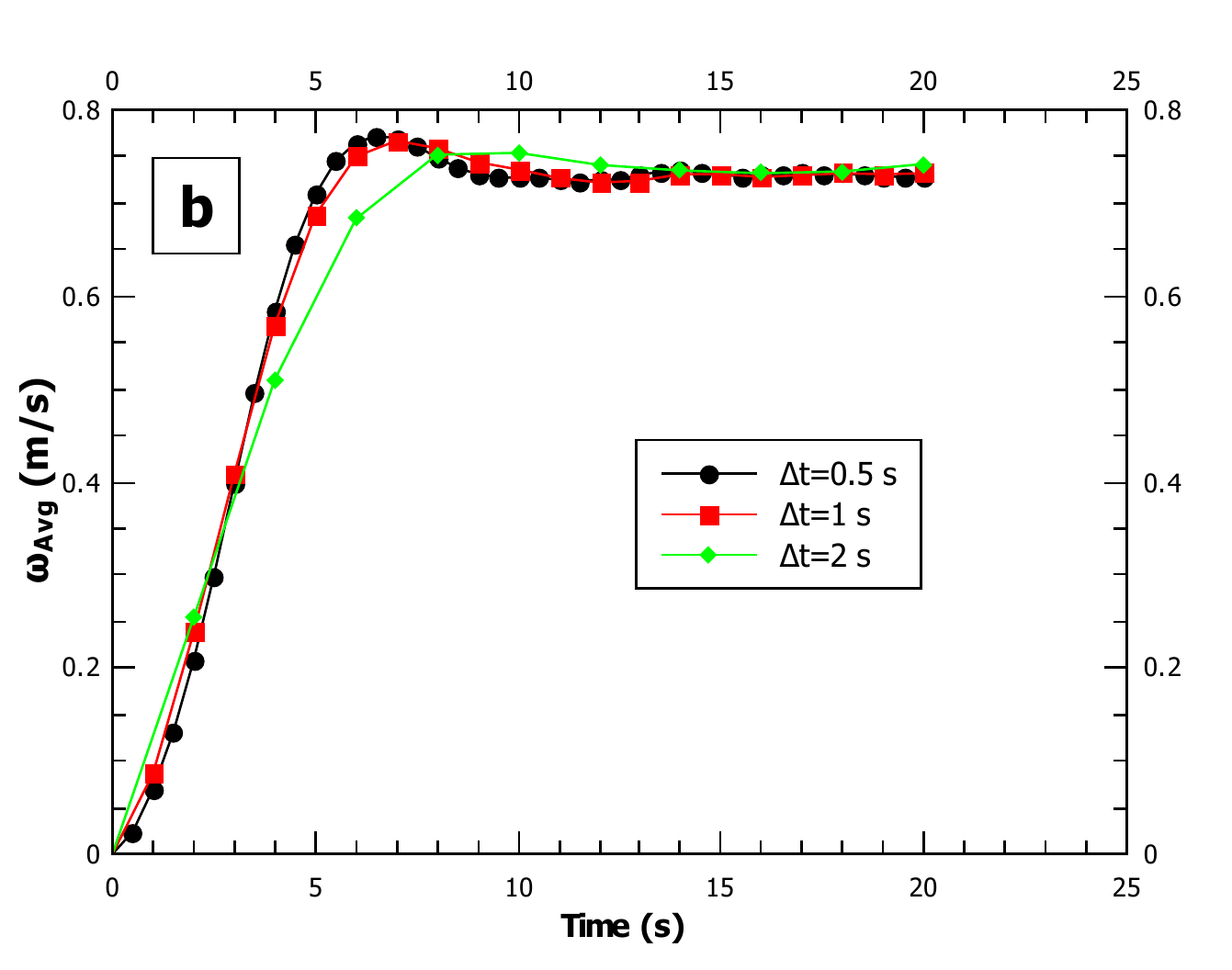}
	\end{subfigure}
	
	\caption{Reliability evaluation of the CFD simulation employing grid and time step independence tests.(a) Grid independence test (b) Time step independence test. }
	\label{Independence tests}
\end{figure}

To confirm the reliability of the results obtained, grid and time step independence tests were performed to ensure that the spatial and temporal discretization errors have a minimal impact on the physics of the study. Figures \ref{Independence tests}a and \ref{Independence tests}b represent the time step and grid independence results, respectively. The tests are considered to be satisfactory when percentage error is less than $5\%$. From the tests, it is observed that a time step of `$1s$' and Mesh A with no of elements  `$=520968$' to be ideal for CFD study for the case when both the loops of the CNCL are filled with FF1. We observe that introducing inflation (Mesh B) has not much effect on the transient behavior, indicating that Mesh A captures all the essential physics of the problem considered. The grid and time step studies were performed for all the cases used for validation with the 1-D semi-analytical model.

\subsection{Flow model selection}

\begin{figure}[!h]
	\centering
	\begin{subfigure}[b]{0.49\textwidth}
		\includegraphics[width=1\linewidth]{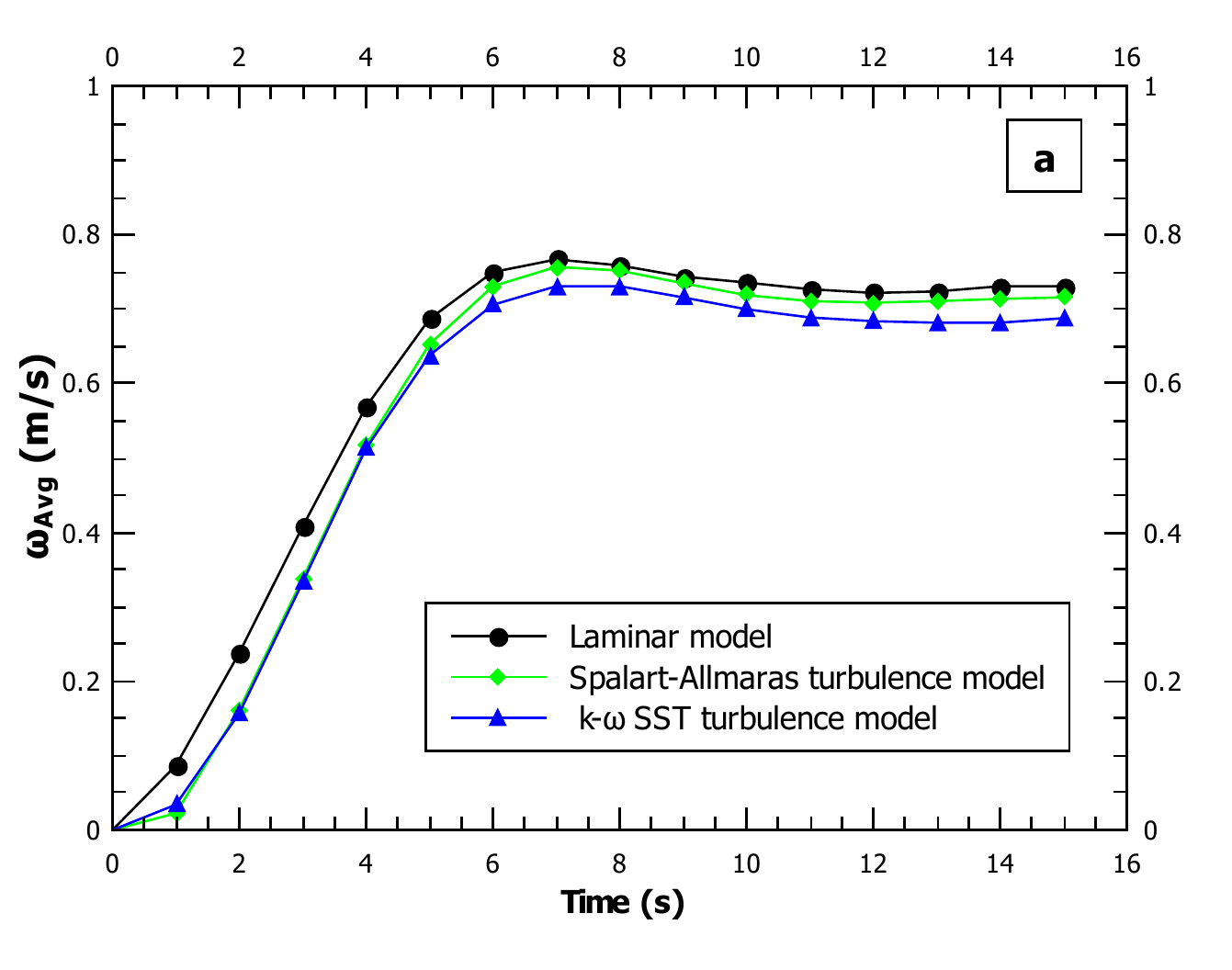}
	\end{subfigure}
	\hspace{\fill}
	\begin{subfigure}[b]{0.49\textwidth}
		\includegraphics[width=1\linewidth]{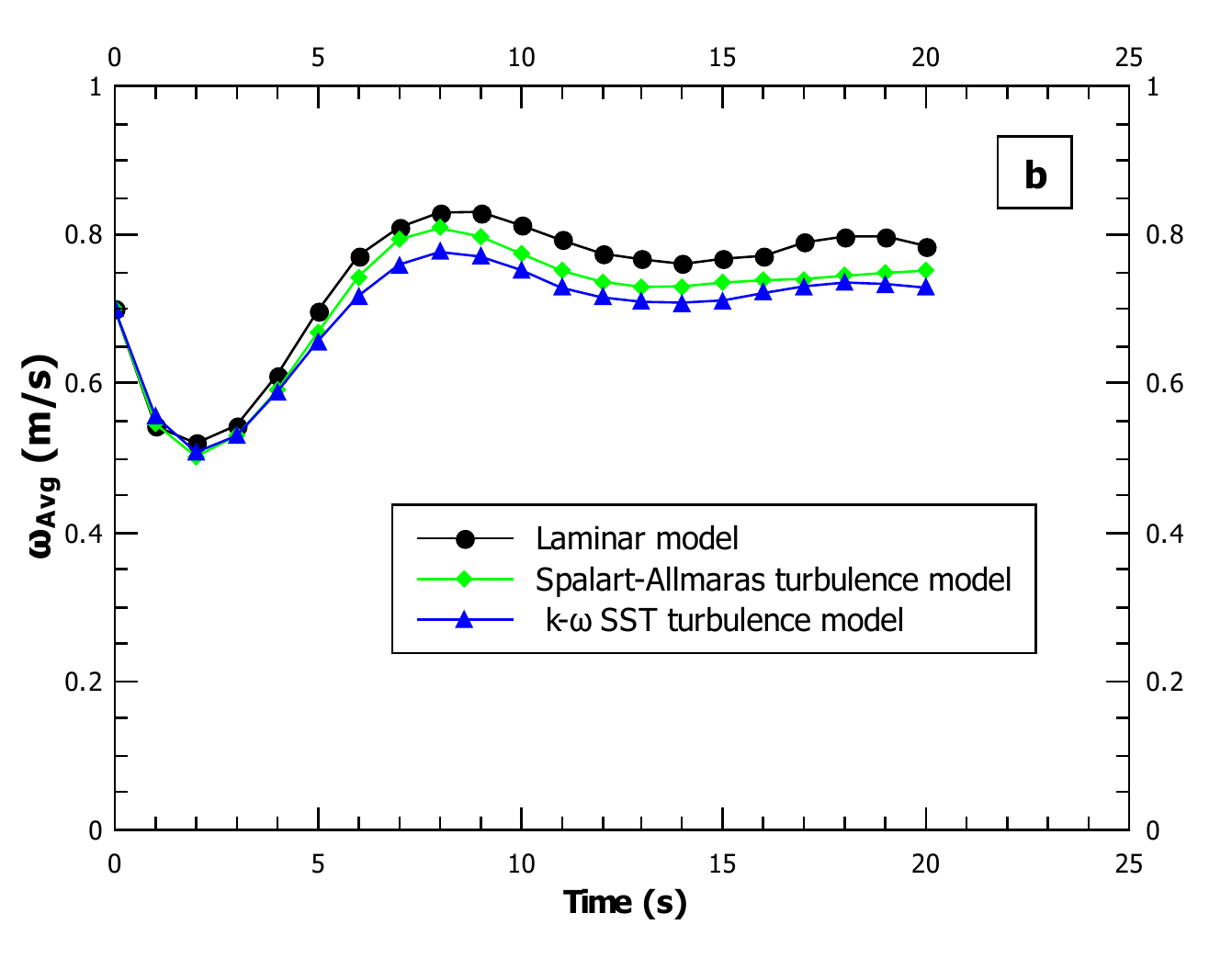}
	\end{subfigure}
	
	\caption{: Effect of different viscosity models on the CNCL system dynamics  for a maximum Reynolds number of ≈ 3000 . (a) Flow model comparison for vertical CNCL for Re = 2920  (b) Flow model comparison for horizontal CNCL for Re = 3160 . }
	\label{fig:fig8viscous-flow-models}
\end{figure}

The  CFD simulations undertaken for the cases CNCL-(a) to CNCL-(d) for the given boundary conditions results in a maximum Reynolds number of $\approx3000$ (corresponding to $0.73$ $ m/s$ for CNCL-(a) and $0.79$ $m/s$ for CNCL-(d)), thus to ensure the validity of selection of the laminar model it is compared with other turbulence models which take care of the transition from laminar to turbulence. From Fig.\ref{fig:fig8viscous-flow-models} we can clearly see that there is an good agreement ($\pm7\%$) among all the flow models considered. Thus from this study one can clearly note that laminar flow model is ideal for capturing the transient dynamics of natural circulation systems which transition across different flow regimes and is in good agreement ($\pm5\%$) with the Spalart-Allmaras turbulence model for the considered cases. Thus the laminar flow model is considered for all the CFD cases for a maximum Reynolds number of approximately 3000 as the transition effects are minimal.

\newpage
\section{CFD results and analysis}

This section provides the validation of the CFD methodology and a brief discussion of the 3-D CFD simulation of the horizontal and vertical CNCL systems. The heat transfer and transient dynamics observed in the CNCL systems are governed by coupled velocity and temperature fields of each of its constituent NCLs as observed in equations (1-4) along with the thermal coupling in the heat exchanger section of the CNCL. Thus the velocity and temperature distribution in both Loop 1 and Loop 2 need to be studied to characterize a CNCL system. 

\subsection{Validation of the CFD methodology}

\begin{figure}[!h]
	\centering
	\includegraphics[width=0.65\linewidth]{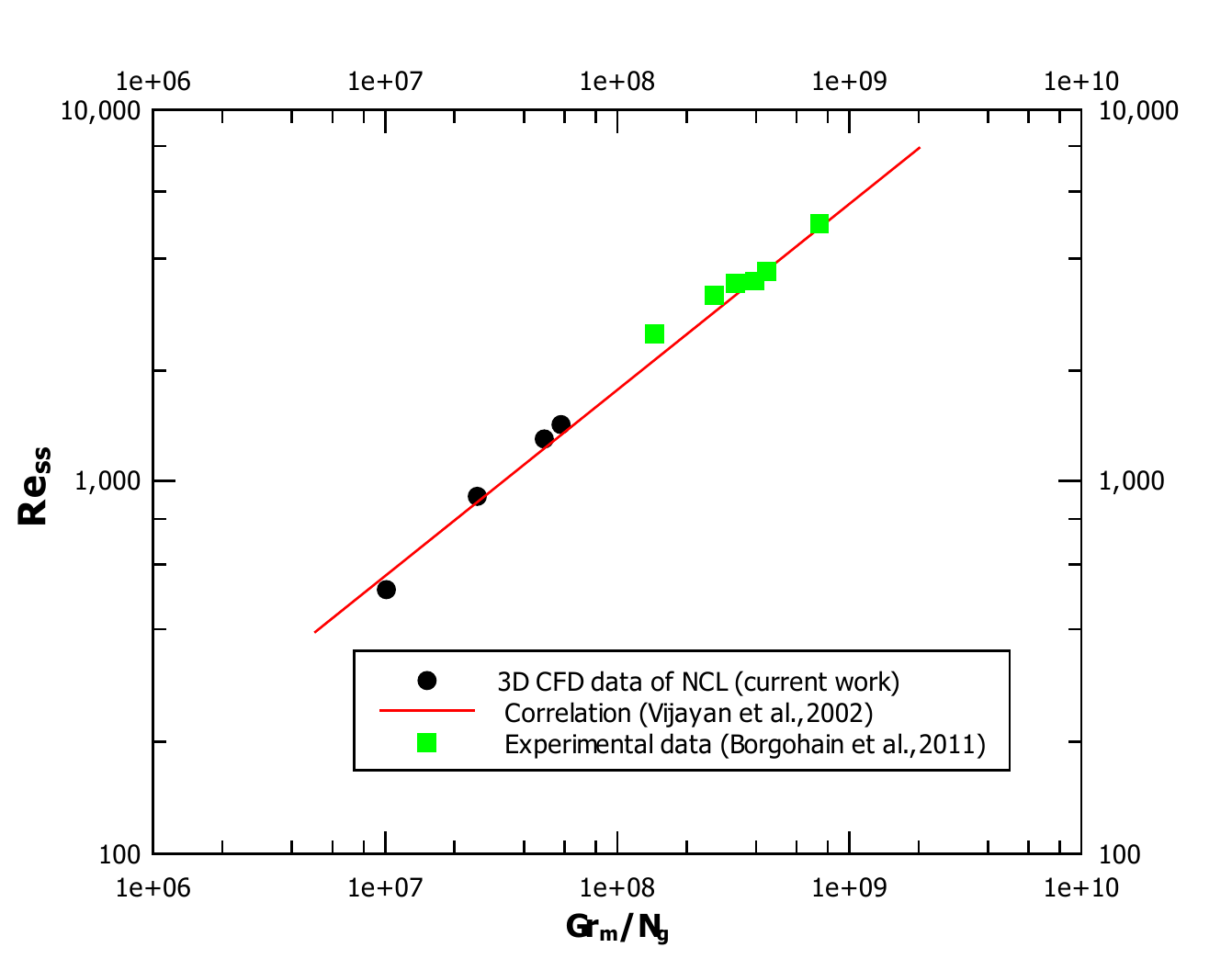}
	\caption{Validation of the CFD methodology with correlation and experimental data.}
	\label{fig:fig15validation-of-cfd-methodology}
\end{figure}

The validation of the CFD methodology is performed by considering an NCL with horizontal heater and cooler sections and having the same geometry as Loop 1 of case CNCL-(a). The CFD simulation is carried out using the same settings and flow models described in section.4. Figure \ref{fig:fig15validation-of-cfd-methodology} indicates an excellent agreement between the CFD simulations and the previous literature, thus successfully validating the CFD study performed. The experimental data is taken from Borgohain et al. \cite{Borgohain2011} and the correlation proposed by Vijayan \cite{Vijayan2002} for the steady state Reynolds number as a function of $Gr_m/N_g$ for an NCL is given as :

\begin{equation}
Re_{ss}=0.1768\bigg(\frac{Gr_m}{N_g}\bigg)^{0.5}
\end{equation}

the definitions of $Gr_m$ and $N_g$ can be found in \cite{Vijayan2002}.

\subsection{Steady state velocity and temperature contours}

\begin{figure}[!h]
	\centering
	\includegraphics[width=\linewidth]{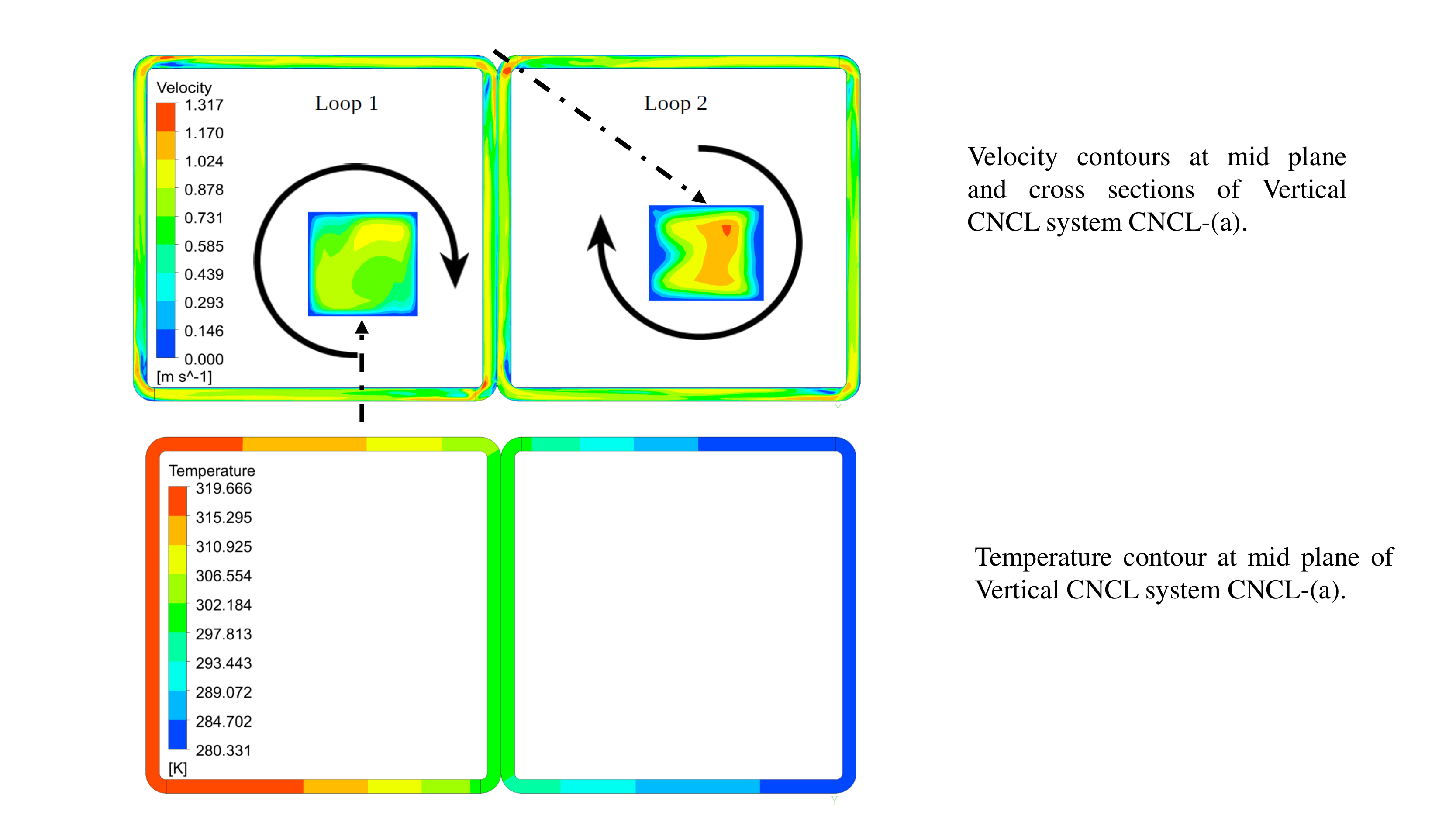}
	\caption{Steady state contours of vertical CNCL considered for study with both the loops containing FF1.}
	\label{fig:fig10verticalcnclvelocitytemperaturecontourrho7070}
\end{figure}

The velocity and temperature contours at steady state are illustrated in Fig.\ref{fig:fig10verticalcnclvelocitytemperaturecontourrho7070}. The velocity of the loop is uniform across the entire domain at the mid-plane except near the bends of the CNCL where the velocity magnitude is relatively lower. This indicates that the 90 $^{\circ}$ elbow bends have a significant impact on the flow. We also observe that the temperature drop or rise occurs only along the direction of flow, with very little radial variation in the temperature profile. This is mainly due to the high thermal conductivity of the fluid considered for the study.

From the velocity contours of the cross-section of Loop 1 and Loop 2 we observe that the velocity is zero at the walls and gradually becomes higher in magnitude at the center, while the peculiar distribution of velocity observed in cross-section contour of Loop 2 is because of the centrifugal forces acting upon on the fluid at the 90$^\circ$ smooth elbow bend.

    \subsection{Transient dynamics of the CNCL system}

\begin{figure}[!h]
	\centering
	\begin{subfigure}[b]{0.85\textwidth}
		\includegraphics[width=1\linewidth]{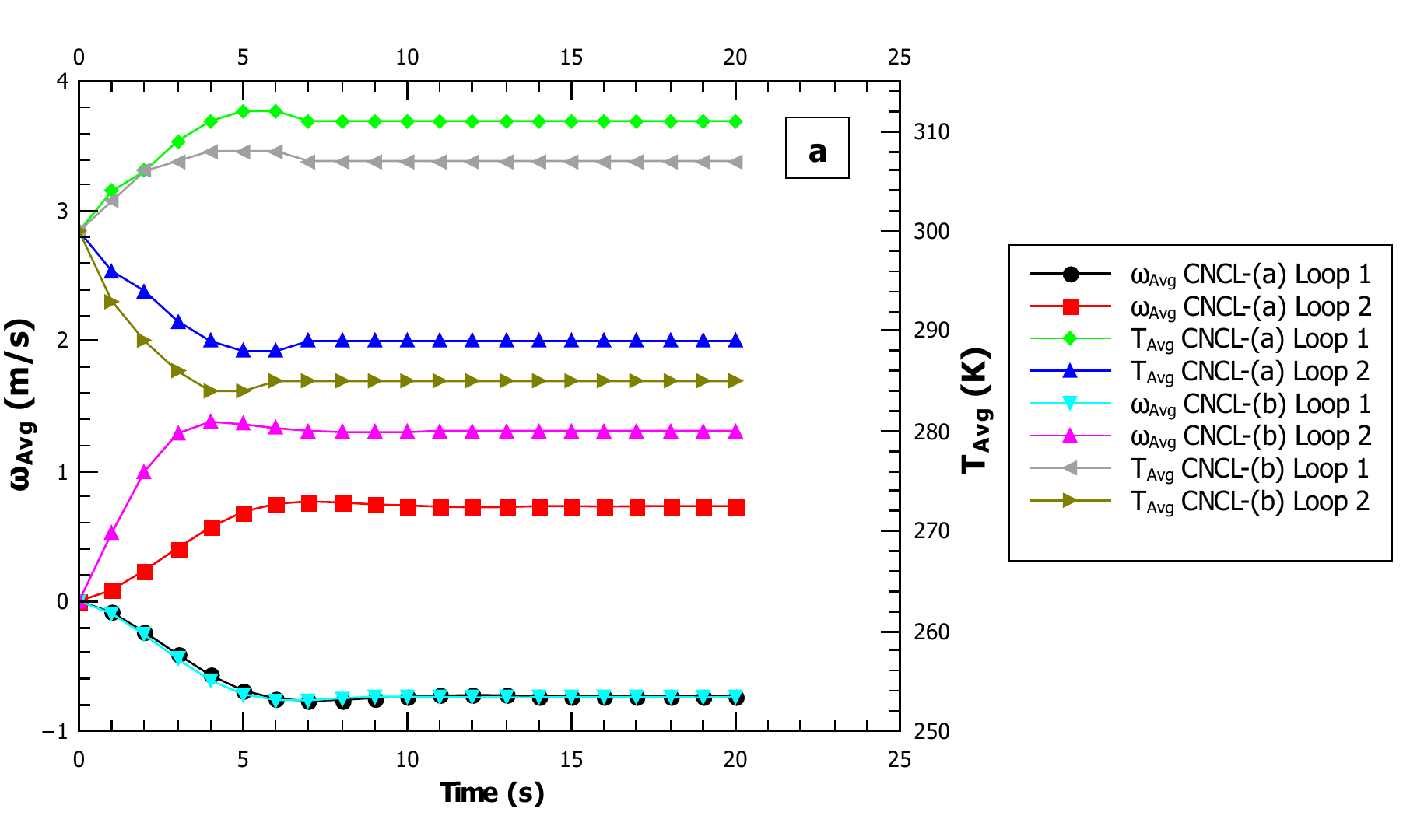}
	\end{subfigure}
	\hspace{\fill}
	\begin{subfigure}[b]{0.85\textwidth}
		\includegraphics[width=1\linewidth]{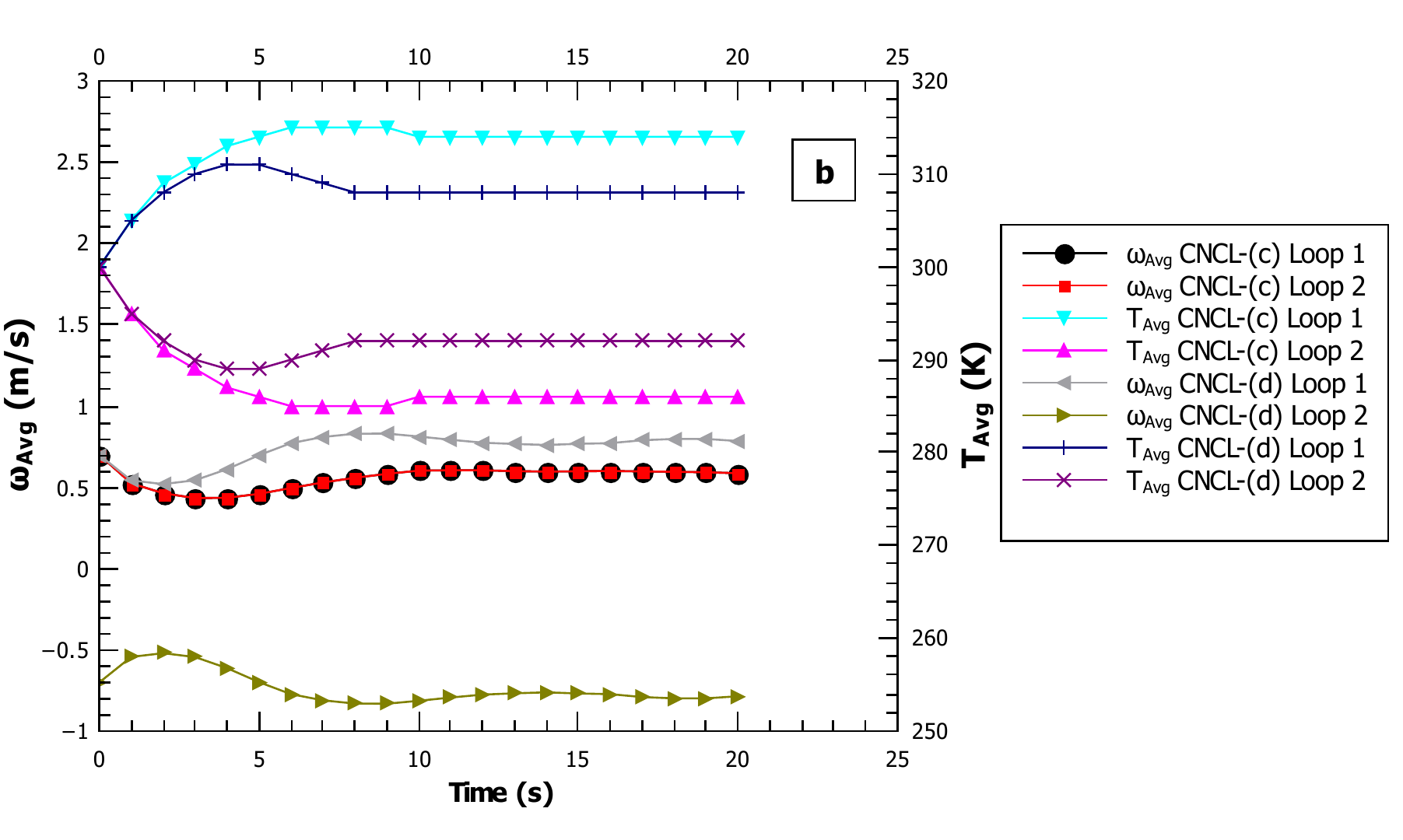}
	\end{subfigure}
	
	\caption{Transient volume averaged plots of velocity and temperature of the vertical and horizontal CNCL system obtained using CFD.(a) Vertical CNCL transient plot (b) Horizontal CNCL transient plot. }
	\label{3D CFD data}
\end{figure}

Average velocity and average temperature are utilized as parameters to compare the 3-D CFD case with the 1-D semi-analytical model. These parameters are averaged over the volume of the respective NCLs, which make up the CNCL system. Figure \ref{3D CFD data} represents the CFD data obtained from the 3-D simulation of the CNCL system for cases CNCL-(a) to CNCL-(d).

From Fig.\ref{3D CFD data} the following observations can be made:
\begin{enumerate}
	\item The introduction of a different fluid in one of the loops( as in the case of CNCL-(b)) does not seem to affect the transient average velocity plot of the other loop, but the transient temperature plot is influenced significantly as observed from Fig.\ref{3D CFD data}a which compares CNCL-(a) and CNCL-(b). 
	\item When both loops contain the same fluid the average transient temperature plot is symmetric about $T_0$ (which is the initial value of the average temperature of the constituent loops of the CNCL system), but when a different fluid (as in the case of CNCL-(b)) is introduced in one of the loops this symmetry is lost.
	\item From Fig.\ref{3D CFD data}b (horizontal CNCL system) we observe that when both the loops are initiated with a certain velocity, depending on the direction of velocity in each of the loops the flow can either be in counter-flow(CNCL-(d)) or parallel-flow configuration(CNCL-(c)). The counter-flow configuration has a greater velocity magnitude in comparison with the parallel-flow configuration for the horizontal CNCL system. The cause behind this effect is elaborated in section 8.2.
	\item For horizontal CNCL, the average temperature of the counter-flow configuration is lower in magnitude than the parallel-flow configuration. This is because of the uniform temperature drop along the length of the heat exchanger section for the counter-flow arrangement in contrast with the parallel-flow arrangement.
\end{enumerate}

The steady state overall heat transfer coefficient ($U$) obtained from CFD simulations for cases CNCL-(a) to CNCL-(c) is about $\approx 19000$  $W/m^2K$ and for CNCL-(d) it is about $\approx 29000$  $W/m^2K$. For the counter flow case (CNCL-(d)) the magnitude of overall heat transfer coefficient is larger due to higher magnitude of velocity at the steady state in comparison with the parallel flow case (CNCL-(c)). The cases considered for the CFD study and the initial and boundary conditions used are shown in Table.\ref*{tab:table2} and Table.\ref{tab:table5} respectively.

\section{Validation of the 1-D model with the 3-D CFD numerical simulation}

An exhaustive 3-D CFD study was performed to properly understand the dynamics and physics of the CNCL system for all the cases mentioned in Table.\ref{tab:table2}. We shall now employ the CFD results to validate the 1-D semi-analytical model developed. 

\begin{figure}[!h]
	\centering
	\begin{subfigure}[b]{0.49\textwidth}
		\includegraphics[width=1\linewidth]{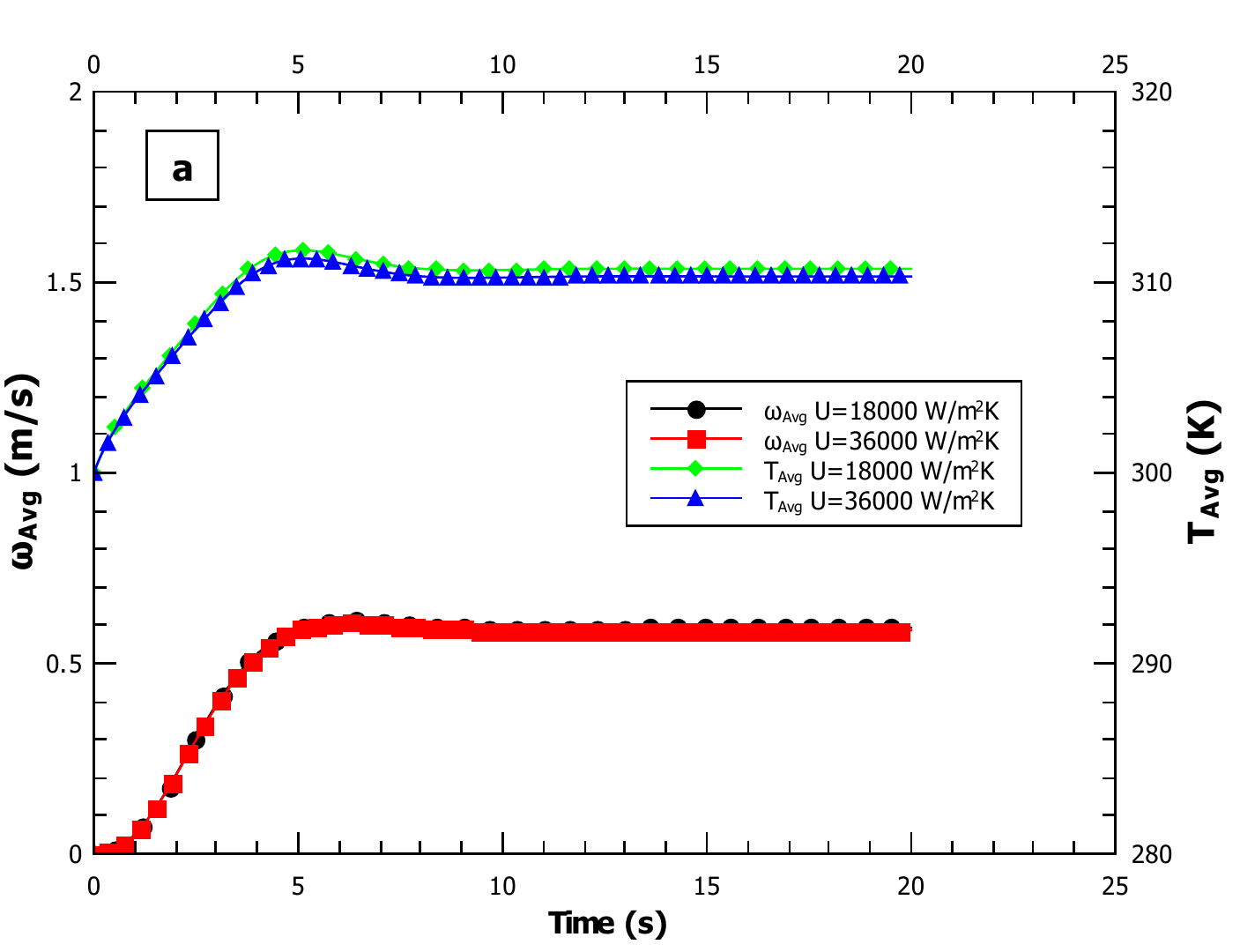}
	\end{subfigure}
	\hspace{\fill}
	\begin{subfigure}[b]{0.49\textwidth}
		\includegraphics[width=1\linewidth]{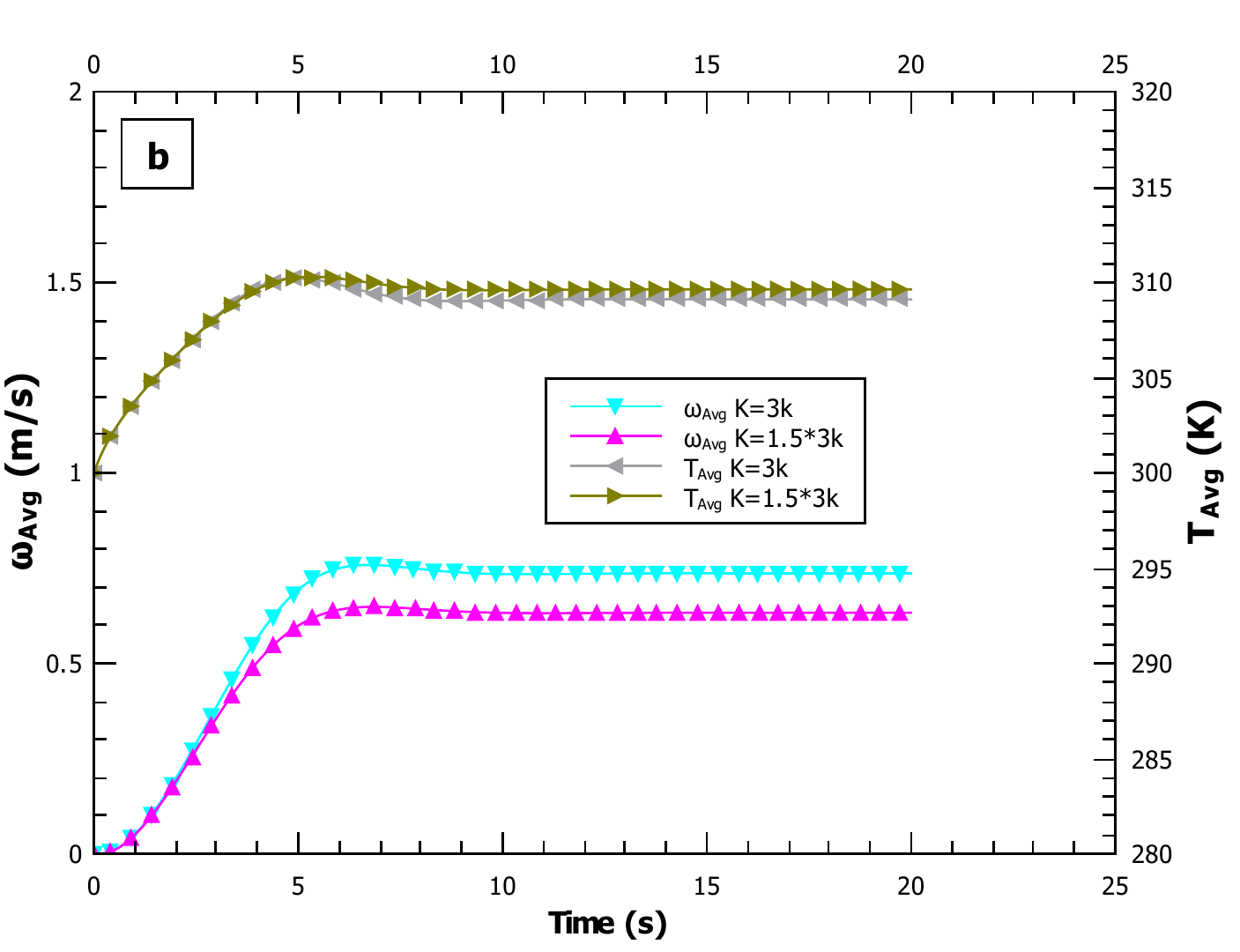}
	\end{subfigure}
	
	\caption{Sensitivity of the CNCL system dynamics to the magnitude of U and K.(a) Sensitivity w.r.t U (b) Sensitivity w.r.t K, U=19000.}
	\label{Senitivity of CNCL with U,K}
\end{figure}

The study conducted in section 5.4  indicates that the magnitude of heat transfer coefficient $U$ for both the horizontal and vertical CNCL system lies in the range: 19,000-29,000 $W/m^2K$. The influence of variation of $U$ on the CNCL system is studied and represented by Fig.\ref{Senitivity of CNCL with U,K}a and it indicates that the system dynamics is not greatly influenced by the heat transfer coefficient. Fig.\ref{Senitivity of CNCL with U,K}b indicates that the loop velocity of CNCL system is significantly affected by the variation in bend loss coefficient magnitude. The sensitivity of the CNCL system to the magnitude of the bend loss coefficient is significant when the Reynolds number of the system is lower, as at very low Reynolds number the value of `K' is much greater than 1. Figure \ref{Bend loss coefficient}b shows that we can expect a $ \pm 50\%$ error in 3K prediction, which explains the deviation presented in Table. \ref{tab:table6}.

To validate the CNCL 1-D model with 3-D CFD results, the length for 1-D model has to be equal to the length of the centerline axis minus two times the radius of curvature of the bend($L=Y-2R,L1=X-2R$) . The bend length needs to be ignored as the bend loss coefficient takes into account the frictional effects and change in momentum at the bend. It is to be noted that this is only for a smooth $90^\circ$ elbow bend considered for the present study. For other fittings the length has to determined after accounting for the loss through the fitting.  
 Thus for validating the present CFD study with 1-D model we employ $L=L1=1-2\times 0.04=0.92$ $m$.

\renewcommand{\arraystretch}{0.7}
\begin{table}[!h]
	\begin{center}
		\caption{Bend loss coefficient (K) used for validation of the 1-D model with CFD.}
		\label{tab:table6}
		\scalebox{0.9}{
			\begin{tabular}{p{3cm} p{3cm} p{2cm} p{2cm} p{4cm}} 
				\textbf{Case} & {\textbf{Reynolds number at steady state }} &{\textbf{3K method prediction}} & {\textbf{K used for validation}} & {\textbf{\% Deviation between 3K prediction and K used for validation}}\\
				\hline 
				CNCL-(a) & 2920 & 0.90 &0.90 &0 \\  
				CNCL-(b) Loop 1 & 2920 & 0.90&0.90 &0 \\ 
				CNCL-(b) Loop 2 & 524 & 2.16& 1.1& 49.07\\
				CNCL-(c) & 2348 & 0.97&  0.97& 0\\ 
				CNCL-(d) & 3160 & 0.88&  0.88&0\\ \hline
		\end{tabular}}
	\end{center}
\end{table}

\begin{figure}[!h]
	\centering
	\begin{subfigure}[b]{0.49\textwidth}
		\includegraphics[width=1\linewidth]{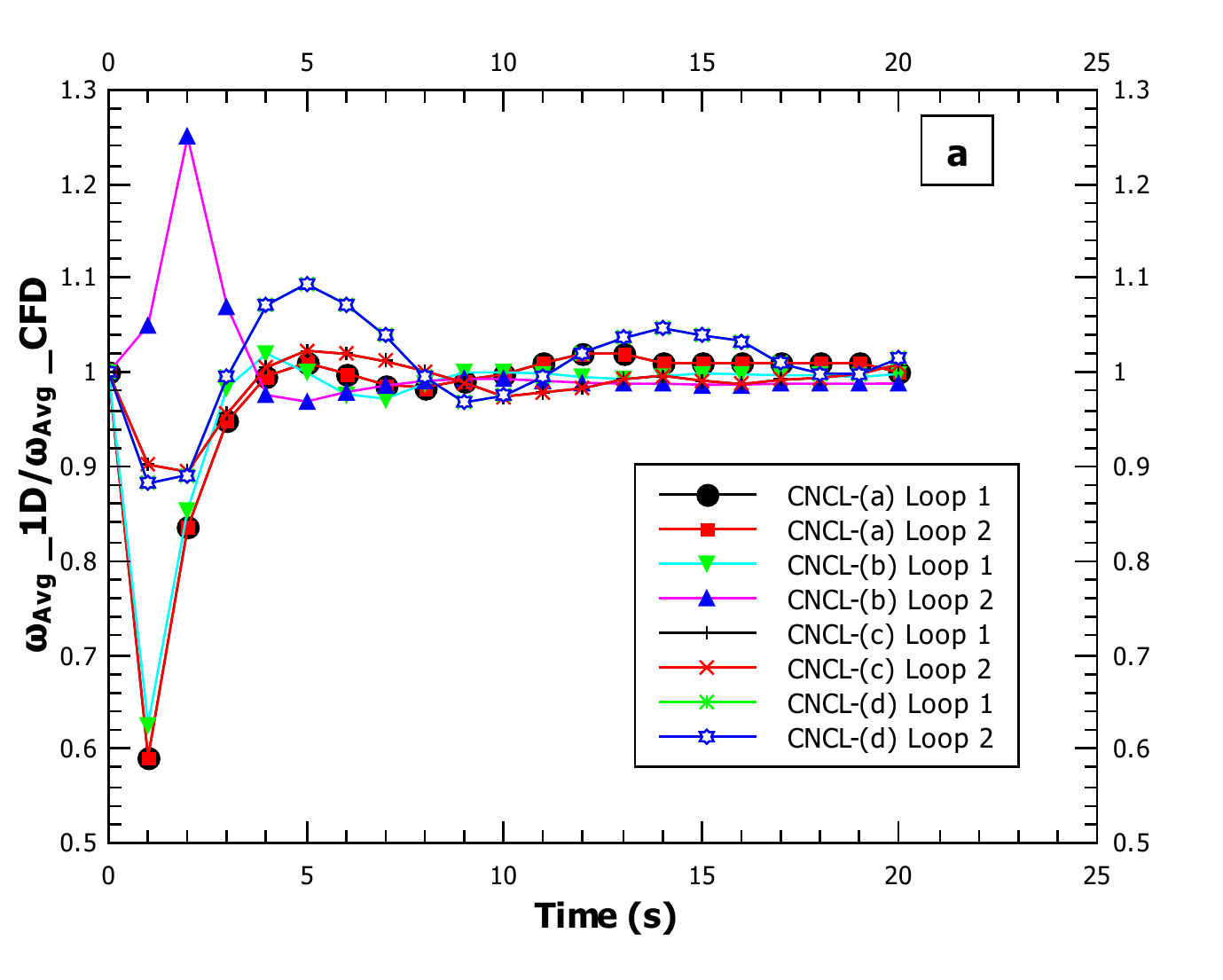}
	\end{subfigure}
	\hspace{\fill}
	\begin{subfigure}[b]{0.49\textwidth}
		\includegraphics[width=1\linewidth]{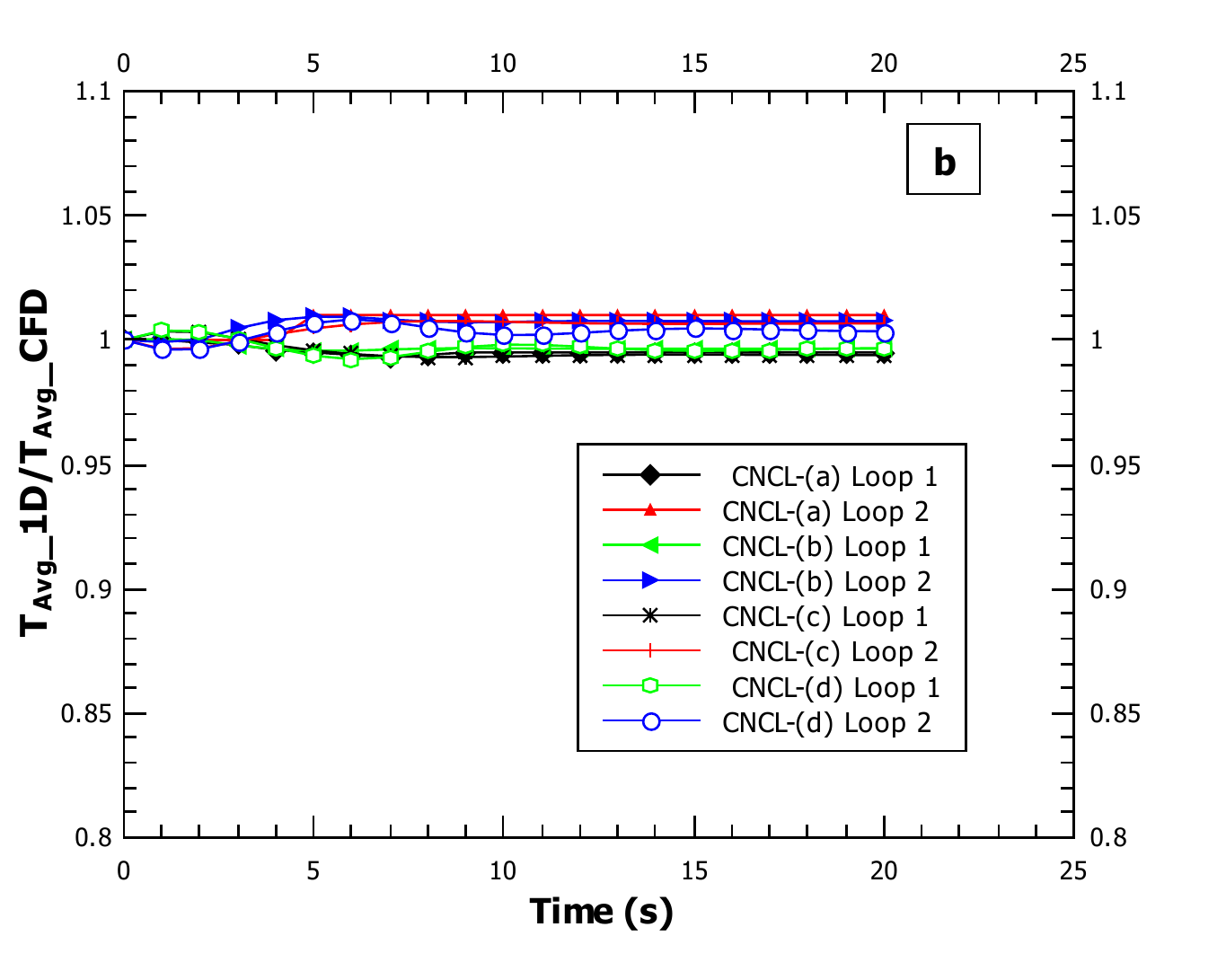}
	\end{subfigure}
	\hspace{\fill}
	\begin{subfigure}[b]{0.49\textwidth}
		\includegraphics[width=1\linewidth]{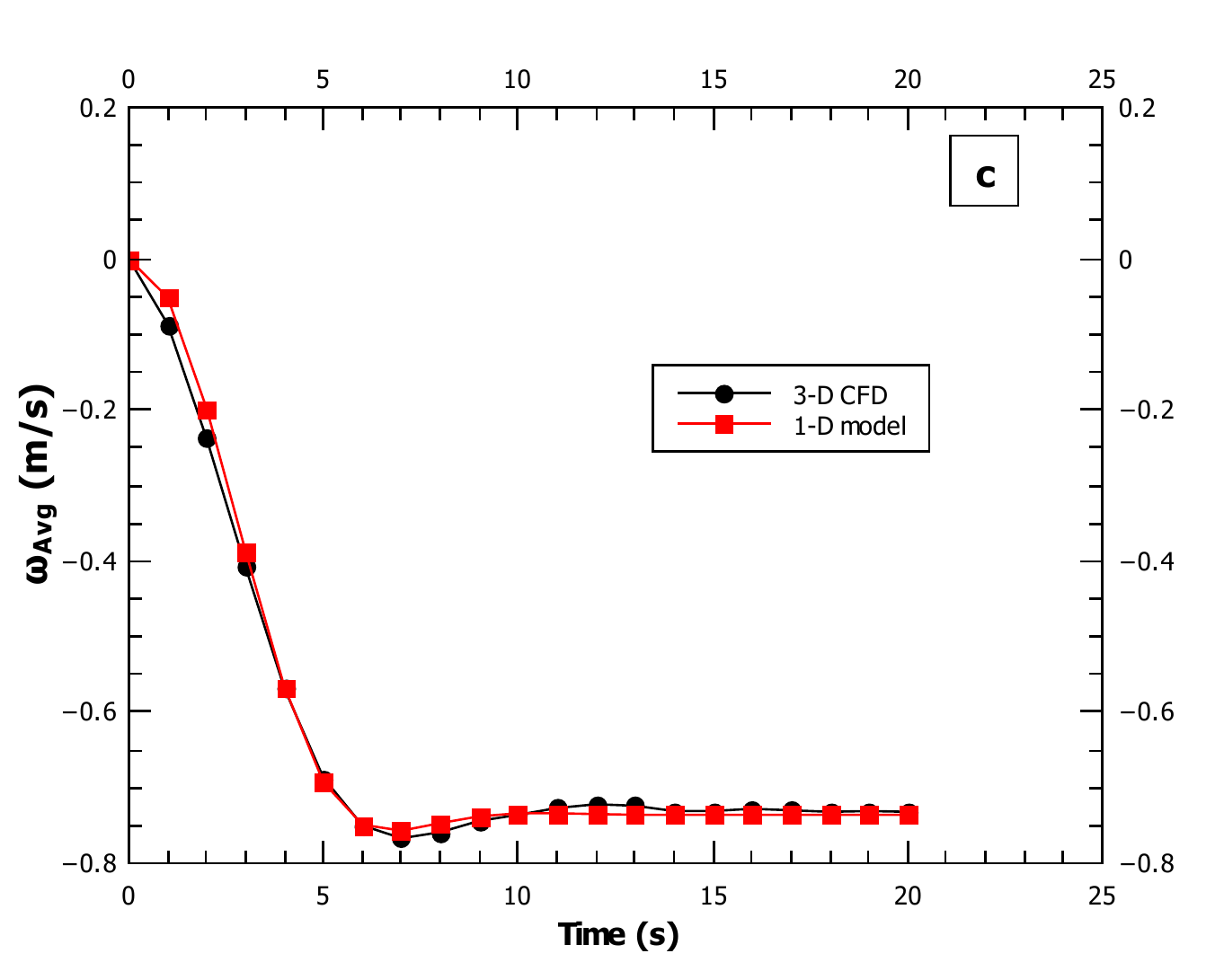}
	\end{subfigure}

    \caption{The validation of the 1-D CNCL model with 3-D CFD simulations are presented in the current figure. (a) The transient plot of ratio average velocity of 1D with CFD (b) Transient plot of ratio average temperature of 1D with CFD (c) Transient plot of the average velocity of 1D and CFD prediction of Loop 1 for case-CNCL-(a).}
\label{validation of 1D with 3D CFD data}
\end{figure}

Figure \ref{validation of 1D with 3D CFD data} indicates a good agreement among the 3D CFD simulations and the 1-D semi-analytical model predictions with $\pm(2-3)\%$ error at steady state for velocity and $\pm1\%$ error at steady state for average temperature. Figure \ref{validation of 1D with 3D CFD data}a indicates that during the initial transience of velocity there is a huge deviation, this is because of the low magnitudes of velocity and because of the fully developed flow assumption at time $t=0$ $s$ (The transient nature of the friction factor in the developing region is not modeled). Figure \ref{validation of 1D with 3D CFD data}c represents the comparison of transient velocity behavior of Loop 1 of CNCL-(a) (which corresponds to the case exhibiting maximum deviation during the initial transience behavior presented in Figure \ref{validation of 1D with 3D CFD data}a), we can clearly see that there is a good agreement between 3-D CFD data and 1D model prediction.

\clearpage

\section{ Non-dimensionalization of the CNCL system}

To perform a general study of the CNCL system and identify the non-dimensional numbers which characterize the CNCL system behavior, the system has been non-dimensionalized.

The non-dimensional CNCL system is represented by the following equations:

\begin{equation}
\frac{d Re_1}{d\zeta}=\Bigg [ Gr_1 \Bigg ] \oint(\theta_1) f(s) ds - \Bigg [ Co_1 \Bigg] (Re_1)^{2-d} -\frac{nK}{4}(Re_1)^2
\end{equation}

\begin{equation}
\frac{d\theta_1}{d\zeta} +  Re_1\frac{d\theta_1}{ds} =
\Bigg[ Fo_1 \Bigg]\frac{d^2\theta_1}{ds^2} + h_1(s) - \Bigg[     St_1 \Bigg]\delta(s)\Bigg(\theta_1- (1/Co_2) \theta_2\Bigg)
\end{equation}

\begin{equation}
\frac{d Re_2}{d\zeta}=\Bigg [ Gr_2 \Bigg ] \oint(\theta_2) f(s) ds - \Bigg [ Co_1 \Bigg] (Re_2)^{2-d} -\frac{nK}{4}(Re_2)^2
\end{equation}

\begin{equation}
\begin{split}
\frac{d\theta_2}{d\zeta} + \Bigg[\frac{\nu_2}{\nu_1}\Bigg] Re_2\frac{d\theta_2}{ds}  =
\Bigg[ Fo_2 \Bigg]\frac{d^2\theta_2}{ds^2} + h_2(s)  + \Bigg[     St_2 \Bigg]\delta(s)\Bigg( Co_2 \theta_1- \theta_2\Bigg)
\end{split}
\end{equation}

where, $\theta_i={T_i-T_0}/{\Delta T_i}$, $\zeta={t}/{t_0}$, $s={x}/{x_0}$, $t_0={x_0D_h}/{\nu_1} $, $\Delta T_i=(4Q"t_0)/(\rho_i Cp_i D_h) $, $x_0=(L+L1)$  and the non-dimensional parameters are defined as follows:

\begin{equation}	
Co_1=\frac{2bx_0}{D_h}
\end{equation}

\begin{equation}	
Co_2=\frac{\Delta T_1}{\Delta T_2}
\end{equation}

	\begin{equation}	
Fo_i=\frac{\alpha_it_{0,i}}{x_0^2}
\end{equation}

\begin{equation}	
St_i=\frac{Ut_{0,i}}{\rho_i Cp_i D_h}
\end{equation}

\begin{equation}		
Re_i=\frac{\omega_iD_h}{\nu_i}
\end{equation}

\begin{equation}	
Gr_i=\frac{g \beta_i \Delta T_i x_0 D_h t_{0,i}}{(L+L1) \nu_i}
\end{equation}

From the above-mentioned equations, we can clearly see that Reynolds number, Grashof number, Fourier number, Stanton number, $Co_1$ and $Co_2$ determine the complete CNCL system behavior. A thorough parametric study is undertaken to understand the physics of the CNCL system considering the same fluid within the two loops. Considering same fluid in both loops leads to the following simplification $Gr_1=Gr_2=Gr$, $Fo_1=Fo_2=Fo$, $St_1=St_1=St$, $\Delta T_1=\Delta T_2 \implies Co_2=1$.The simplified governing equations are represented as follows:

\begin{equation}
\frac{d Re_1}{d\zeta}=\Bigg [ Gr \Bigg ] \oint(\theta_1) f(s) ds - \Bigg [ Co_1 \Bigg] (Re_1)^{2-d} -\frac{nK}{4}(Re_1)^2
\end{equation}

\begin{equation}
\frac{d\theta_1}{d\zeta} +  Re_1\frac{d\theta_1}{ds} =
\Bigg[ Fo \Bigg]\frac{d^2\theta_1}{ds^2} + h_1(s) - \Bigg[     St \Bigg]\delta(s)\Bigg(\theta_1- \theta_2\Bigg)
\end{equation}

\begin{equation}
\frac{d Re_2}{d\zeta}=\Bigg [ Gr \Bigg ] \oint(\theta_2) f(s) ds - \Bigg [ Co_1 \Bigg] (Re_2)^{2-d} -\frac{nK}{4}(Re_2)^2
\end{equation}

\begin{equation}
\begin{split}
\frac{d\theta_2}{d\zeta} +  Re_2\frac{d\theta_2}{ds}  =
\Bigg[ Fo\Bigg]\frac{d^2\theta_2}{ds^2} + h_2(s)  + \Bigg[     St \Bigg]\delta(s)\Bigg(\theta_1- \theta_2\Bigg)
\end{split}
\end{equation}

\section{Results and Discussion}

\begin{figure}[!h]
	\centering
	\includegraphics[width=0.8\linewidth]{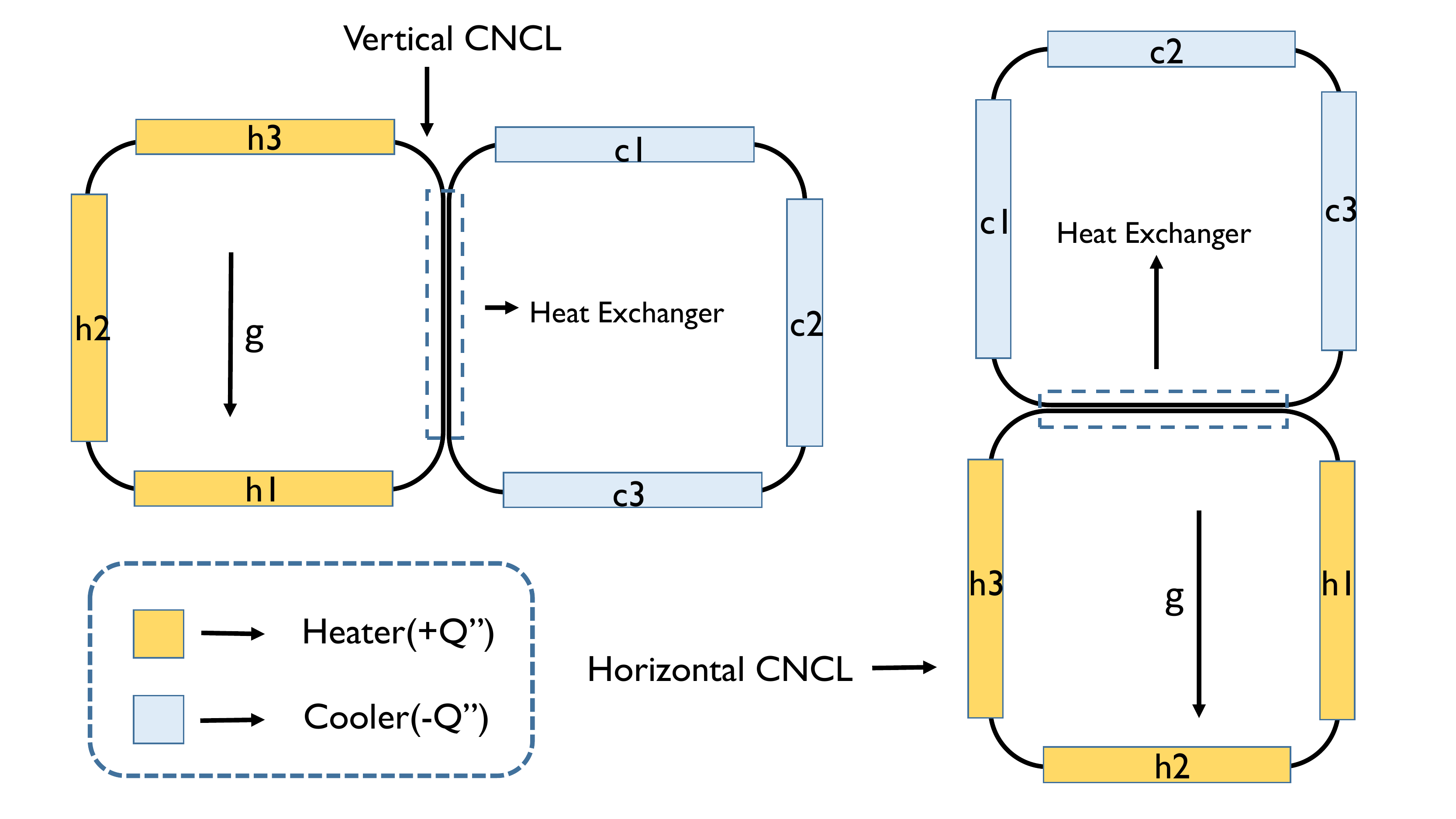}
	\caption{Heater and Cooler notations used for the current study.The configuration `Vertical CNCL h1c1' refers to a vertical CNCL system where, only heater-`h1' and cooler-`c1' are switched on and the remaining heaters(`h2' and `h3') and coolers (`c2' and `c3') are set to $Q^{\prime\prime}=0$ $W/m^2$. The horizontal CNCL system is obtained by rotating the vertical CNCL system anticlockwise w.r.t gravity as represented in the figure.}
	\label{fig:fig15heater-cooler-orientations}
\end{figure}

A detailed parametric study is undertaken to understand the physics of the CNCL system. The CNCL system containing the same fluid in its component loops is considered as it leads to symmetrical behavior of transient velocity and average temperature plots about the initial conditions when heater and cooler positions similar to cases CNCL-(a) to CNCL-(d) are used. This allows us to describe the complete CNCL system using $Re=\frac{|\omega_{1,2}|D_h}{\nu}$ and $\theta_{Avg}=|\theta_{Avg,1,2}|$ unless explicitly stated. For simplifying the specification of the heater-cooler location on both the vertical and horizontal CNCL systems we use the notation represented in Fig.\ref{fig:fig15heater-cooler-orientations} from here onwards. In this study, we are solely interested in studying the effect of one heater and cooler per CNCL system. 

\subsection{Comparison of vertical and horizontal CNCL systems}

\begin{figure}[!htb]
	\centering
	\includegraphics[width=0.6\linewidth]{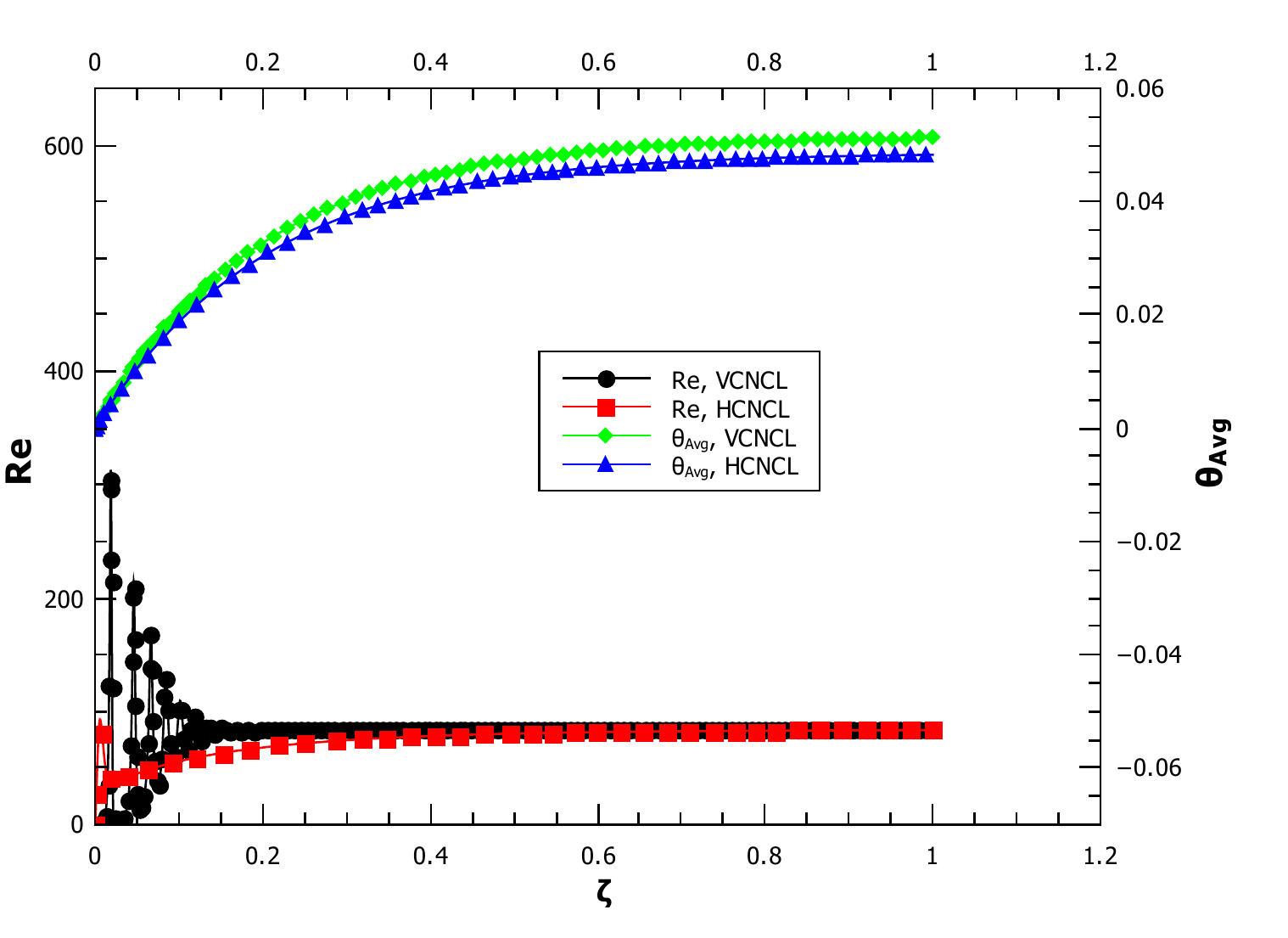}
	\caption{Comparison of transient behaviour of vertical and horizontal CNCL system for heater cooler arrangement h1c1 for $Gr$=$10^8$, $Fo$=1, $As$=1, $St$=10, $Co1$=1000.}
	\label{fig:fig16comparison-of-vertical-and-horizontal-system-transience-for-h1c1}
\end{figure}

Figure \ref{fig:fig16comparison-of-vertical-and-horizontal-system-transience-for-h1c1} compares the  horizontal and vertical CNCL systems for heater cooler configuration h1c1. We can observe a distinct difference in the behavior exhibited by the vertical and horizontal systems, as the heater h1 and cooler c1 are located on the horizontal legs (w.r.t gravity) of the vertical CNCL system (VCNCL) and on the vertical legs (w.r.t gravity) of the horizontal CNCL system (HCNCL).

The horizontal orientation of the heater and cooler on the VCNCL system implies that the energy supplied at the heater and cooler must propagate through conduction (within the fluid) to the vertical legs and only then can the buoyancy forces begin to act to drive the fluid within the loops of the system. This initial delay causes the vertical CNCL to have lesser oscillations in the `$Re$ vs $\zeta$' plot and a higher $\theta_{Avg}$ at steady state.

\subsection{Comparison of counter flow and parallel flow behavior }

The HCNCL system with h2c2 heater cooler orientation is the only configuration where both the counter flow and parallel flow at the common heat exchanger section are exhibited just by changing the initial velocity conditions. The parallel flow configuration is obtained when the initial velocity at time $t=0$ $s$ is in positive direction in both Loop 1 and Loop 2 (i.e. both loops have clockwise velocity w.r.t origin `O'. The counter flow configuration can be obtained when the initial velocity at time $t=0$ $s$ is in the positive direction in Loop 1 and negative direction in Loop 2 (i.e. Loop 1 has clockwise velocity and Loop 2 has anti-clockwise velocity w.r.t origin `O').

\begin{figure}[!htb]
	\centering
	\includegraphics[width=0.6\linewidth]{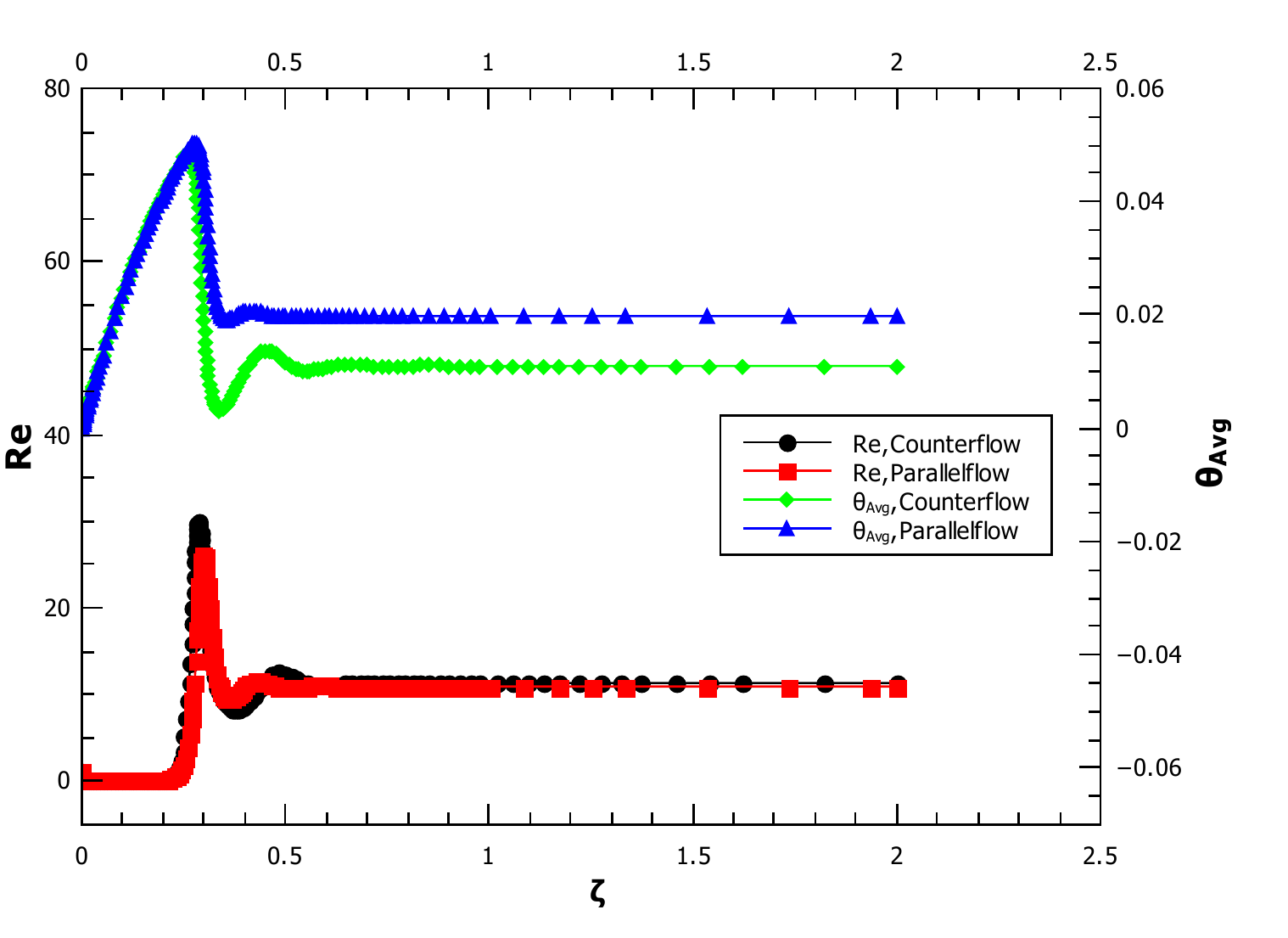}
	\caption{Effect of parallel and counter flow configurations on horizontal CNCL with h2c2 heater cooler configuration for $Gr$=$10^6$, $Fo$=1, $As$=1, $St$=100, $Co1$=1000.}
	\label{fig:fig17hcncl-h2c2-parallel-vs-counter-flow-transience}
\end{figure}

From Fig.\ref{fig:fig17hcncl-h2c2-parallel-vs-counter-flow-transience} we observe that for the given conditions the steady state $\theta_{Avg}$ for counterflow is lower than parallel flow, this is because of the temperature distribution within the fluid at the common heat exchange section for Loop 1 and Loop 2.
For both the counterflow and parallel flow conditions we observe from the `$Re$ vs $\zeta$' plot that for the h2c2 heater cooler configuration, both the heater and cooler are on the horizontal legs (w.r.t gravity) of the HCNCL, thus no flow is initiated until the energy is transferred to the vertical legs. This is the reason we can initialize the flow conditions to attain the parallel flow or counterflow arrangement at the common heat exchange section.

The non-dimensional average temperature for both counter and parallel flow condition increases at the same rate till the flow is initiated. After the flow is initiated by the buoyancy forces, the magnitude of $\theta_{Avg}$ begins to drop for both parallel and counter flow condition, thus the case which attains steady state quicker has a higher magnitude of $\theta_{Avg}$.
For the parallel flow condition the temperature difference between the hot and cold fluids along the flow direction decreases, but it is nearly constant for the counterflow condition. This implies that the heat transfer along the flow direction for parallel flow condition at the common heat exchange section decreases, while it is nearly constant along the flow direction for counterflow condition. This promotes a faster heat transfer rate for the parallel flow condition compared to the counterflow condition, which results in the parallel flow case attaining the steady state faster. This in turn leads to a larger magnitude of $\theta_{Avg}$ and lower value of $Re$ for the parallel flow case compared to the counterflow case.

 \subsection{Effect of Grashof number on the CNCL system}

 \begin{figure}[!h]
 	\centering
 	\begin{subfigure}[b]{0.49\textwidth}
 		\includegraphics[width=1\linewidth]{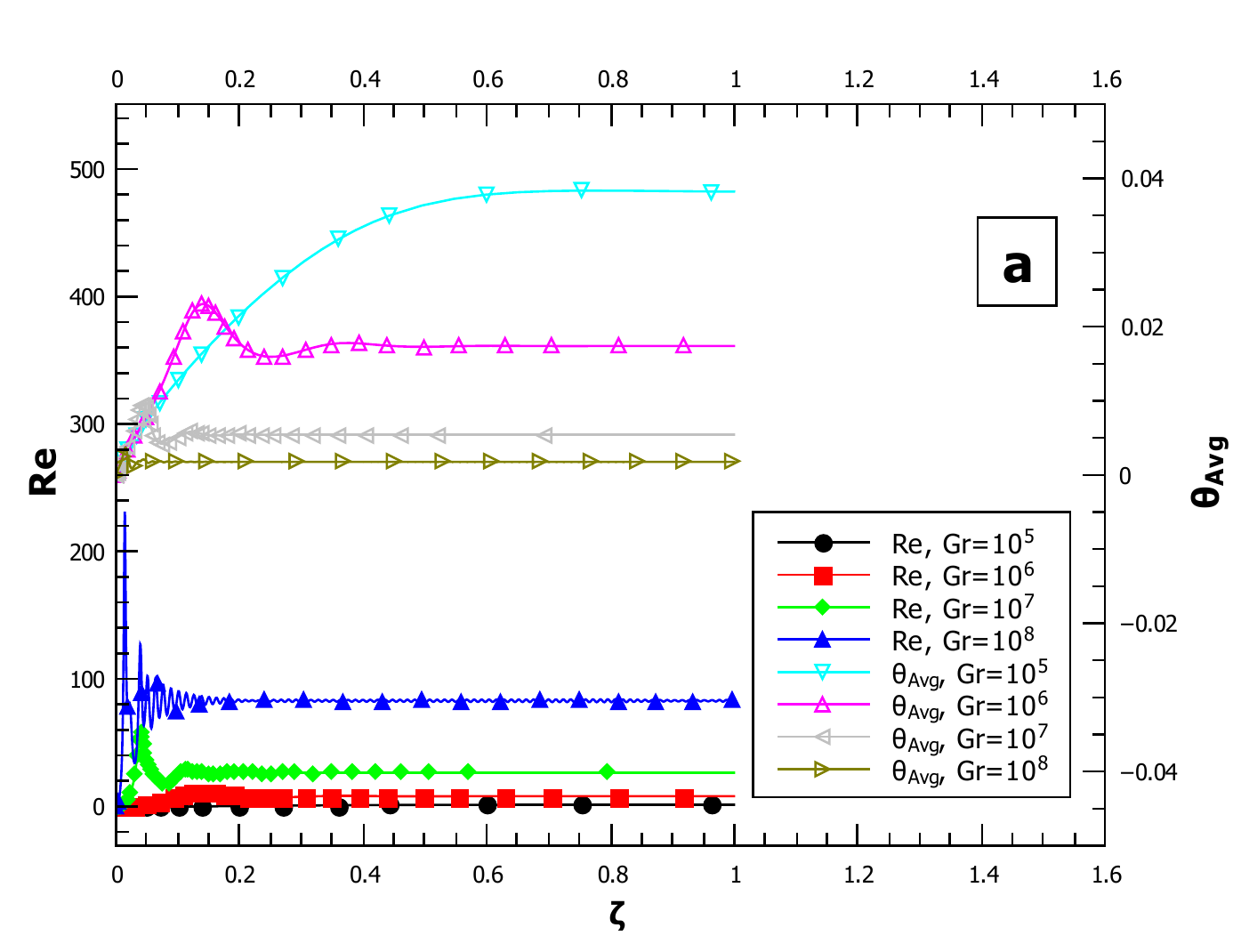}
 	\end{subfigure}
 	\hspace{\fill}
 	\begin{subfigure}[b]{0.49\textwidth}
 		\includegraphics[width=1\linewidth]{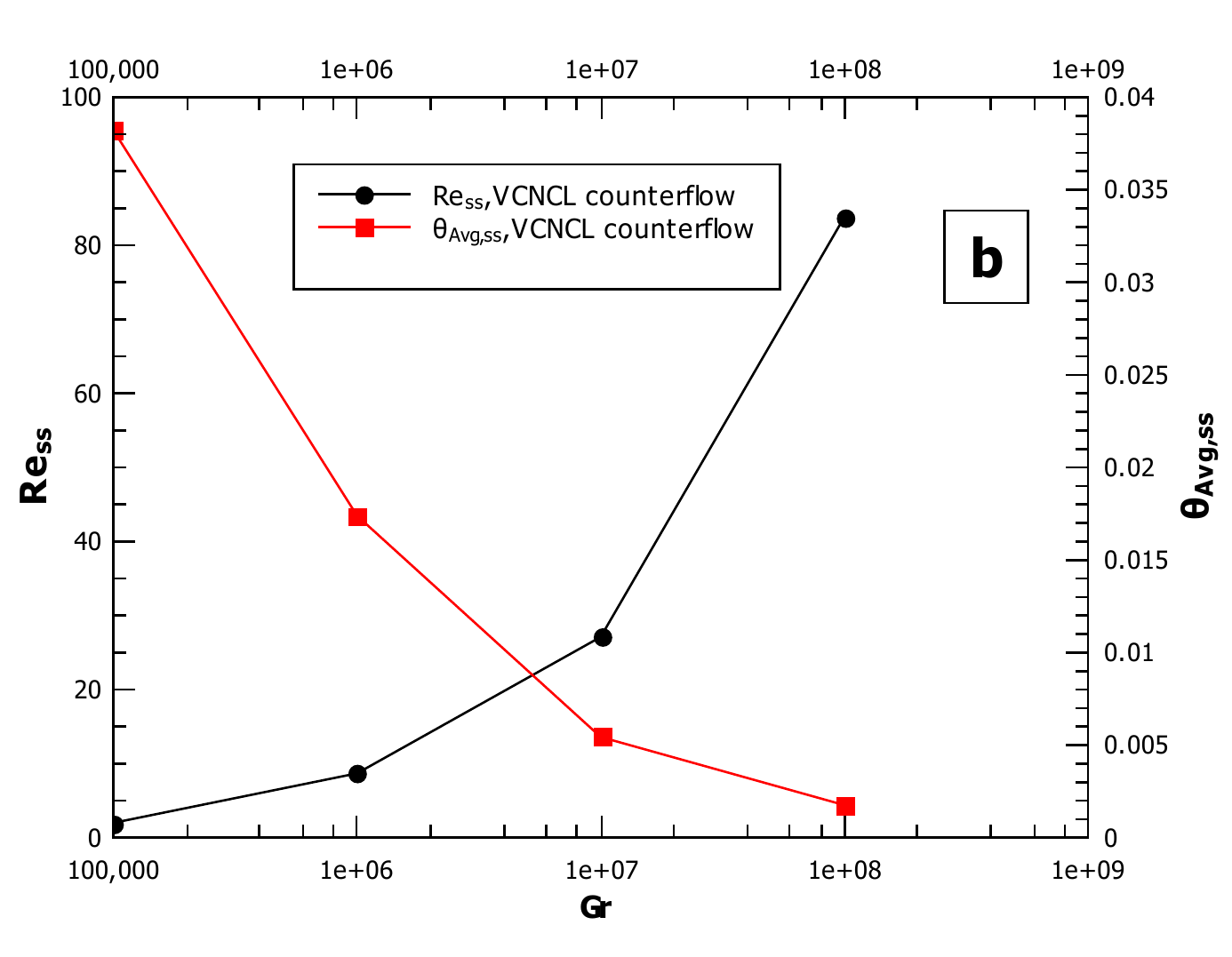}
 	\end{subfigure}
 	
 	\caption{Effect of Grashof number ($Gr$) on the VCNCL system h1c1 configuration for $As$=1, $St$=3000, $Fo$=1, $Co1$=1000.(a) Transient behaviour (b) Steady state trend.}
 	\label{Effect of Gr}
 \end{figure}

From Fig.\ref{Effect of Gr}a, we observe that with an increase in Grashof number the flow oscillations increase along with an increase in the steady-state Reynolds number and decrease in $\theta_{Avg}$. 

The flow in the component loops of the CNCL system is induced by $\oint(\theta_i) f(s) ds$, which represents the temperature distribution in the vertical legs (w.r.t gravity) of the CNCL system and determines the magnitude of the buoyancy forces \big($Gr\oint(\theta_i) f(s) ds$\big).

With the increase in Grashof number, the magnitude of the buoyancy forces increases which push the fluid at a faster rate resulting in higher fluid velocity. The forces resulting from friction and momentum change at the bend act opposite the buoyancy force and try to restrain the fluid motion. Because the fluid is flowing at a faster rate within the loop, this leads to a quicker attainment of a temperature distribution along the loop such that the magnitude of $\oint(\theta_1) f(s) ds$ reduces which in-turn leads to the decrease in buoyancy forces. The decrease in buoyancy forces causes a decrease in the flow velocity and a simultaneous increase in the viscous forces (as it is inversely propotional to $Re$ according to equation-11 ), which further tries to decrease the flow velocity. As the flow velocity drops, the temperature distribution ($\oint(\theta_1) f(s) ds$)  increases leading to an increase in the buoyancy forces yet again. The process repeats itself till all the forces balance each other, which satisfies the following condition:
\begin{equation}
Gr\oint(\theta_i) f(s) ds =  Co_1(Re_{i,ss})^{2-d} +\frac{nK}{4}(Re_{i,ss})^2
\end{equation}

This explains the oscillations observed in the '$Re$ vs $\zeta$' plot. Equation 54 explains why the magnitude of the steady-state Reynolds number increases with $Gr$.

Fig.\ref{Effect of Gr}b depicts the steep rate of increase in $Re$ and decrease in the rate of drop of $\theta_{Avg}$ with increasing $Gr$. The steep rate of increase in $Re_ss$ is due to the steep decline in the viscous forces with increasing Reynolds number. Increase in $Gr$ results from an increase in the magnitude of $\Delta T$. Since $\theta_{Avg} \propto \frac{1}{\Delta T}$ with increasing $Gr$ the magnitude of $\theta_{Avg}$ drops, this also explains the rate of drop of $\theta_{Avg}$.

\subsection{Effect of Fourier number on the CNCL system}

\begin{figure}[!h]
	\centering
	\begin{subfigure}[b]{0.49\textwidth}
		\includegraphics[width=1\linewidth]{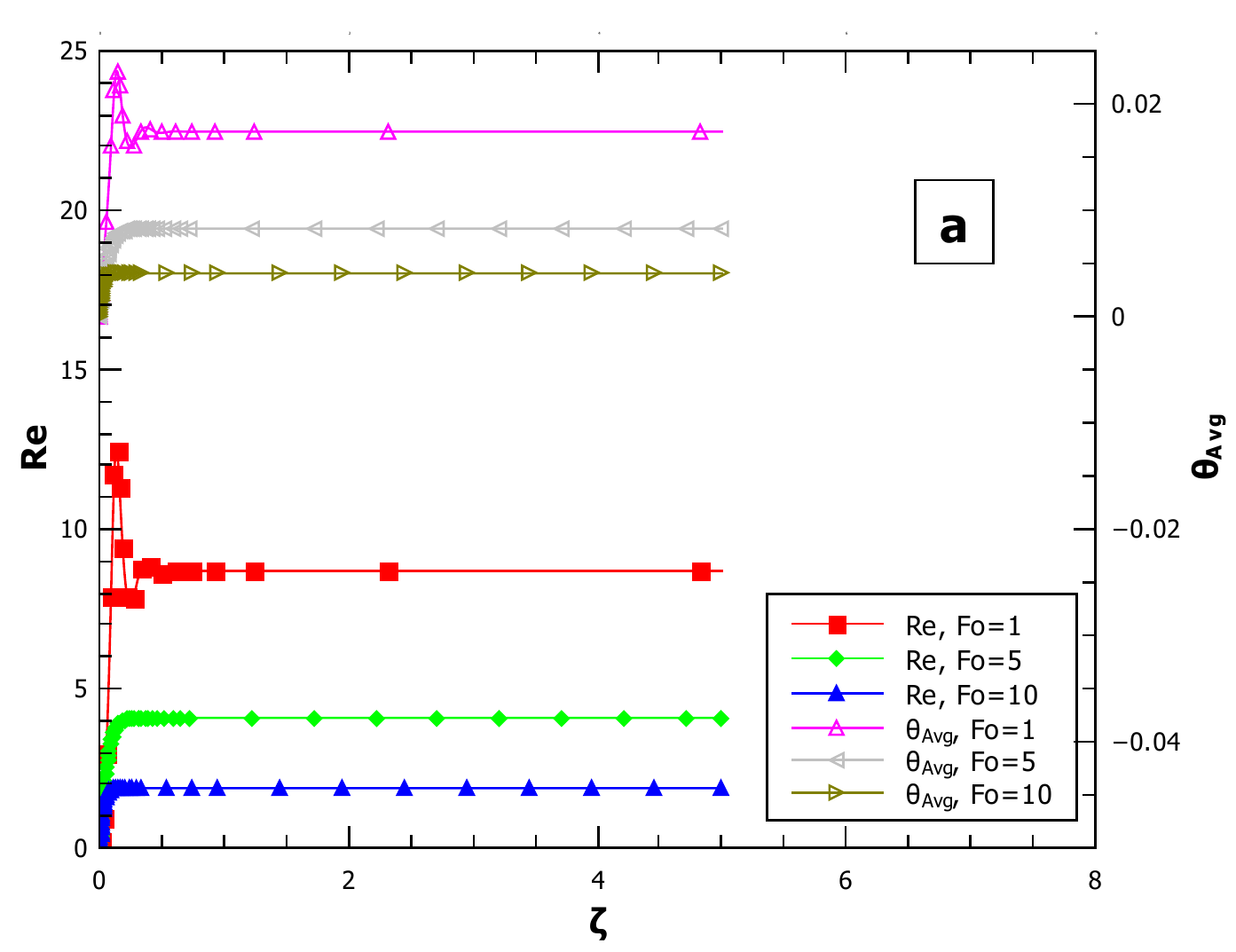}
	\end{subfigure}
	\hspace{\fill}
	\begin{subfigure}[b]{0.49\textwidth}
		\includegraphics[width=1.0\linewidth]{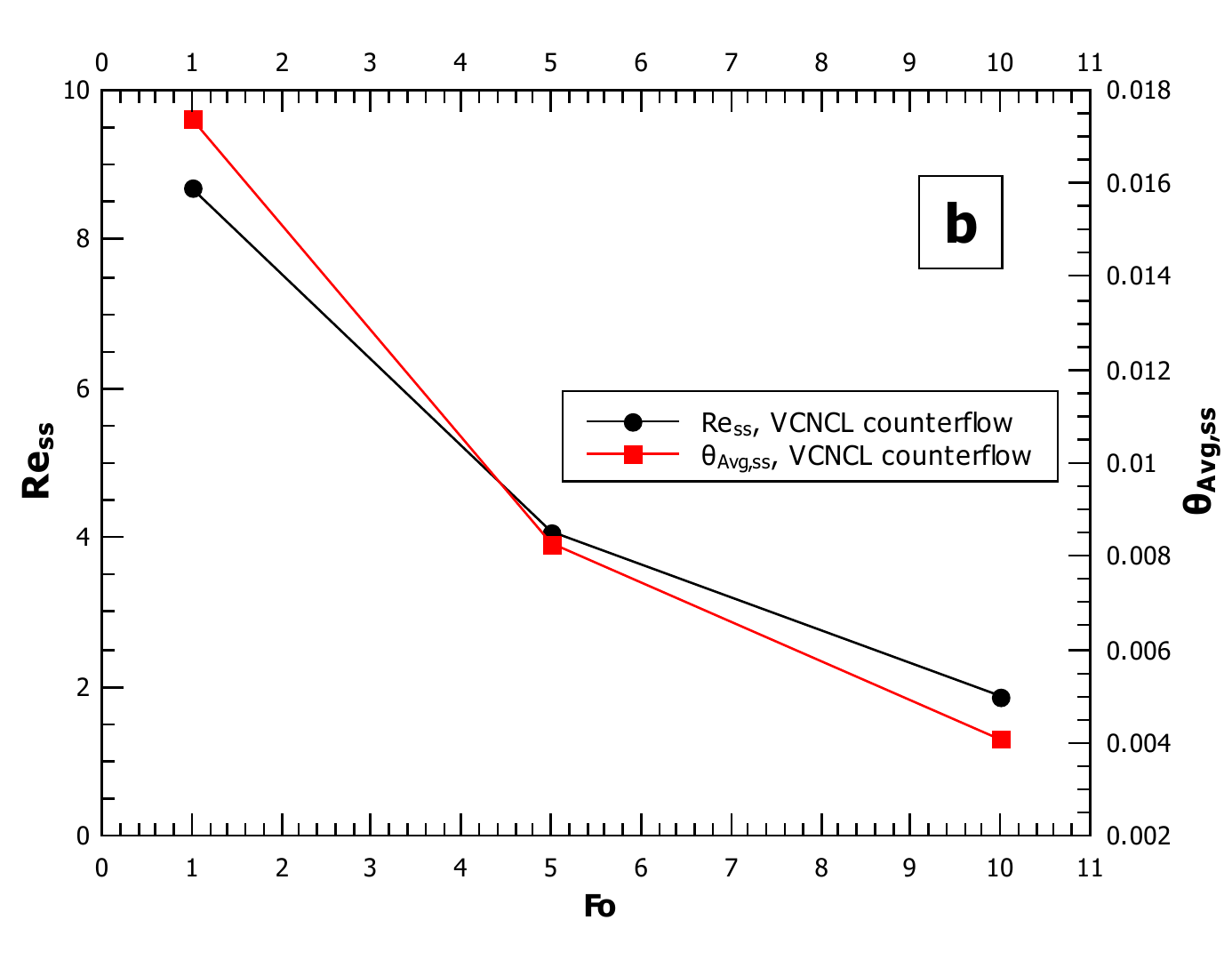}
	\end{subfigure}
	
	\caption{Effect of Fourier number ($Fo$) on the VCNCL system with h1c1 configuration for $Gr$=$10^6$, $St$=3000, $As$=1, $Co1$=1000.(a) Transient behaviour (b) Steady state trend.}
	\label{Effect of Fo}
\end{figure}

From Fig.\ref{Effect of Fo}a and Fig.\ref{Effect of Fo}b we observe that both $Re$ and $\theta_{Avg}$ decrease with increase in $Fo$. With the increase in $Fo$, we are effectively increasing the thermal diffusivity of the fluid considered for the study. With an increase in thermal diffusivity the temperature gradient across the entire CNCL system drops which leads to decrease in the buoyancy forces that drive the system. This explains the decrease in magnitude of $Re$ with increasing $Fo$.  

The increase in $Fo$ can also be due to the increase in magnitude of $t_0$ which leads to an increase in $\Delta T$. With rise in magnitude of $\Delta T$ the magnitude of $\theta_{Avg}$ decreases (as per the definition). This explains the decrease in $\theta_{Avg}$ with increasing $Fo$.

\subsection{Effect of aspect ratio on the CNCL system}

$As$ of the CNCL is defined as :

\begin{equation}
As=\frac{L}{L1}
\end{equation}

To understand the effect of variation of $As$ ($0<As\leq1$) on the CNCL system, we assume $L1=1$ and $0<L\leq 1$.

\begin{figure}[!h]
	\centering
	\begin{subfigure}[b]{0.49\textwidth}
		\includegraphics[width=1\linewidth]{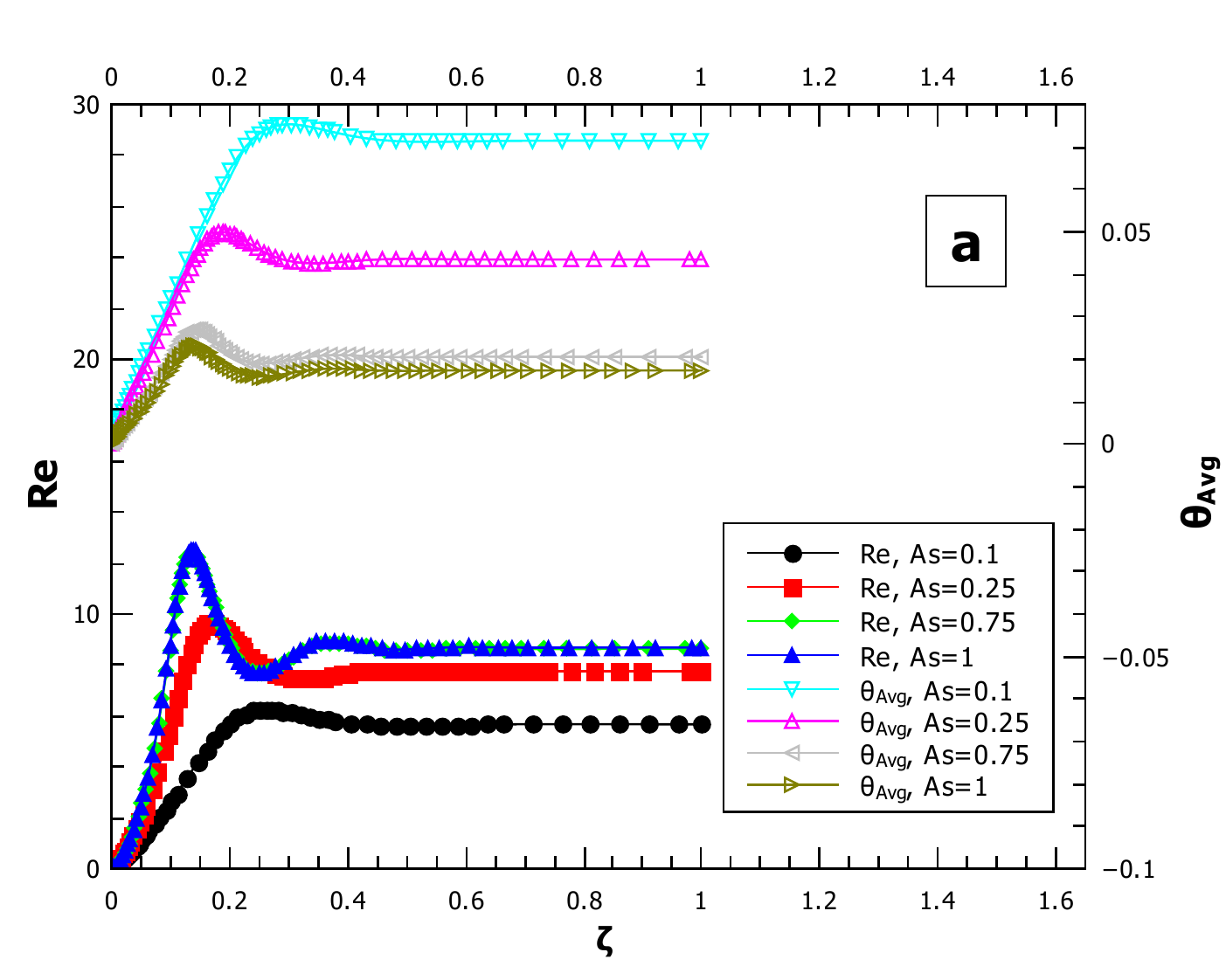}
	\end{subfigure}
	\hspace{\fill}
	\begin{subfigure}[b]{0.49\textwidth}
		\includegraphics[width=1\linewidth]{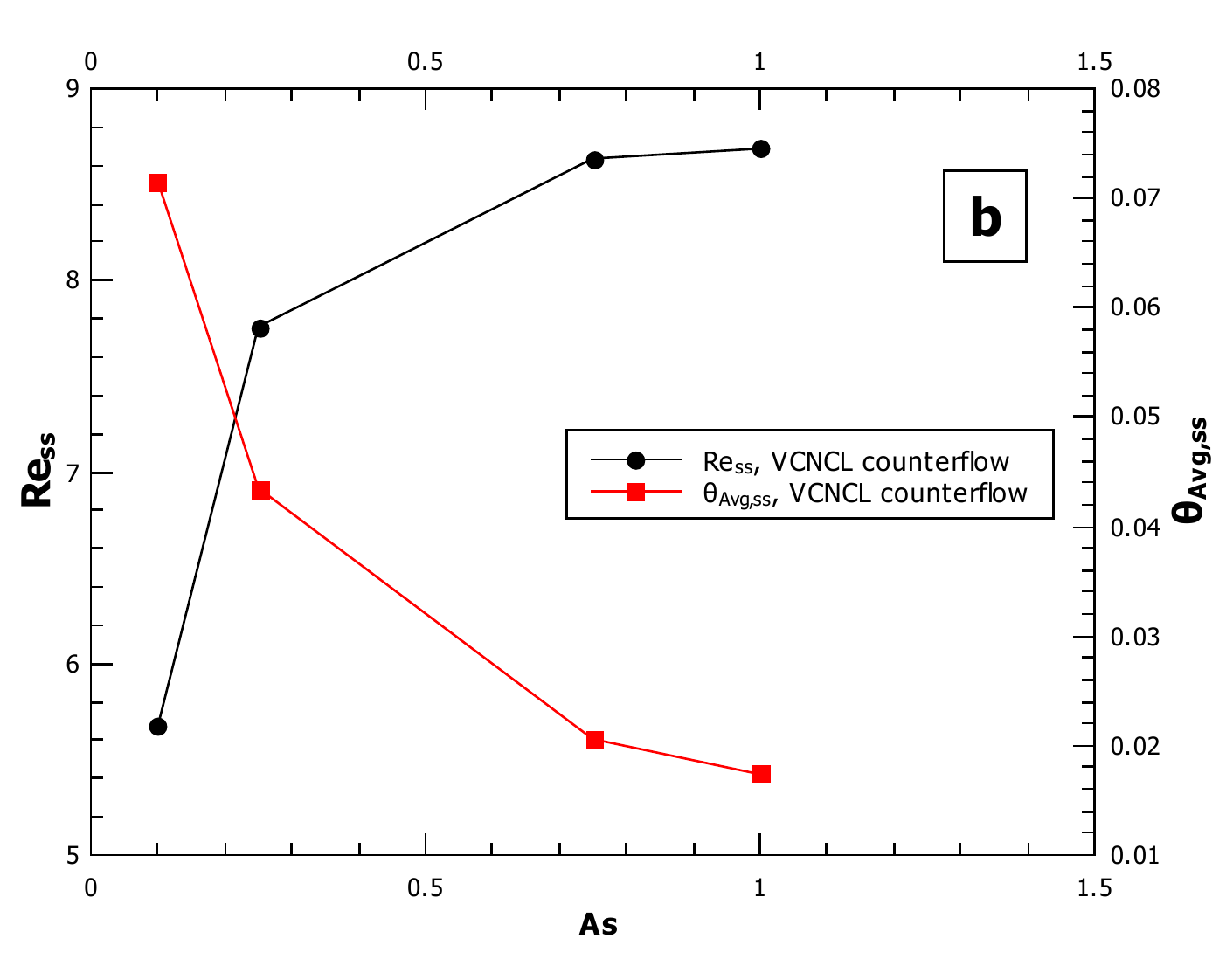}
	\end{subfigure}
	
	\caption{Effect of aspect ratio ($As$) on the VCNCL system h1c1 configuration for $Gr$=$10^6$, $St$=3000, $Fo$=1, $Co1$=1000.(a) Transient behaviour (b) Steady state trend.}
	\label{Effect of As}
\end{figure}

 With the increase in the magnitude of $As$, the height of the VCNCL (L) increases which leads to the following effects:
\begin{enumerate}
	\item An increase in the volume of fluid which is propelled by buoyancy forces represented by $D_h^2\oint f(s) ds$. 
	\item An increase in the length ($2x_0$) of the component loops of the CNCL which results in an increase in $Co_1$ and $Gr$ both of which are directly proportional to $x_0$. This leads to an increase in the viscous forces$\big(Co_1 Re^{2-d} \big)$ and buoyancy forces($Gr\oint(\theta_1) f(s) ds$). With an increase in $x_0$ the Fourier number decreases which results in an increase in $\oint(\theta_1) f(s) ds$ that leads to lesser uniformity in temperature across the loop. 
	\item An overall increase in the volume of the system: $ 2D_h^2 x_0 $. This results in an increase in the thermal inertia of the system, causing the system to resist change in $\theta_{Avg} $. This along with the fact that $\theta_{Avg} \propto \frac{1}{x_0}$ explains the decrease in magnitude with increase in $As$.
	\item An increase in the area of the common heat exchange section of the VCNCL system. This results in an increase in the rate of heat transfer leading to faster attainment of steady state for $\theta_{Avg}$ as observed in Fig.\ref{Effect of As}a.
	\item From equation 56 we can infer that although the magnitude of $\theta_{Avg}$ decreases the magnitude of buoyancy forces has to be relatively greater than the viscous forces, which results in the increase in $Re$ with increase in $As$ and the rate of drop of slope of $Re$ vs $As$ as observed in Fig.\ref{Effect of As}b can be attributed to the reduction in the rate of increase in buoyancy forces relative to viscous forces.
\end{enumerate}

\begin{equation}
Gr(\uparrow)\big(\oint(\theta_1) f(s) ds(\uparrow)\big) =  Co_1(Re_{1,ss})^{2-d}(\uparrow) +\frac{nK}{4}(Re_{1,ss})^2
\end{equation}

The $As$ of a system is a complex parameter, as any change in its value effects all the other parameters of the CNCL system. Thus it is difficult to gauge the effect of variation of $As$ on the system behavior.

\subsection{Effect of Stanton number on the CNCL system}

\begin{figure}[!h]
	\centering
	\begin{subfigure}[b]{0.49\textwidth}
		\includegraphics[width=1\linewidth]{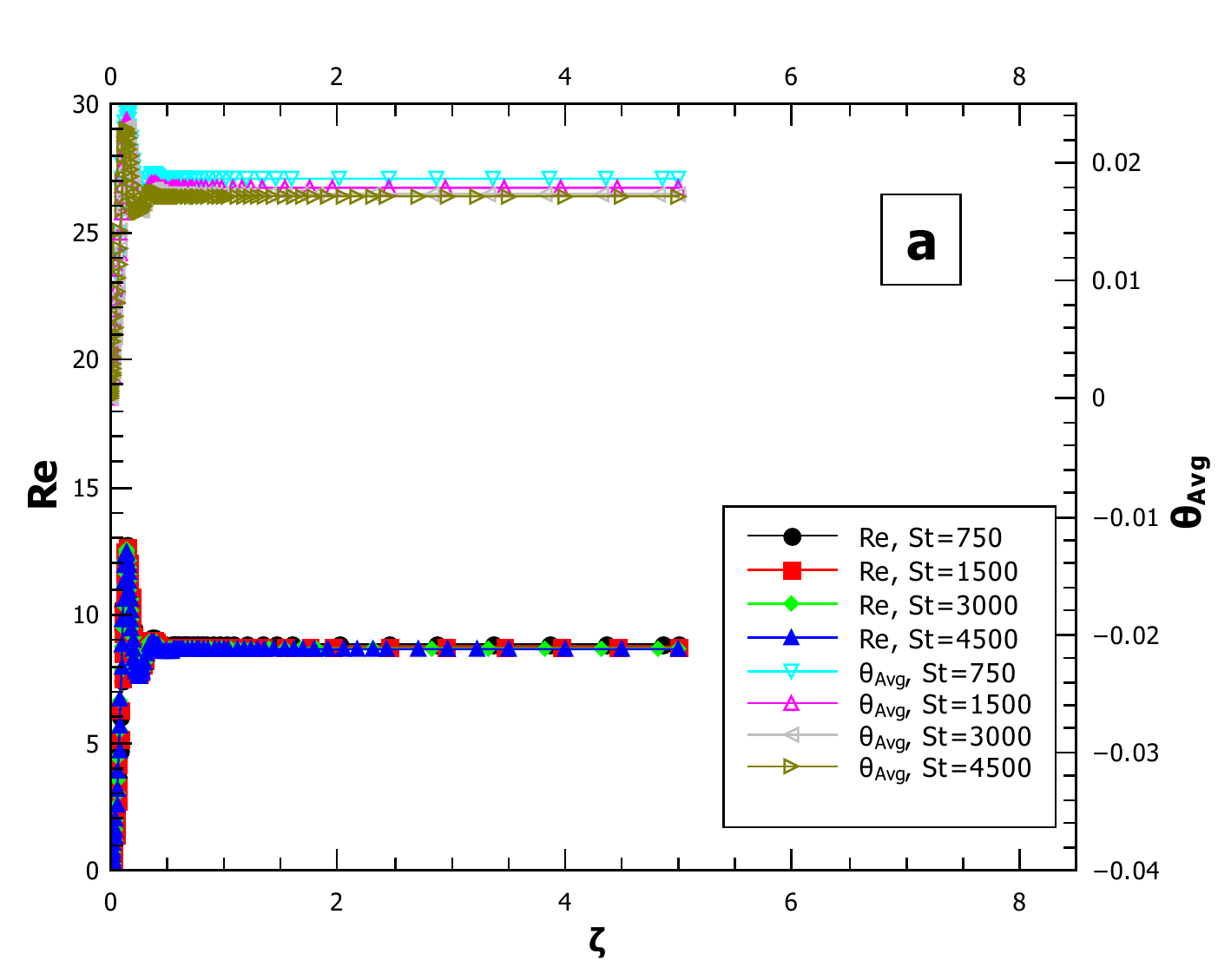}
	\end{subfigure}
	\hspace{\fill}
	\begin{subfigure}[b]{0.49\textwidth}
		\includegraphics[width=1\linewidth]{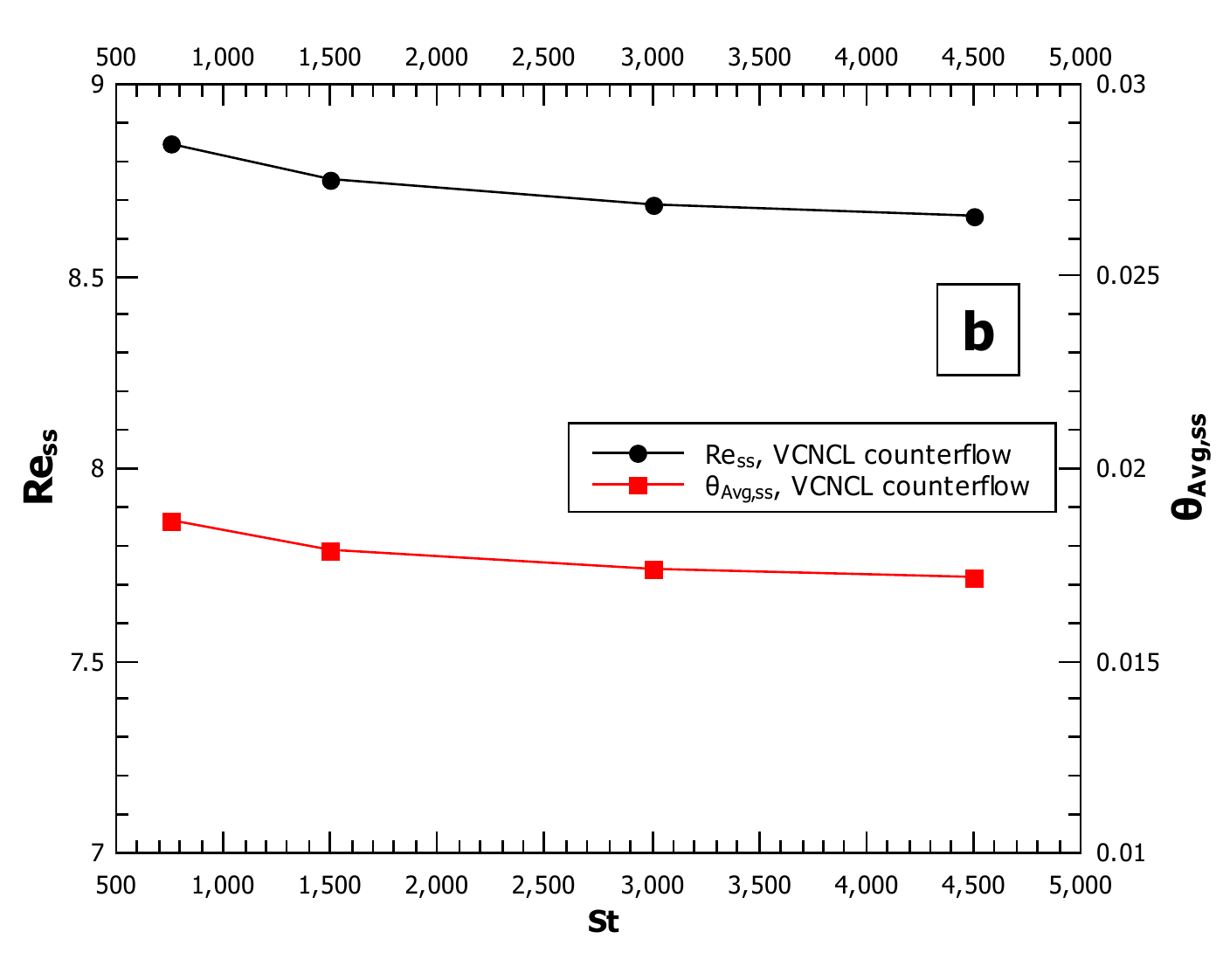}
	\end{subfigure}
	
	\caption{Effect of Stanton number ($St$) on the CNCL system for $Gr$=$10^6$, $Fo$=1, $As$=1, $Co1$=1000.(a) Transient behaviour (b) Steady state trend.}
	\label{Effect of St}
\end{figure}

As observed in Fig.\ref{Effect of St} we notice that varying the magnitude of $St$ does not affect the transient and steady-state behavior significantly at least for the parameters considered for the study. This study corroborates the sensitivity study of the CNCL system considered in Fig.\ref{Senitivity of CNCL with U,K}. This leads us to conclude that the correlations used to predict the heat transfer need not be very precise to give a decent prediction of the CNCL behavior.

\vspace{5cm}

\subsection{Effect of Co1 on the CNCL system}

\begin{figure}[!h]
	\centering
	\begin{subfigure}[b]{0.49\textwidth}
		\includegraphics[width=1\linewidth]{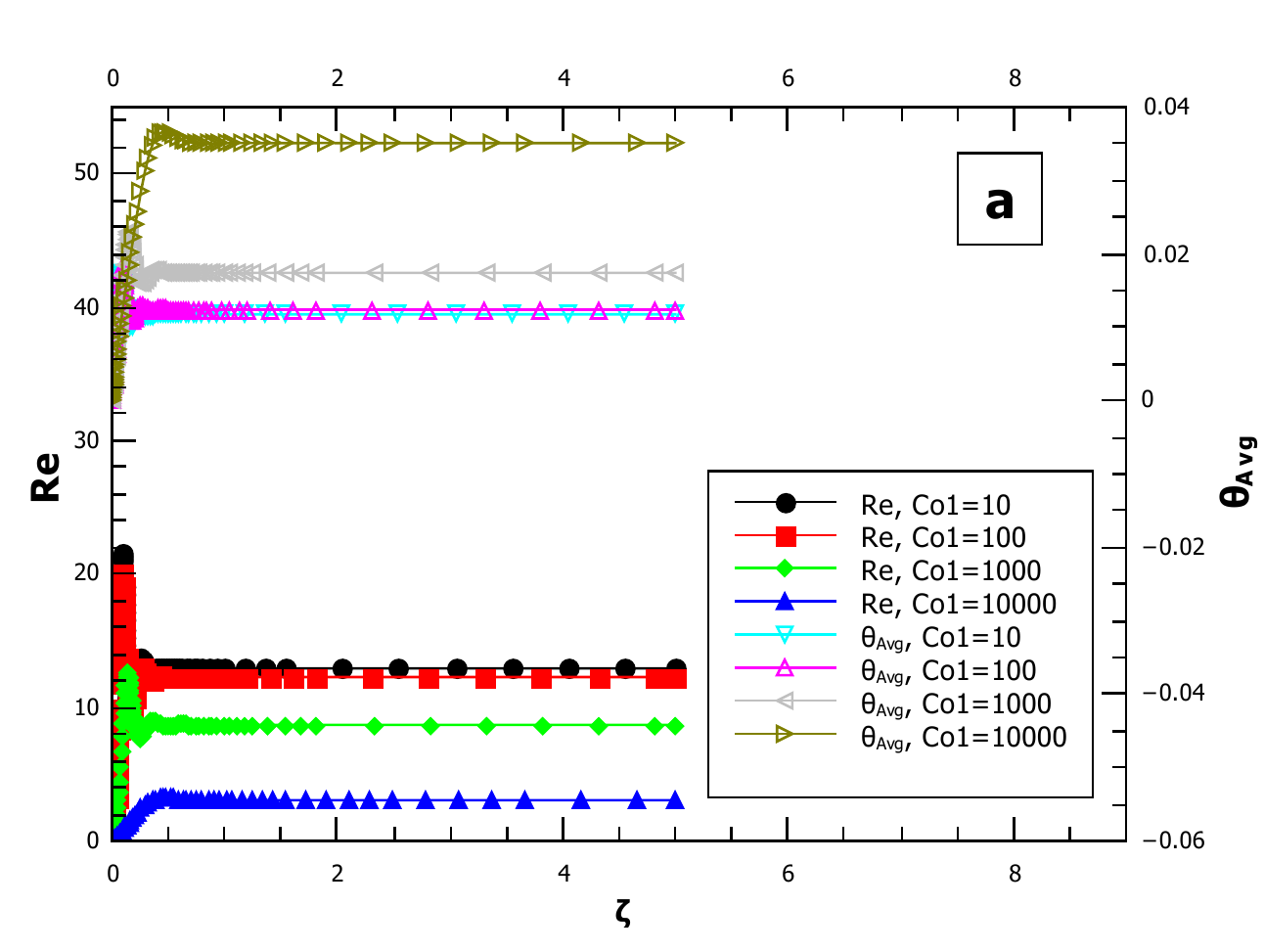}
	\end{subfigure}
	\hspace{\fill}
	\begin{subfigure}[b]{0.49\textwidth}
		\includegraphics[width=1\linewidth]{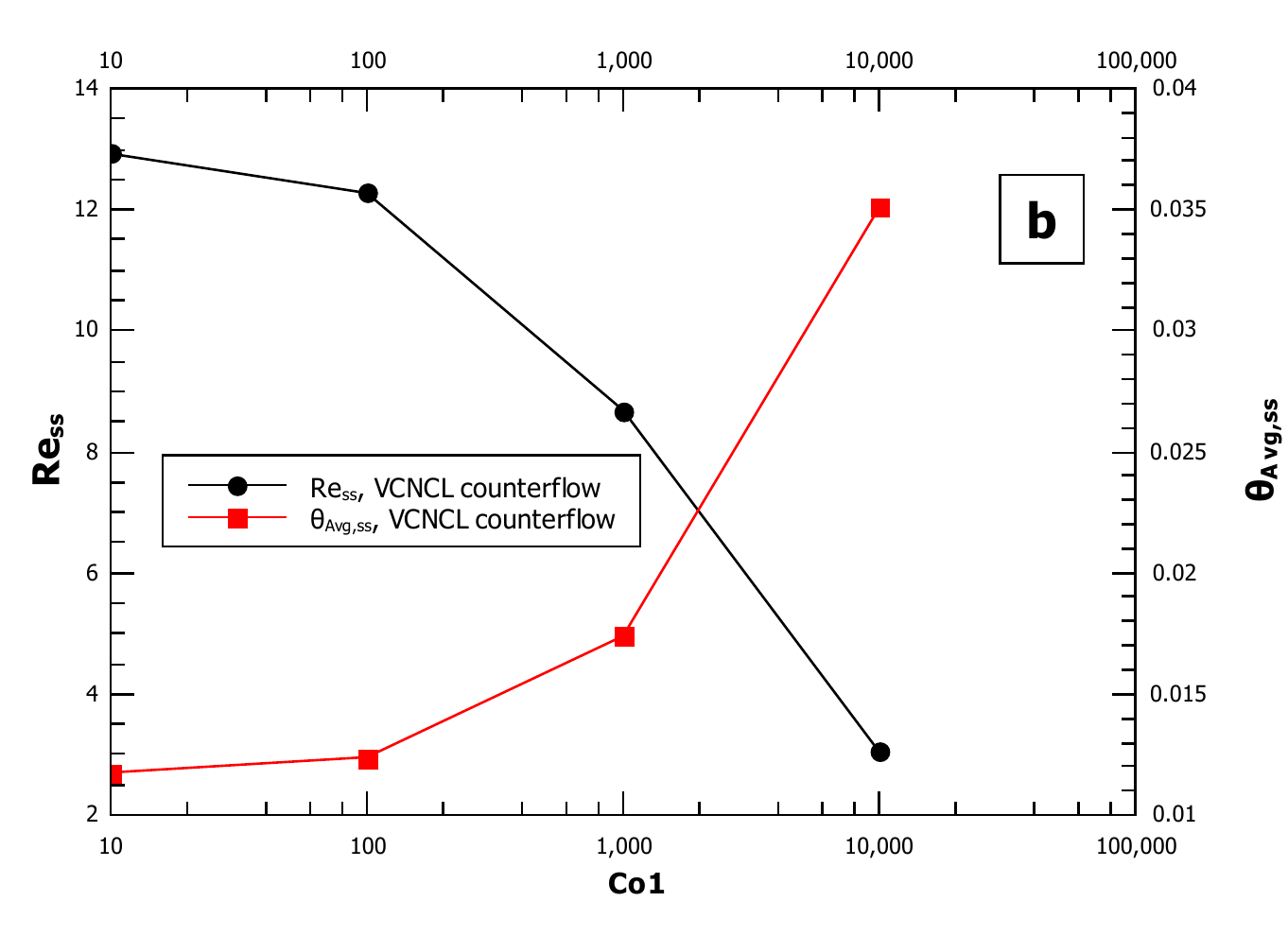}
	\end{subfigure}
	
	\caption{Effect of Geometric factor ($Co1$) on the VCNCL system with h1c1 configuration for $Gr$=$10^6$, $Fo$=1, $As$=1, $St$=3000.(a) Transient behaviour (b) Steady state trend.}
	\label{Effect of Co1}
\end{figure}

From Fig.\ref{Effect of Co1} we observe that with an increase in $Co1$ the magnitude of $Re$ drops, while the magnitude of $\theta_{Avg}$ rises. From Fig.\ref{Effect of Co1}a, we also observe the decrease in the oscillatory behavior exhibited by the CNCL system. With the increase in $Co1$ the magnitude of viscous forces increases relative to the buoyancy forces which explains the decrease in the magnitude of $Re$. As the velocity in the loop decreases, the temperature drop across the heating and cooling section increases which leads to an increase in the magnitude of $\theta_{Avg}$ with increasing $Co1$.

\begin{equation}
\big( Gr\oint(\theta_1) f(s) ds\big)(constant) =  Co_1(Re_{1,ss})^{2-d}(\uparrow) +\frac{nK}{4}(Re_{1,ss})^2
\end{equation}

\subsection{Effect of initial conditions on the CNCL system}

\begin{figure}[!h]
	\centering
	\begin{subfigure}[b]{0.49\textwidth}
		\includegraphics[width=1\linewidth]{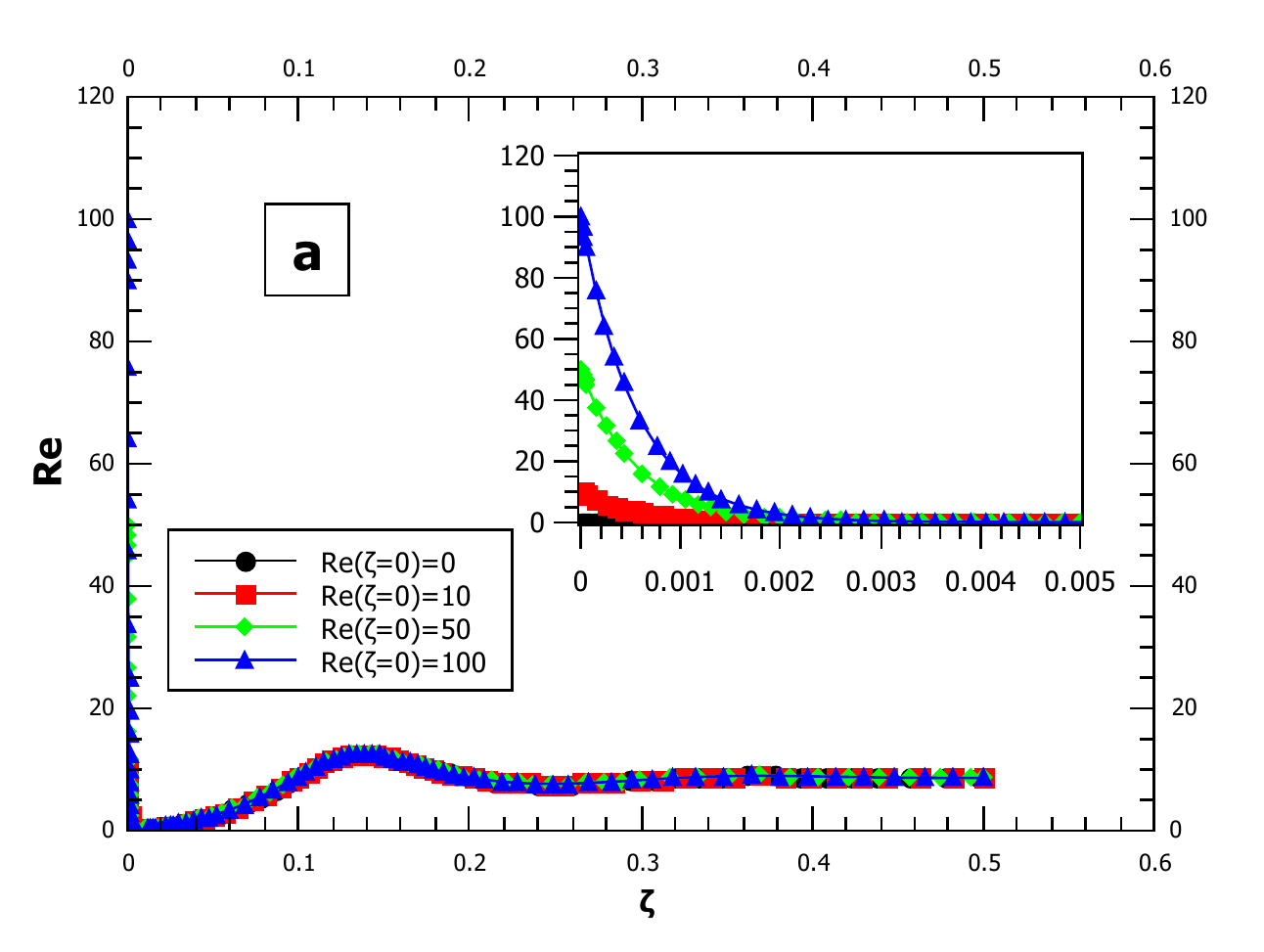}
	\end{subfigure}
	\hspace{\fill}
	\begin{subfigure}[b]{0.49\textwidth}
		\includegraphics[width=1\linewidth]{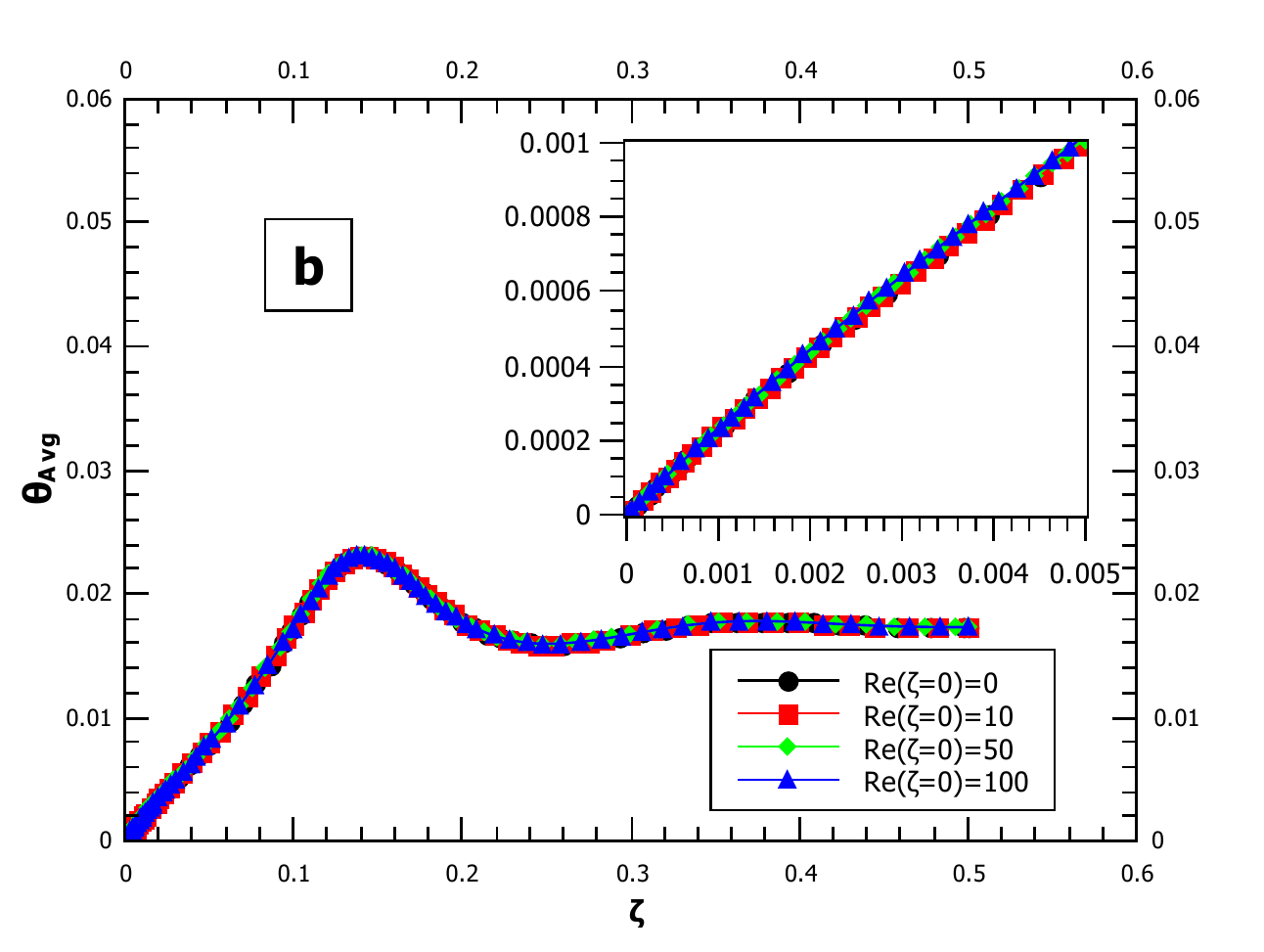}
	\end{subfigure}
	
	\caption{Effect of initial velocity represented by $Re(\tau=0)$ on the VCNCL system with h1c1 configuration for $Gr$=$10^6$, $Fo$=1, $As$=1, $St$=3000, $Co1$=1000.(a) Transient behaviour (b) Steady state trend.The inset graphs represent the transient behaviour of the respective plots for $\zeta \to (0-0.005)$.}
	\label{Effect of Re(tau=0)}
\end{figure}

The initial condition of the CNCL system which can be changed is $Re(\zeta=0)$. From Fig.\ref{Effect of Re(tau=0)} we observe that the transient behavior of the VCNCL system is largely unaffected by the variation in initial conditions. From Fig.\ref{Effect of Re(tau=0)}a it is to be noted that any transience associated with is $Re(\zeta=0)$ dies out very quickly and the VCNCL system response remains unaltered.

\subsection{Effect of parallel flow vs counter flow arrangement at the heat exchanger section of HCNCL system}

The horizontal CNCL system with h2c2 heater cooler configuration is the only condition which can exhibit both counter and parallel flow arrangement at the common heat exchanger section of the CNCL system. This is because for the h2c2 configuration of the HCNCL system all the energy transfer in the CNCL system occurs on the horizontal legs of the component loops. This implies that the buoyancy forces generated at the heat transfer sections does not dictate the flow direction. Thus by choosing appropriate initial flow conditions as described in section 8.2, the parallel flow or counter flow can be induced within the HCNCL system by directing the heated or cooled fluid sections to the vertical legs of the system.

\begin{figure}[!h]
	\centering
	\begin{subfigure}[b]{0.49\textwidth}
		\includegraphics[width=1.1\linewidth]{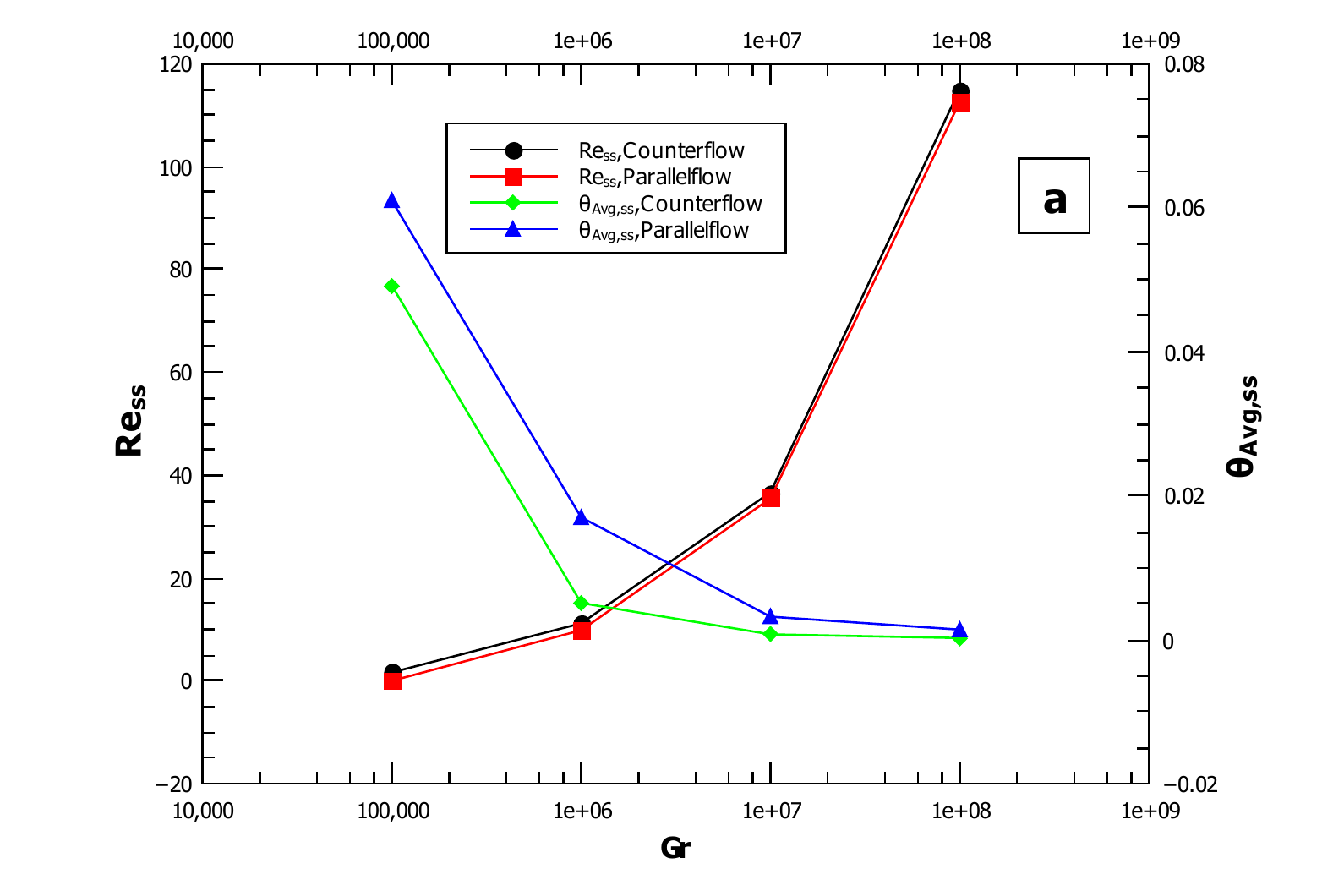}
	\end{subfigure}
	\hspace{\fill}
	\begin{subfigure}[b]{0.49\textwidth}
		\includegraphics[width=1.1\linewidth]{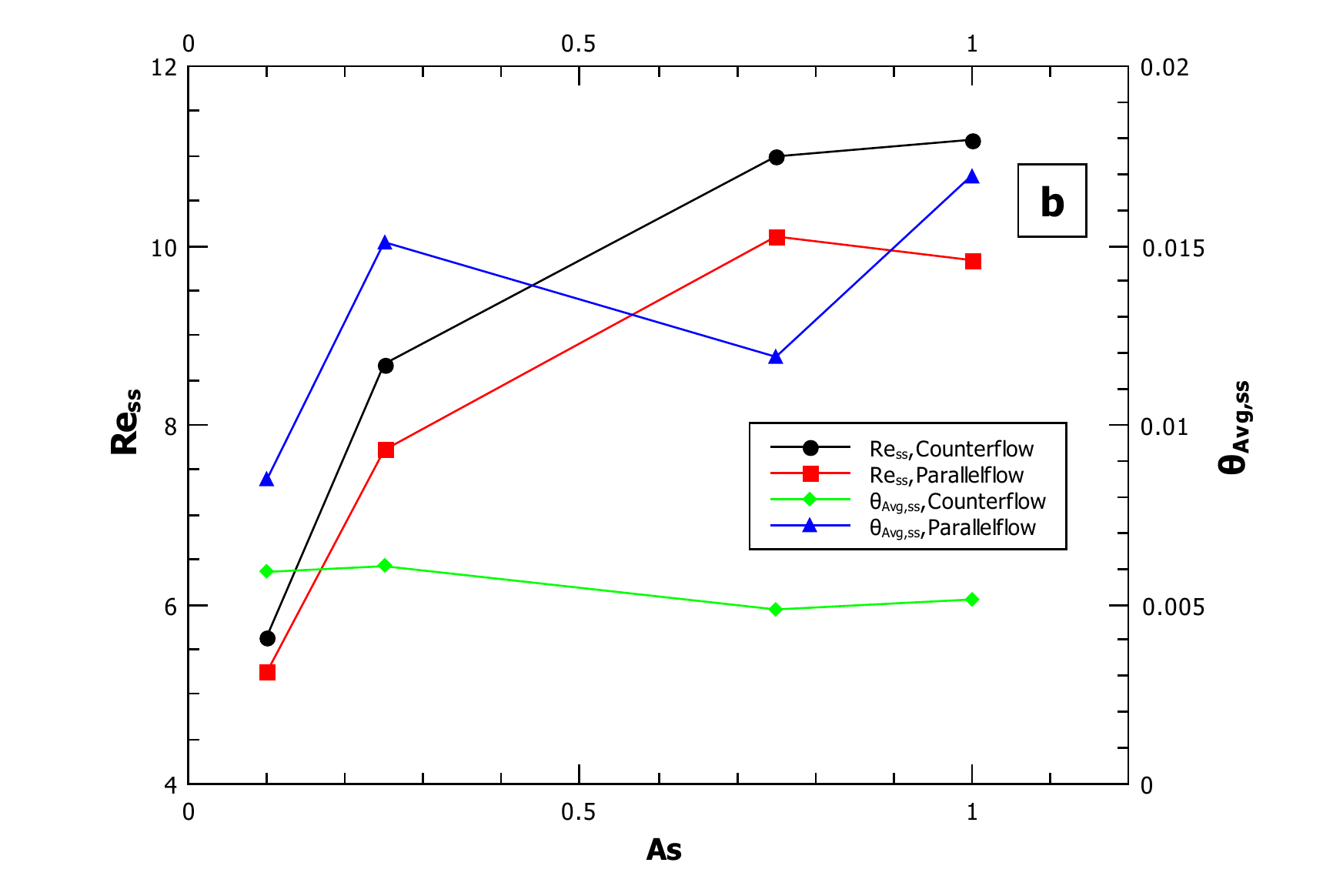}
	\end{subfigure}
	\hspace{\fill}
	\begin{subfigure}[b]{0.49\textwidth}
		\includegraphics[width=1.1\linewidth]{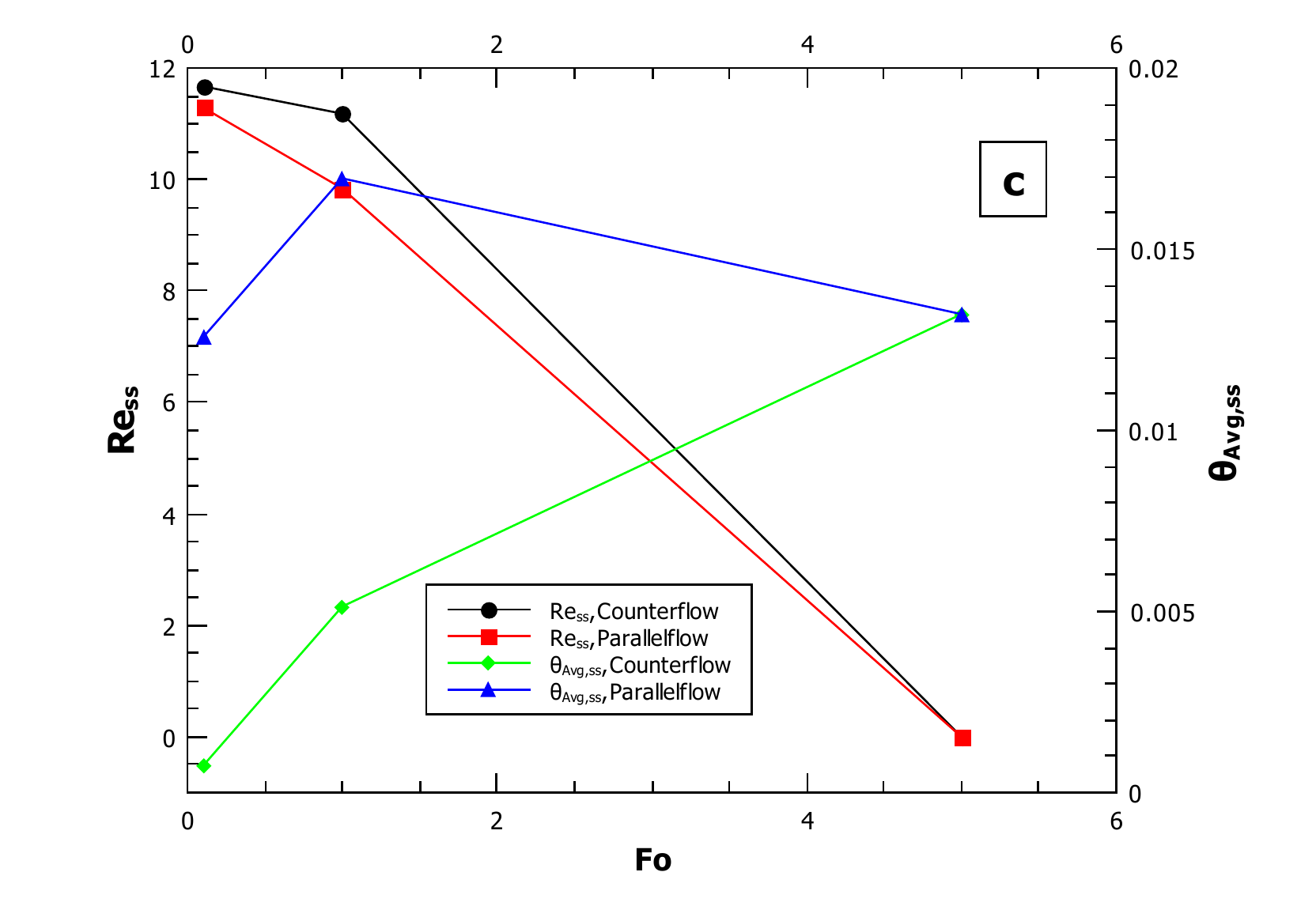}
	\end{subfigure}
	\hspace{\fill}
	\begin{subfigure}[b]{0.49\textwidth}
		\includegraphics[width=1.1\linewidth]{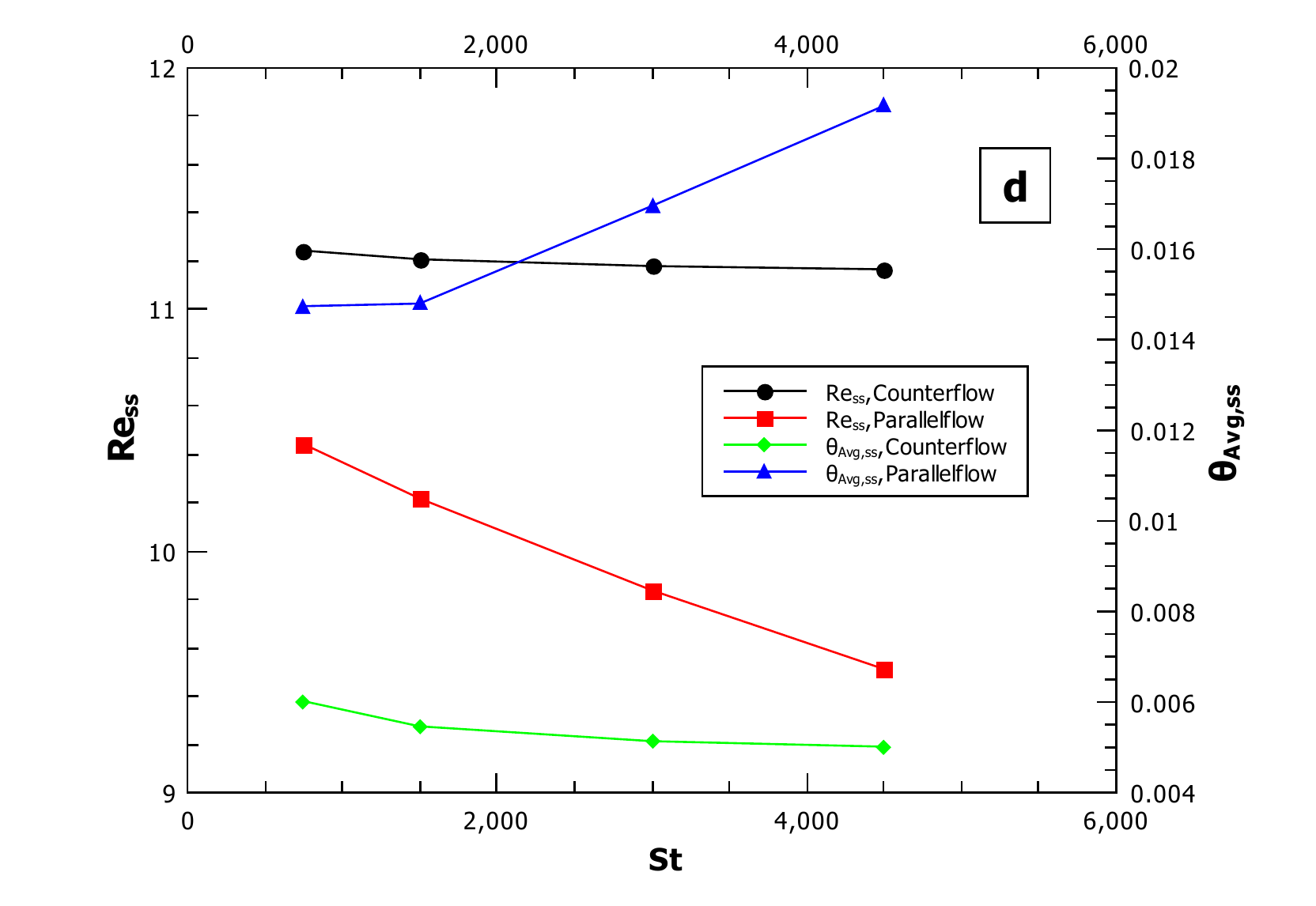}
	\end{subfigure}
	\hspace{\fill}
	\begin{subfigure}[b]{0.49\textwidth}
		\includegraphics[width=1.1\linewidth]{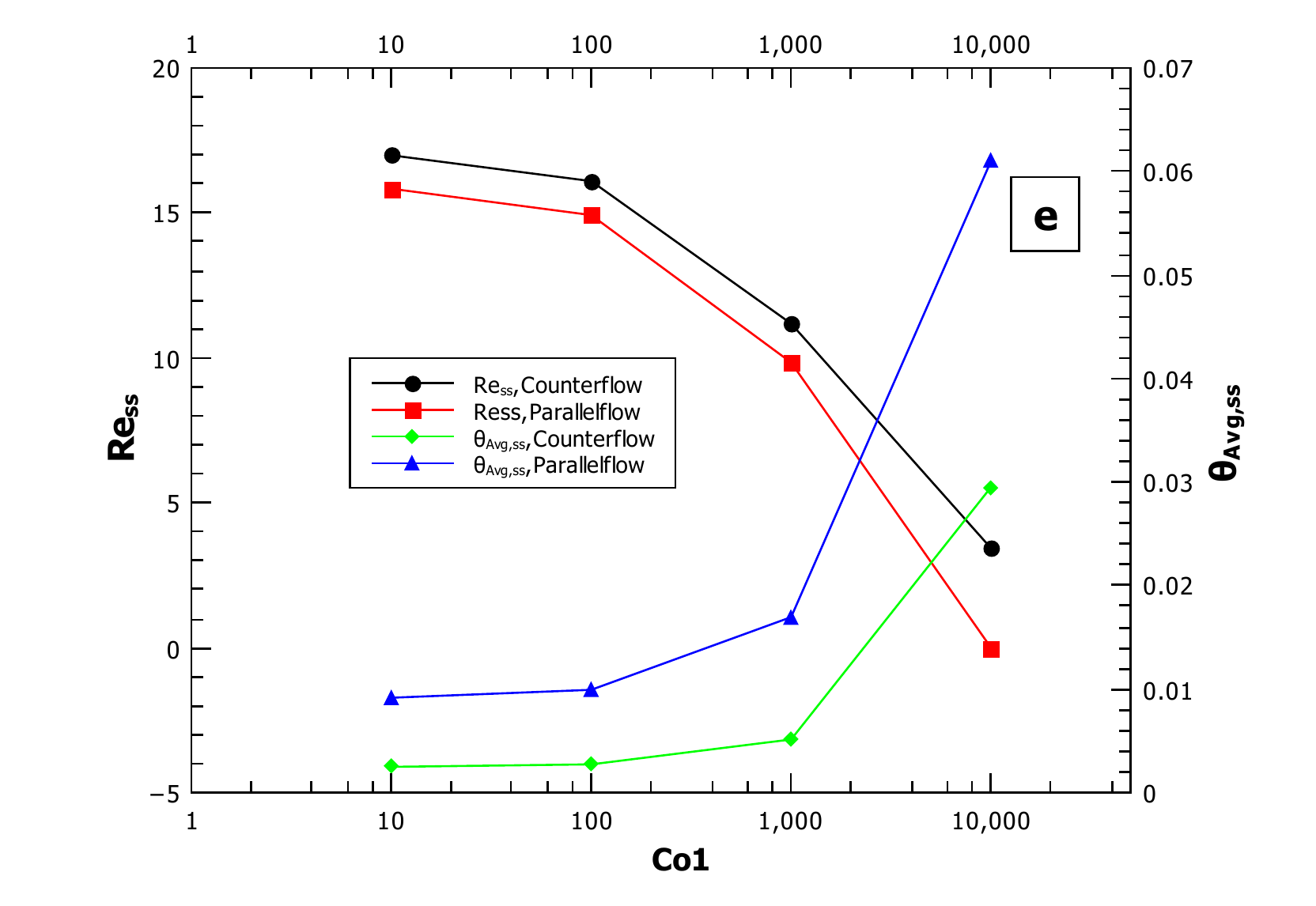}
	\end{subfigure}
	\caption{Parametric study of the parallel and counterflow configurations of horizontal CNCL system with h2c2 heater cooler configuration. The values of the non-dimensional numbers considered are listed in Table.7.}
	\label{fig:Effect-of-non dimensional numbers-on-hcncl-for-counter-and-parallel-flow-configurations}
\end{figure}

\renewcommand{\arraystretch}{0.8}
\begin{table}[!h]
	\begin{center}
		\caption{Values of non-dimensional numbers utilised for the parametric study of HCNCL system.}
		\label{tab:table7}
		\scalebox{0.9}{
			\begin{tabular}{p{2cm} p{2cm} p{2cm} p{2cm} p{2cm} p{2cm}} 
				\textbf{Figure} & {\textbf{Gr }} &{\textbf{As}} & {\textbf{Fo}}  & {\textbf{ St}} & {\textbf{Co1}}\\
				\hline 
				Fig.22(a) & $10^5 \rightarrow 10^8$  & 1 & 1 & 3000 & 1000\\  
				Fig.22(b) & $10^6$ & $0.1 \rightarrow 1$ & 1 &3000 & 1000\\ 
				Fig.22(c) & $10^6$ & 1& $1 \rightarrow 10$ &3000 & 1000\\
				Fig.22(d) & $10^6$ & 1 & 1 &$750 \rightarrow 4500$ & 1000\\ 
				Fig.22(e) & $10^6$ & 1 & 1 &3000 & $10 \rightarrow 10^4$\\ \hline
		\end{tabular}}
	\end{center}
\end{table}

 For all the other heater cooler positions for both the HCNCL and VCNCL systems atleast one of the heat transfer sections is vertical w.r.t gravity and the buoyancy forces generated there determine the flow direction irrespective of the initial flow conditions.

Table.\ref{tab:table7} indicates the ranges in which parametric studies were conducted on the HCNCL system for different non-dimensional parameters. The only difference between the parallel flow and counter flow systems is the temperature distribution at the common heat exchanger section of the CNCL system, which leads to a difference in trends observed in Fig.\ref{fig:Effect-of-non dimensional numbers-on-hcncl-for-counter-and-parallel-flow-configurations}.

\clearpage

\subsection{Effect of heater cooler configuration on the CNCL system}

A transient and steady-state analysis on the effect of the heater and cooler location on the CNCL dynamics of both horizontal and vertical systems is presented.

\subsubsection{Effect of heater cooler configuration on the vertical CNCL system}

\begin{figure}[!h]
	\centering
	\begin{subfigure}[b]{0.49\textwidth}
		\includegraphics[width=1\linewidth]{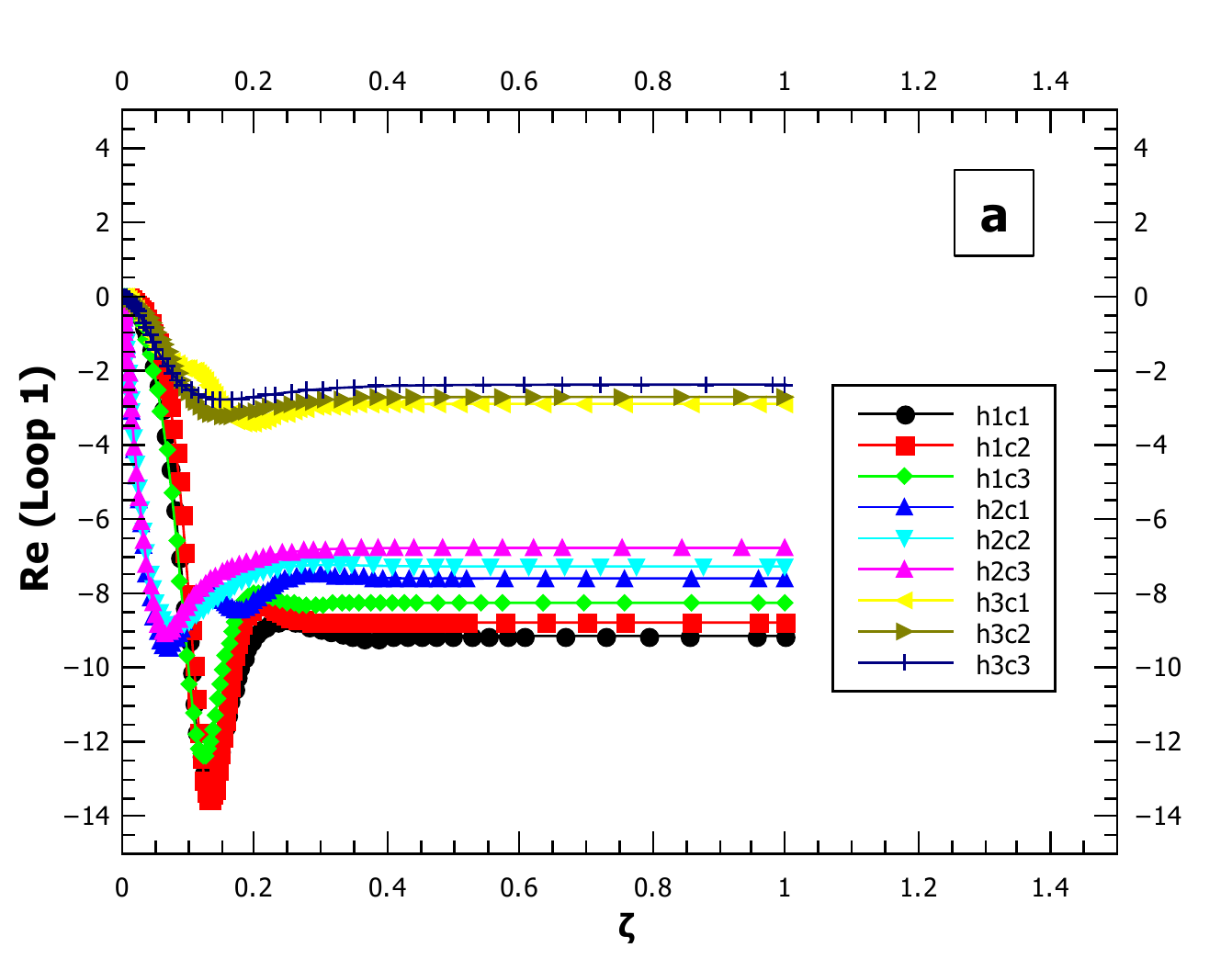}
	\end{subfigure}
	\hspace{\fill}
	\begin{subfigure}[b]{0.49\textwidth}
		\includegraphics[width=1\linewidth]{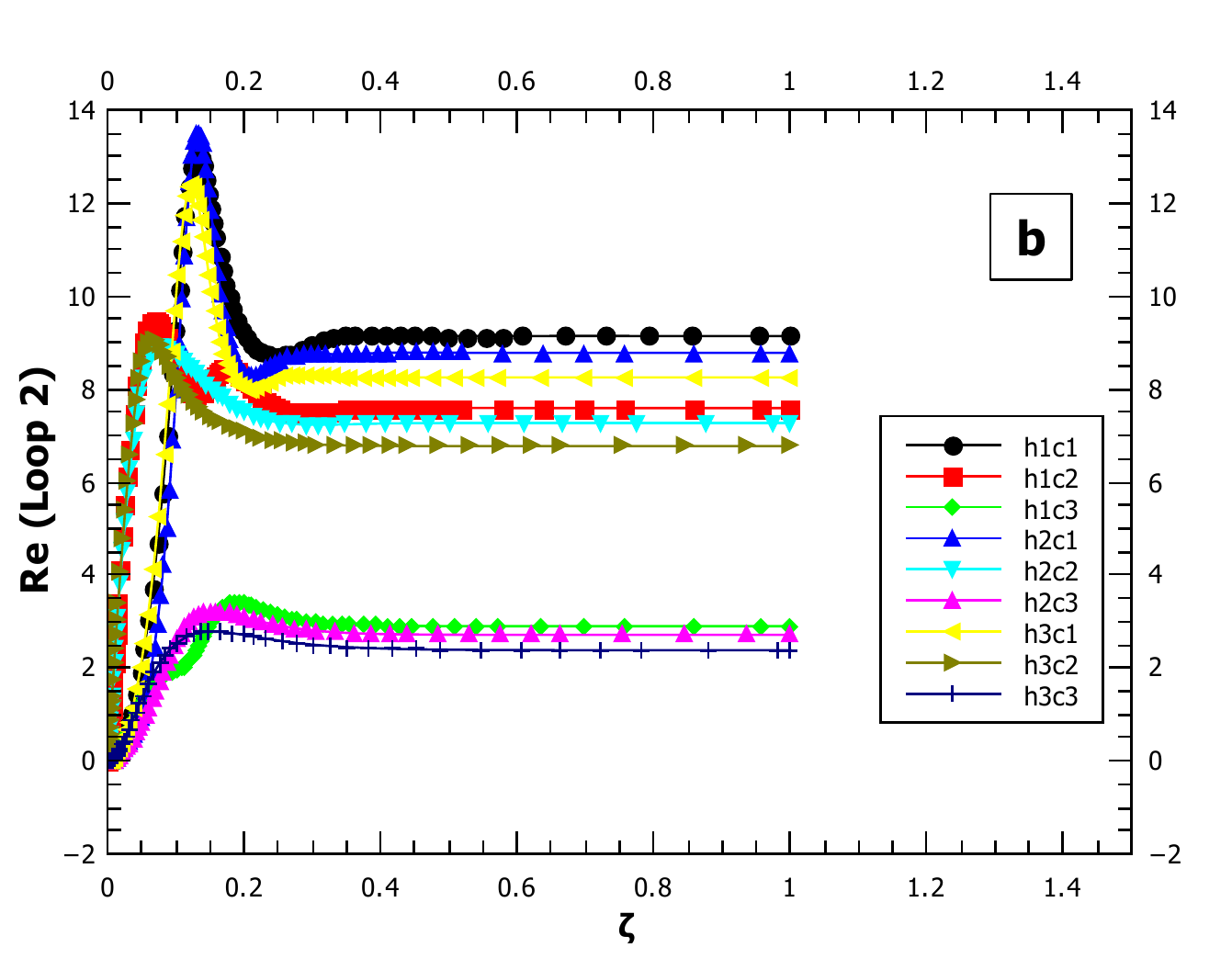}
	\end{subfigure}
	\hspace{\fill}
	\begin{subfigure}[b]{0.49\textwidth}
		\includegraphics[width=1\linewidth]{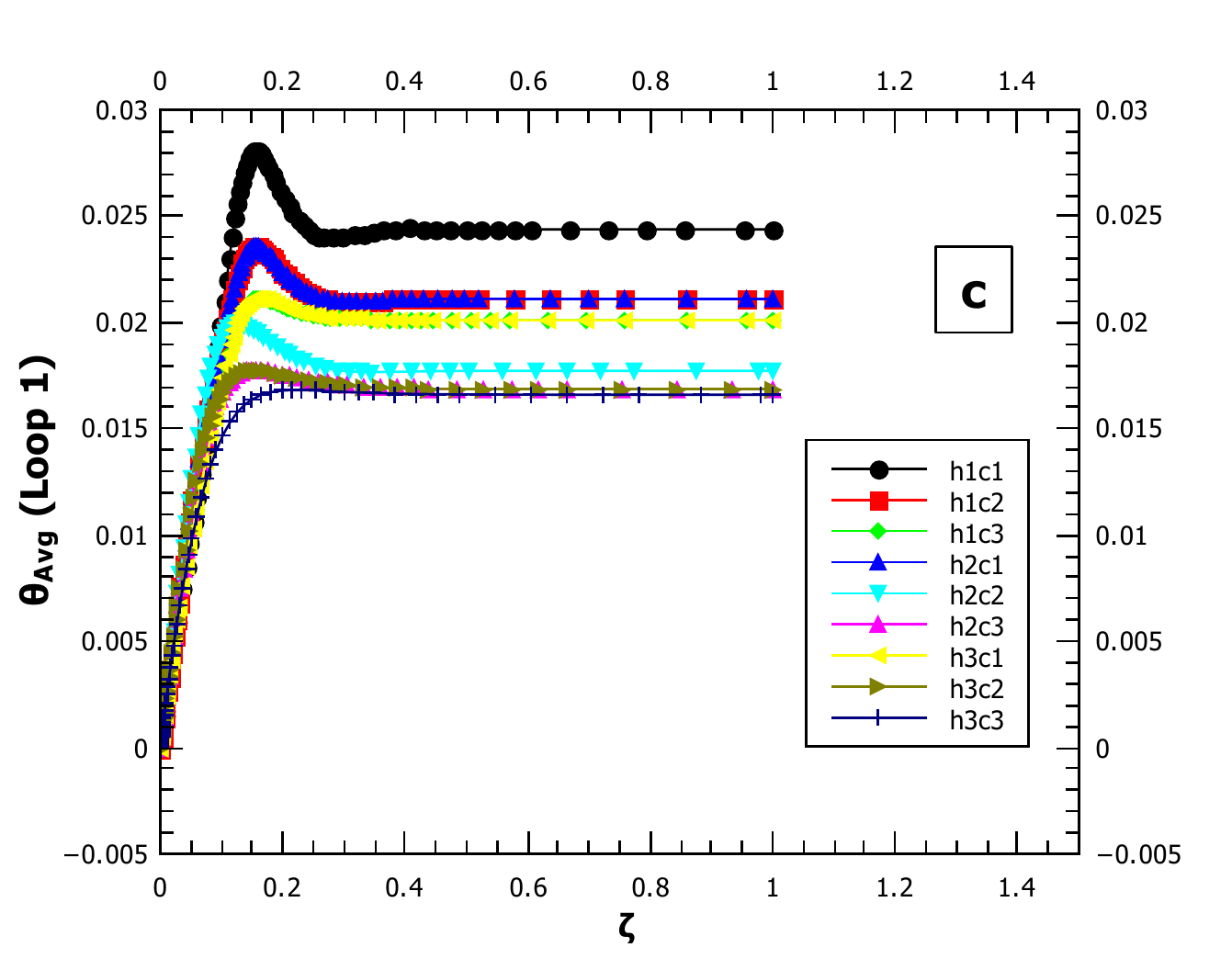}
	\end{subfigure}
	\hspace{\fill}
	\begin{subfigure}[b]{0.49\textwidth}
		\includegraphics[width=1\linewidth]{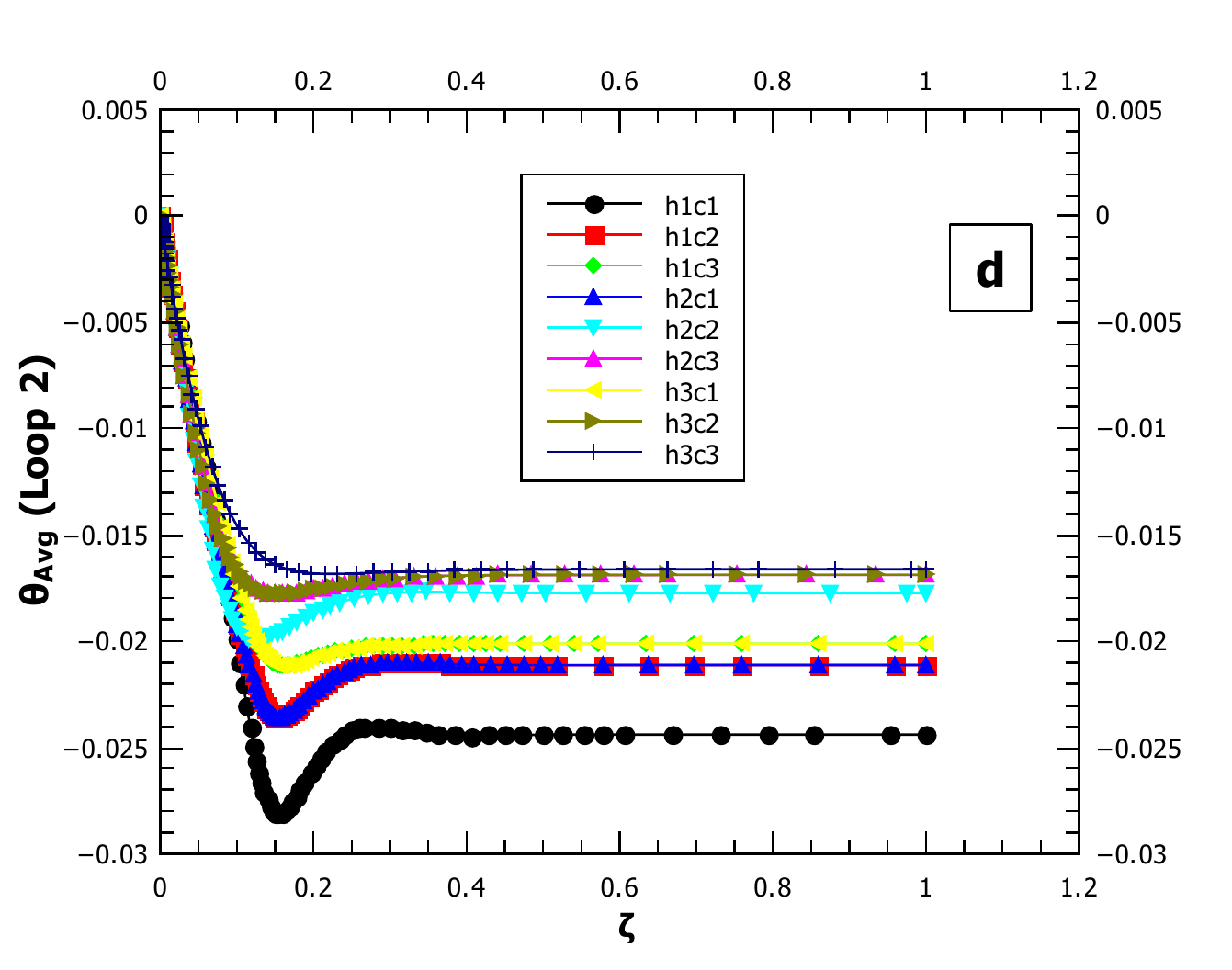}
	\end{subfigure}
	\hspace{\fill}
	\begin{subfigure}[b]{0.49\textwidth}
		\includegraphics[width=1\linewidth]{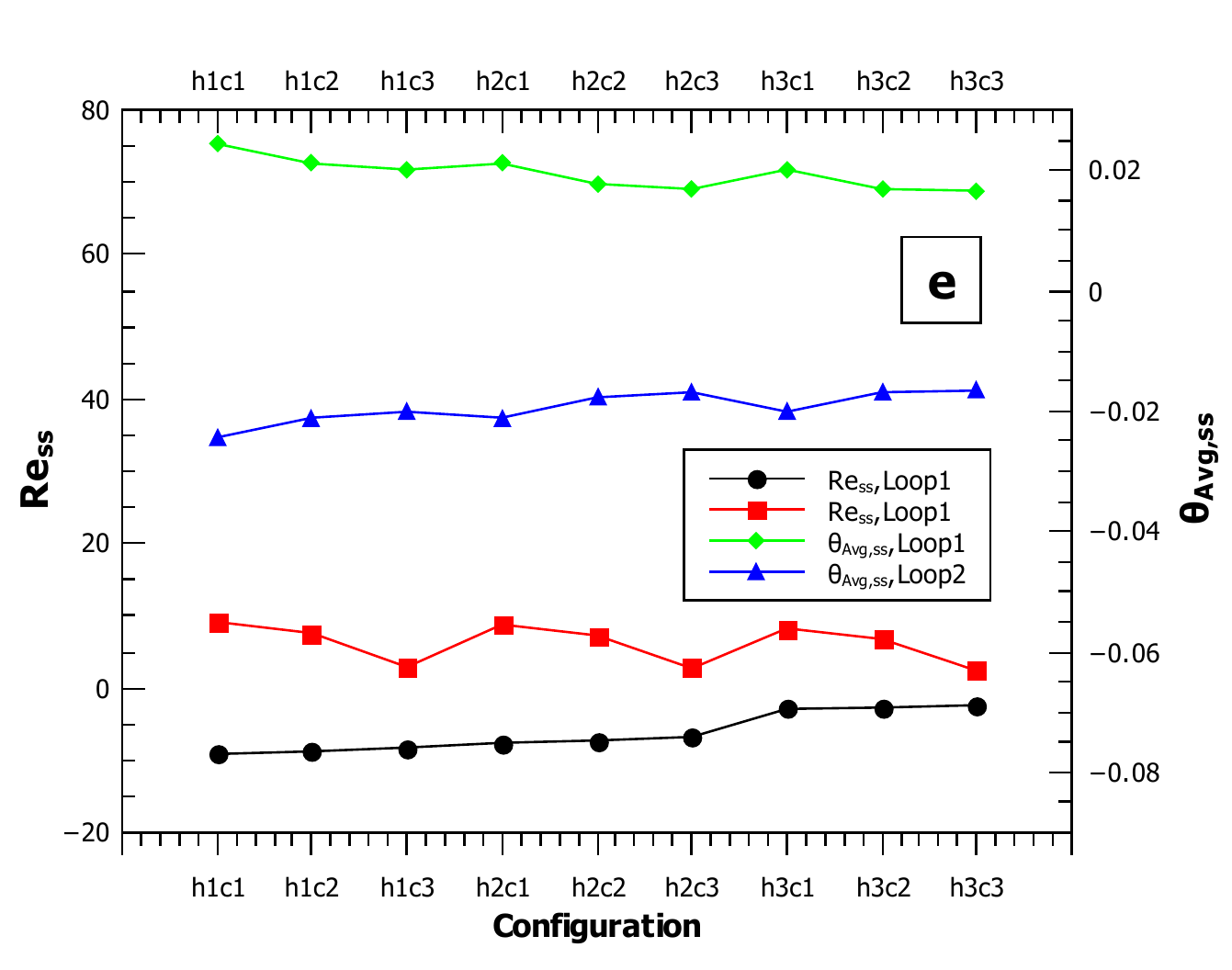}
	\end{subfigure}
	
	\caption{Effect of heater cooler orientation on the Vertical CNCL system for $Gr$=$10^6$, $Fo$=1, $As$=1, $St$=100, $Co1$=1000.}
	\label{Heater cooler configuration study VCNCL}
\end{figure}

The following observations can be made from Fig.\ref{Heater cooler configuration study VCNCL} :
\begin{enumerate}
	\item The magnitude of $\theta_{Avg}$ for all the heater-cooler configurations at steady state is symmetric about $\theta_{Avg}(\zeta=0)$ (which is equal to zero). This implies ${\theta_{Avg,1,ss}}/{\theta_{Avg,2,ss}}=1$.
	\item The magnitude of $Re_{ss}$ is not equal in the component loops of the VCNCL system for all the heater cooler configurations. 
	\item Table.\ref{tab:table8} presents the ratio of Reynolds numbers of Loop 1 to Loop 2 at steady state. We observe that only for h1c1, h2c2, h3c3 configurations do we have symmetric transient and steady state behavior. The negative sign denotes that the flow arrangement at the common heat exchanger is counterflow.
	\item From table.\ref{tab:table8} we also observe that :
	
	\begin{equation}
	\frac{Re_{1,ss}(h\textbf{i}c\textbf{j})}{Re_{2,ss}(h\textbf{i}c\textbf{j})}\times \frac{Re_{1,ss} (h\textbf{j}c\textbf{i})}{Re_{2,ss} (h\textbf{j}c\textbf{i})}=1
	\end{equation}
	
	where `\textbf{i}',`\textbf{j}' represent the position of the heater and cooler for the 'h\textbf{i}c\textbf{j}' configuration. 
	\item All the heater-cooler configurations of the VCNCL system exhibit counterflow condition at the common heat exchanger section for all stable convective flow cases.
	\item h1c1 configuration corresponds to the maximum magnitude of $Re_{ss}$ and $\theta_{Avg}$ and the h3c3 configuration corresponds to the minimum values of the same parameters for the vertical CNCL system respectively.
\end{enumerate}

\renewcommand{\arraystretch}{1}
\begin{table}[!h]
	\begin{center}
		\caption{Ratio of \bigg($\frac{Re_{1,ss}}{Re_{2,ss}}\bigg)$ of the VCNCL system for all heater cooler configurations.}
		\label{tab:table8}
		\scalebox{0.9}{
			\begin{tabular}{p{1cm}| p{2cm} p{2cm} p{2cm} } 
				 & {\textbf{c1 }} &{\textbf{c2}} & {\textbf{c3}} \\
				\hline 
				{\textbf{h1 }} & -1  & -1.15 & -2.85\\  
				{\textbf{h2 }} & -1/1.15 & -1 & -2.5 \\ 
				{\textbf{h3 }} & -1/2.85 & -1/2.5 & -1\\
			
			\end{tabular}}
	\end{center}
\end{table}

\subsubsection{Effect of heater cooler configuration on the horizontal CNCL system}

\begin{figure}[!h]
	\centering
	\begin{subfigure}[b]{0.49\textwidth}
		\includegraphics[width=1\linewidth]{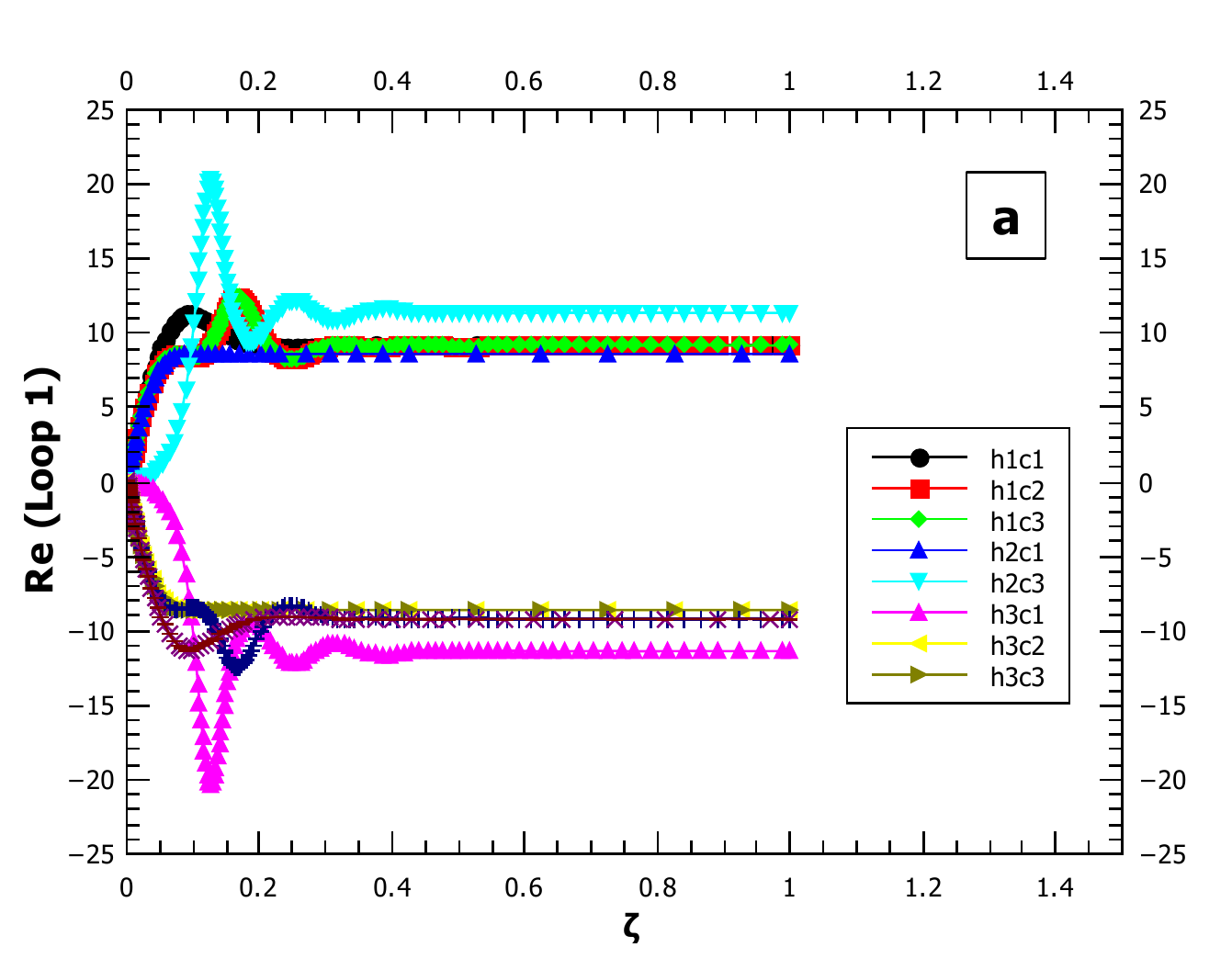}
	\end{subfigure}
	\hspace{\fill}
	\begin{subfigure}[b]{0.49\textwidth}
		\includegraphics[width=1\linewidth]{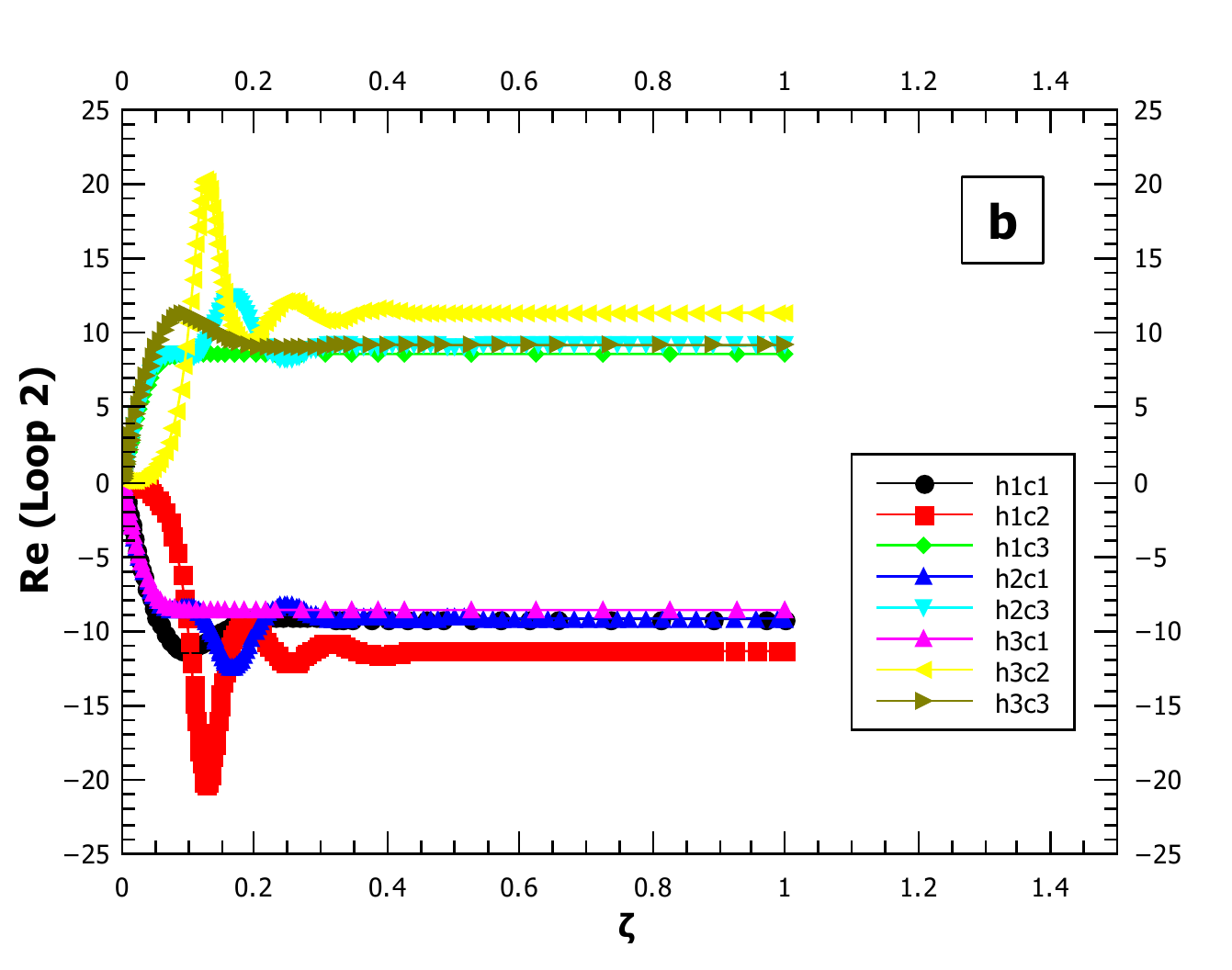}
	\end{subfigure}
	\hspace{\fill}
	\begin{subfigure}[b]{0.49\textwidth}
		\includegraphics[width=1\linewidth]{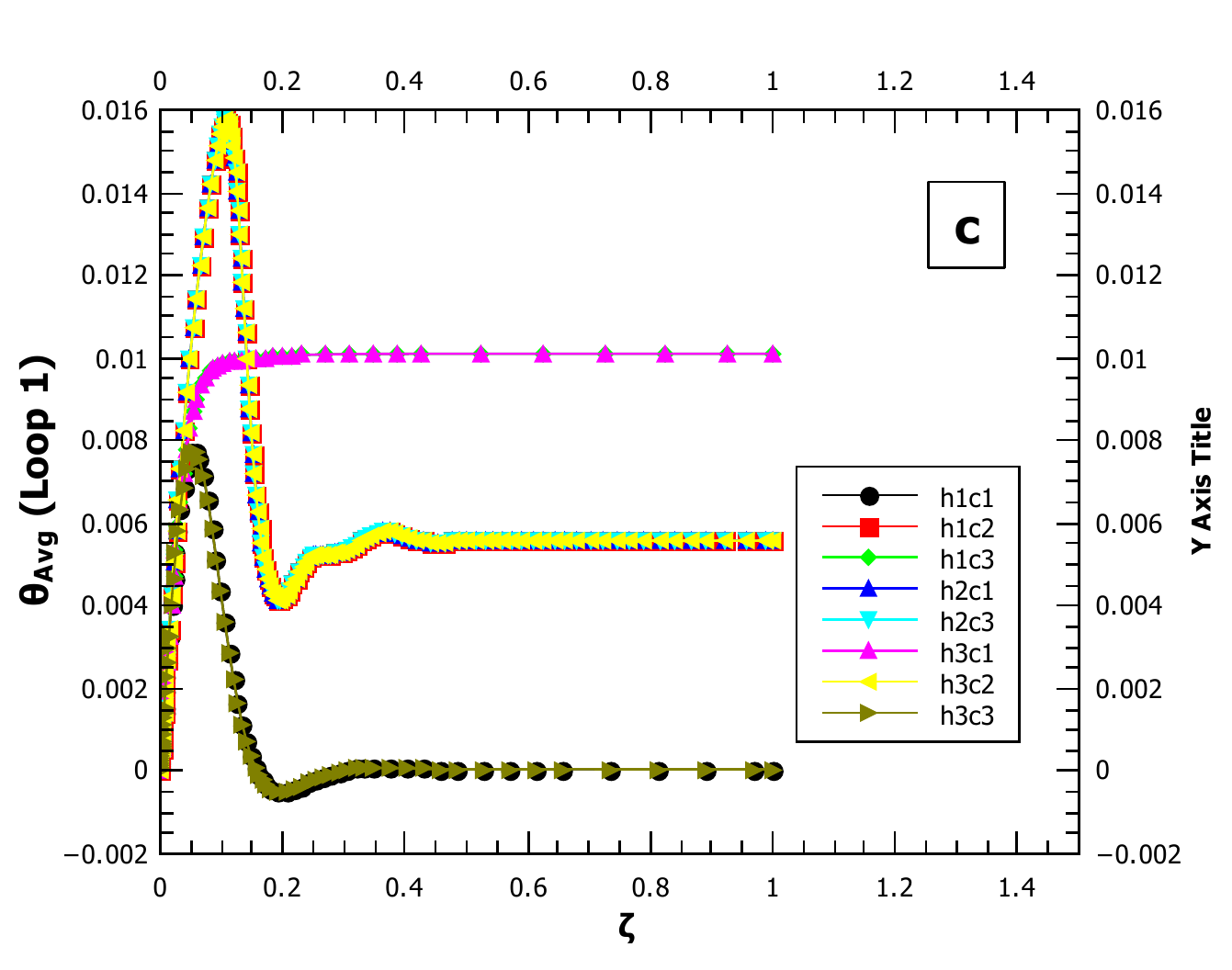}
	\end{subfigure}
	\hspace{\fill}
	\begin{subfigure}[b]{0.49\textwidth}
		\includegraphics[width=1\linewidth]{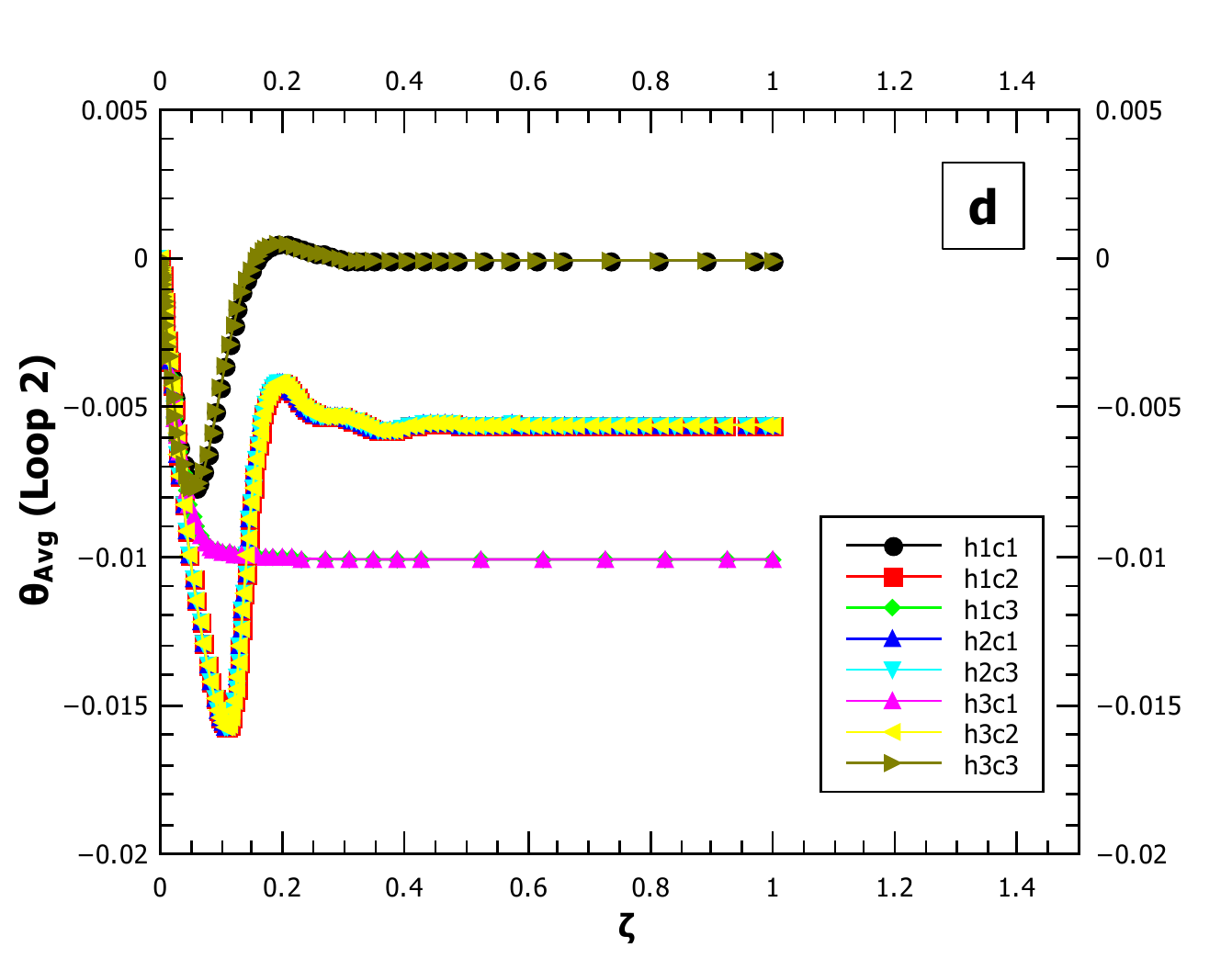}
	\end{subfigure}
	\hspace{\fill}
	\begin{subfigure}[b]{0.49\textwidth}
		\includegraphics[width=1\linewidth]{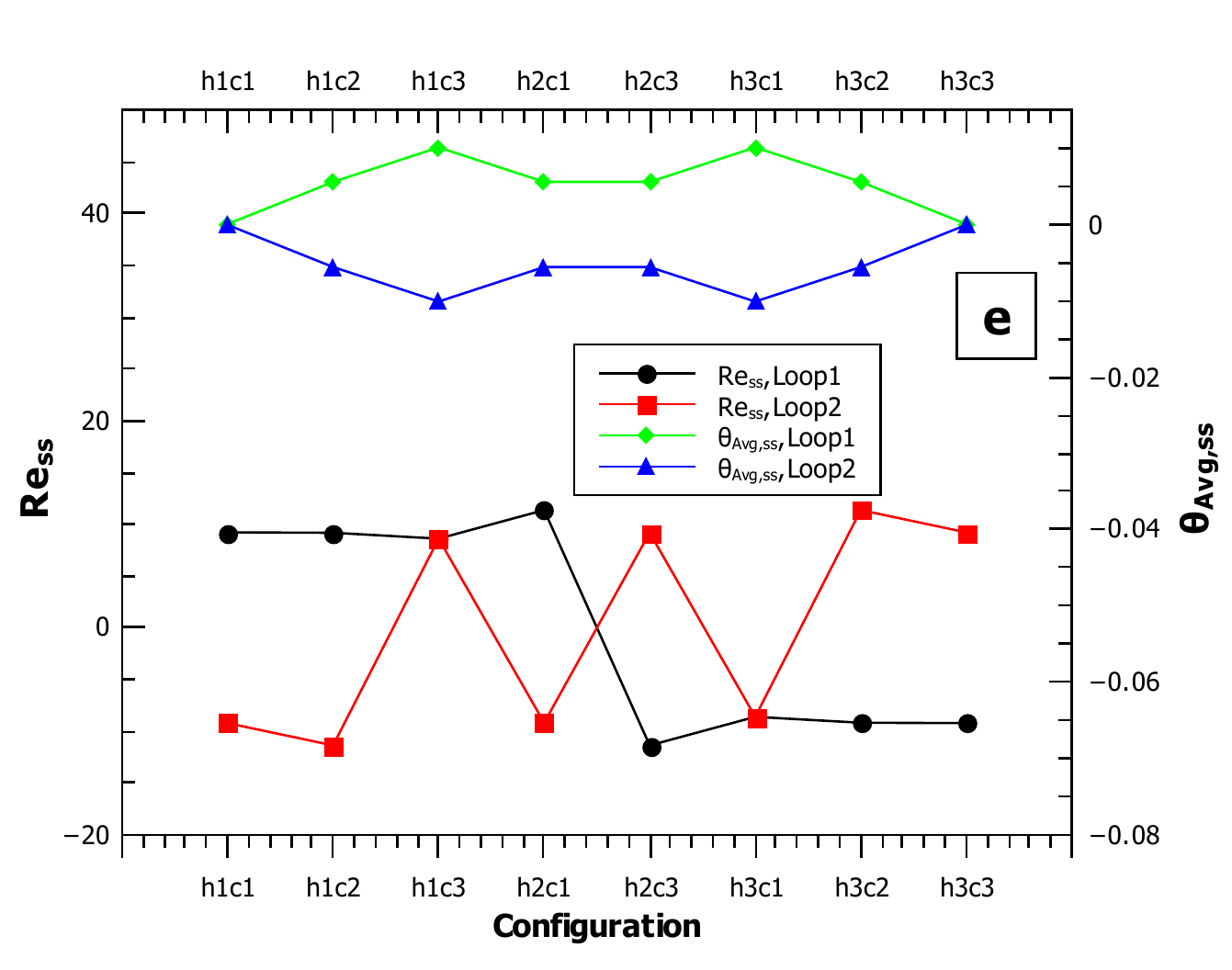}
	\end{subfigure}
	
	\caption{Effect of heater cooler orientation on the Horizontal CNCL system for $Gr$=$10^6$, $Fo$=1, $As$=1, $St$=100, $Co1$=1000.}
	\label{Heater cooler configuration study HCNCL}
\end{figure}

For the study of heater cooler configuration of the HCNCL system, all configurations apart from h2c2 are studied. The h2c2 configuration of the HCNCL system is peculiar as it exhibits both parallel and counter flow depending on the initial conditions.

The following observations can be made from Fig.\ref{Heater cooler configuration study HCNCL}:

\begin{enumerate}
	\item The magnitude of $\theta_{Avg}$ for all the heater-cooler configurations at steady state is symmetric about $\theta_{Avg}(\zeta=0)$, which is equal to zero. This implies ${\theta_{Avg,1,ss}}/{\theta_{Avg,2,ss}}=1$.
	\item The HCNCL systems shifts between the parallel flow and counter flow configuration depending on the heater cooler location on the HCNCL system.
	\item The magnitude of $Re_{ss}$ is not equal in the component loops of the HCNCL system for all the heater cooler configurations. 
	\item Table.\ref{tab:table9} presents the ratio of Reynolds numbers of Loop 1 to Loop 2 at steady state.The sign denotes whether the flow arrangement at the common heat exchanger is parallel flow or counterflow. Negative sign denotes the counterflow arrangement.
	\item From table.\ref{tab:table9} we also observe that :
	
	\begin{equation}
\frac{Re_{1,ss}(h\textbf{i}c\textbf{j})}{Re_{2,ss}(h\textbf{i}c\textbf{j})}\times \frac{Re_{1,ss} (h\textbf{j}c\textbf{i})}{Re_{2,ss} (h\textbf{j}c\textbf{i})}=1
\end{equation}
	
	where `\textbf{i}',`\textbf{j}' represent the position of the heater and cooler for the 'h\textbf{i}c\textbf{j}' configuration.
	
\end{enumerate}

\renewcommand{\arraystretch}{1}
\begin{table}[!h]
	\begin{center}
		\caption{Ratio of \bigg($\frac{Re_{1,ss}}{Re_{2,ss}}\bigg)$ of the HCNCL system for all heater cooler configurations.}
		\label{tab:table9}
		\scalebox{0.9}{
			\begin{tabular}{p{1cm}| p{2cm} p{2cm} p{2cm} } 
				& {\textbf{c1 }} &{\textbf{c2}} & {\textbf{c3}} \\
				\hline 
				{\textbf{h1 }} & -1  & -0.81 & 1\\  
				{\textbf{h2 }} & -1/0.81 & $\pm1$ & -1.24 \\ 
				{\textbf{h3 }} &  1 & -1/1.24 & -1\\
				
		\end{tabular}}
	\end{center}
\end{table}

\section{Conclusions}

Thus far, the literature available on transient single-phase CNCL systems has been focused only on studying simple point contact CNCL which are of little practical value for industrial and engineering requirements. Thus to bridge this gap and fulfill the chosen objectives a 1-D single phase semi-analytical method to model the transient CNCL system having flat plate heat exchanger formed via the coupling of constituent square NCLs which have a square cross-section is proposed. To validate the proposed 1-D model a 3-D transient study is carried out using ANSYS Fluent 16.1 after the grid and time-step independence is established and the CFD methodology is validated. The CNCL with flat plate heat exchanger section is studied for both the vertical and horizontal configurations, thus ensuring a thorough and in-depth study. The following conclusions can be drawn from the present work:

\begin{enumerate}
	\item The 1-D model of the CNCL system is an ideal tool that greatly minimizes the time required to understand the physics of the system, and it is robust enough to handle all the possible heat transfer boundary conditions (constant temperature, heat flux and volumetric heat generation), transient heat inputs and heat loss to the ambient surroundings. The 1-D model accurately predicts the trend of the CNCL system with minimal computational effort, thus the current work greatly simplifies and aids in the design of the CNCL system. Table.\ref{tab:table10} provides a qualitative comparison of the time taken for the 3-D CFD study vs 1-D semi-analytical model. 

\renewcommand{\arraystretch}{1.2}
\begin{table}[!htb]
	\centering
	\caption{Qualitative comparison of the time required for each of the parameters for a 3-D CFD case vs 1-D semi-analytical model.}
	\label{tab:table10}
	\scalebox{0.9}{
		\begin{tabular}{  l  l  l  }
			\hline
			\textbf{Parameter under consideration}     &  \textbf{3-D CFD Study}           &   \textbf{1-D semi analytical model}         \\ \hline
			Geometry creation   &      minimal    &  not required    \\
			Structured mesh generation      &      substantial    &   not required    \\
			Case setup       &      substantial          &   minimal    \\
			Simulation     &      significantly large   &   minimal   \\
			Post processing      &  significantly large    &  substantial \\
			Grid and time step indpendence & significantly large &  minimal \\ \hline
	\end{tabular}}
\end{table}

	\item The non-dimensional numbers - $Gr$,$Re$,$St$,$Fo$,$Co1$,$Co2$ and $As$ - determine the transient dynamics of the CNCL system.
	\item The heater and cooler configurations on both the horizontal and vertical CNCL systems effect the system behavior significantly and the following equation represents the correlation of steady state Reynolds numbers of the component loops of the CNCL system:

\begin{equation}
\frac{Re_{1,ss}(h\textbf{i}c\textbf{j})}{Re_{2,ss}(h\textbf{i}c\textbf{j})}\times \frac{Re_{1,ss} (h\textbf{j}c\textbf{i})}{Re_{2,ss} (h\textbf{j}c\textbf{i})}=1
\end{equation}

		where `\textbf{i}',`\textbf{j}' represent the position of the heater and cooler for the 'h\textbf{i}c\textbf{j}' configuration. 
	 	\item The vertical CNCL always exhibits counterflow arrangement at the common heat exchanger section for all the heater and cooler configurations, but the horizontal CNCL shuffles between the counter and parallel flow depending on the heater-cooler configuration for stable convective flows.
	 	\item The horizontal CNCL with h2c2 heater cooler configuration is the only CNCL system which can exhibit both counter and parallel flow condition at the common heat exchanger section by setting the appropriate initial flow conditions.
	 	\item The CNCL system seems to be very mildly sensitive to the variation in Stanton number, implying that even an un-precise heat transfer correlation can be considered to model the coupling with reasonable accuracy.
	 	\item The limitation of the current modeling approach is that it does not take the wall effects into consideration. This may lead to deviation in the transient prediction, but by modifying the Stanton number to incorporate the conjugate, accurate steady state results can be obtained.
	 	\item The study of chaotic CNCL systems is not considered in the present study and will be part of the future work.
	 	
\end{enumerate}

\clearpage

\section*{\hfil \Large Nomenclature \hfil}

\begin{tabular}{ll}
	$T_{1, Avg}$ & Circuit average temperature of loop 1 ($K$)\\
	$T_{2, Avg}$ &  Circuit average temperature of loop 2 ($K$)\\
	$L$ & CNCL height used for 1-D model ($m$)\\
	$Y$ & CNCL height used for CFD study ($m$)\\
		$L1$ &  CNCL width used for 1-D model ($m$)\\
	$X$ &  CNCL width used for CFD study ($m$)\\
	$x$ &  Distance from origin '$O$'($m$)\\
	$\omega_{1}$ &  Fluid velocity of loop 1 ($m/s$)\\
	$\omega_{2}$ &  Fluid velocity of loop 2 ($m/s$)\\
	$g$ & Gravitational constant ($m/s^2$)\\
	$Q^{\prime\prime}$ & Heat flux ($W/m^2$)\\
		$Q^{\prime \prime}_{CHX}$ & Heat transfer across the common heat exchanger section ($W/m^2$)\\	
	$D_h$ & Hydraulic diameter of both loop 1 \&\ 2 ($m$)\\
	$n$ & No of bends on the component NCL of the CNCL system\\
		$R$ &  Radius of curvature of the bend ($m$)\\
	$U$ & Overall heat transfer coefficient at the heat exchanger section ($W/m^2K$)\\
	$T_{0}$ & Reference temperature of loop 1 \&\ 2 ($K$)\\
	$C_{p}$ & Specific heat capacity ($J/KgK$)\\
	$a$ & Thermal diffusivity ($m^2/s$)\\
	$T_{1}$ & Temperature of Loop 1 ($K$)\\
	$T_{2}$ & Temperature of Loop 2 ($K$)\\
	$t$ &  Time ($s$)\\

\end{tabular}

\section*{Greek letters}

\begin{tabular}{ll}
	$\beta$ & Coefficient of thermal expansion ($1/K$)\\
	$\rho$ & Density ($kg/m^3$)\\
	$\nu$ & Kinematic viscosity ($m^2/s$)\\
	$\rho_0$ & Reference density ($kg/m^3$)\\
	$\kappa$ & Thermal conductivity ($W/(mK)$)\\
	$\tau$ & Wall shear stress exerted on fluid ($Pa$)\\
\end{tabular}

\section*{Non-dimensional numbers}

\begin{tabular}{ll}
	$K$ & 	Bend losses coefficient  ($K=\Delta P_{bend}/(\frac{1}{2}\rho\omega_{i}^2)$)\\
	$f_{F}$ & Fanning friction factor ($f_{F}=\tau_{i}/(\frac{1}{2}\rho\omega_{i}^2)=b/Re^d$)\\
	$Gr_m$ & Modified Grashof number ,Vijayan(2002) \cite{Vijayan2002} \\
    $N_g$ & Geometric parameter ,Vijayan(2002) \cite{Vijayan2002} \\
    
\end{tabular}

\section*{Constants}

\begin{tabular}{ll}
$to$ & ( $t_0={x_0D_h}/{\nu_1} $) \\
$\Delta T$ & ( $\Delta T_i=(4Q^{\prime\prime}t_0)/(\rho_i Cp_i D_h) $) \\
$x_0$ & ( $x_0=(L+L1)$) \\
$b$ & 14.23 (for fully developed flow in laminar regime)\\
$d$ & 1 (for fully developed flow in laminar regime) \\
\end{tabular}

\section*{Non dimensional parameters}

\begin{tabular}{ll}
	$Re$ & Reynolds number ($Re_i=\frac{\omega_iD_h}{\nu_i}$)\\
	$Gr$ & Grashof number ($Gr_i=\frac{g \beta_i \Delta T_i x_0 D_h t_{0,i}}{(L+L1) \nu_i}$)\\
	$Fo$ & Fourier number ($Fo_i=\frac{\alpha_it_{0,i}}{x_0^2}$)\\
	$St$ & Stanton number ($St_i=\frac{Ut_{0,i}}{\rho_i Cp_i D_h}$)\\
	$Co1$ & Non dimensional constant 1  ($Co_1=\frac{2bx_0}{D_h}$)\\
	$Co2$ & Non dimensional constant 2 ($Co_2=\frac{\Delta T_1}{\Delta T_2}$)\\
	$\theta$ & Non dimensional temperature ($\theta_i={T_i-T_0}/{\Delta T_i}$)\\
	$\zeta$ & Non dimensional time ($\zeta={t}/{t_0}$)\\
	$s$ & Non dimensional length ( $s={x}/{x_0}$)\\
\end{tabular}

\section*{Subscripts}

\begin{tabular}{ll}
	$0$ & Any parameter at time $t=0$ $s$ \\
	$1$ & Any parameter referring to Loop 1 \\
	$2$ &  Any parameter referring to Loop 2 \\
	$ss$ &  Any parameter considered at steady state\\
	$Avg$ & Average value of the parameter\\
	$i$ & Refers to subscript `1' or subscript `2' according to relevance\\
\end{tabular}

\section*{Abbreviations}

\begin{tabular}{ll}
	$CFD$ &  Computational Fluid Mechanics \\
	$CNCL$ & Coupled Natural Circulation Loop \\
	$HCNCL$ & Horizontal Coupled Natural Circulation Loop \\
	$NCL$ & Natural Circulation Loop \\
	$PRHRS$ &  Passive Residual Heat Removal system\\
	$VCNCL$ & Vertical Coupled Natural Circulation Loop \\
\end{tabular}

\clearpage

\section{References}

\bibliography{elsarticle-template}

\end{document}